\documentclass[final,3p,times]{elsarticle}

\usepackage{amssymb}

\pdfoutput=1

\journal{X}

\usepackage{lipsum}
\usepackage{amsfonts}
\usepackage{graphicx}
\usepackage{epstopdf}
\usepackage{algorithmic}
\ifpdf
  \DeclareGraphicsExtensions{.eps,.pdf,.png,.jpg}
\else
  \DeclareGraphicsExtensions{.eps}
\fi

\usepackage{natbib}
\usepackage{graphicx}
\usepackage{float}
\usepackage{subfig}
\usepackage{amsopn}
\usepackage{amsmath}
\usepackage{hyperref}
\usepackage{cleveref}
\usepackage{color,xcolor}
\usepackage{algorithm}
\usepackage{ifpdf}
\usepackage{multirow}
\usepackage{makecell}
\usepackage{color,xcolor}
\usepackage{graphicx}
\usepackage{subfig}
\usepackage{subeqnarray}
\usepackage{enumitem}

\newcommand{\dudx}[2]{\frac{\partial{#1}}{\partial{#2}}}
\newcommand{\DuDx}[3]{\frac{D_{#1}{#2}}{D{#3}}}

\newcommand{\dero}[1]{{\textrm{d}{#1}}}

\newcommand{\dvv}[1]{\nabla\cdot\left(#1\right)}
\newcommand{\dvvo}[1]{\nabla\cdot{#1}}
\newcommand{\grad}[1]{\nabla\left(#1\right)}
\newcommand{\grado}[1]{\nabla{#1}}
\newcommand{\vc}[1]{{\mathbf{#1}}}
\newcommand{\bracein}[1]{{\left({#1}\right)}}

\newcommand{\tensor}[1]{{\overline{\overline{#1}}}}

\begin{document}

\begin{frontmatter}



\title{Multi-component multi-scale hydrodynamic plasma flow models in mechanical and thermal disequilibria}

%
%
%
%
%
%
%

\author{Chao Zhang}

\address[org1]{Institute of Applied Physics and Computational Mathematics, Beijing 100094, China}

\begin{abstract}
The present work is motivated by the mixing mechanism in inertial confinement fusion (ICF), which remarkably degrades the ignition performance. The mixing is a direct result of velocity disequilibrium and affected by temperature disequilibrium.
   To deal with these disequilibria, we propose a fully-disequilibrium hydrodynamic model (termed as Baer-Nunziato-Zeldovich model or BNZ model) with 9 equations, 4 temperatures (two for ions, and two for electrons), 4 pressures (two for ions, and two for electrons) and 2 ion velocities for two-component dense plasma flows. The model can be used for describing both grain and atomic mixing by choosing corresponding relaxation mechanism. The derivation starts from a multi-component conservative entropy-dissipative Bhatnagar–Gross–Krook (BGK) model to obtain a 14-equation model. It is then reduced to more practical 9-equation BNZ model and further to simpler 8-equation model, 6-equation model (the Kapila-Zeldovich model or the KZ model), and the 5-equation model. The reductions are performed on the basis of various relaxation time evaluations, the plasma neutrality and the smallness of the electron mass. The models are thermodynamics-compatible, and capable of dealing with mechanical and thermal disequilibria at both atomic and grain scales. Moreover, the corresponding hydrodynamic subsystems are shown to be hyperbolic and solved with the Godunov finite volume method. The BNZ model and solution methods are verified and validated by some benchmark problems and 1D-2D comparisons. We then consider the velocity, pressure and temperature disequilibria during the passage of an ablation shock in ICF.
\end{abstract}

%
\newpage
\begin{keyword}
Multi-component plasma \sep mixing \sep Baer-Nunziato-Zeldovich model \sep Kapila-Zeldovich model
\end{keyword}

\end{frontmatter}


\section{Introduction}
It is well known that the mixing of the ablator into the fuel and the hot spot is one of the most adverse factors that lead to ignition degradation in inertial confinement fusion (ICF) \citep[][]{zylstra2018diffusion}.
Mathematical modelling serves as an important tool to study the physics of mixing and its effect on ignition. However, there is still a noticeable gap in the mathematical model for mixing that incoporates important effects such as inter-penetration, ion-temperature separation and scale variation. On one hand the classical two-fluid (or multi-fluid) plasma model \citep[][]{Braginskii1965ReviewsOP} is only applicable to the case of atomic mixing. They do not apply to the scenario where the plasmas are separated by material interfaces at the macroscopic or mesoscopic grain scale.  On the other hand, up to date, the models for grain mixing with interfaces \citep[][]{murrone2005five,allaire2002five} are only built for hydrodynamic flows under  non-HED (High Energy Density) common conditions and extensions to HED plasma flows are still absent in literature.

\begin{figure}
\centering
{\includegraphics[width=0.4\textwidth]{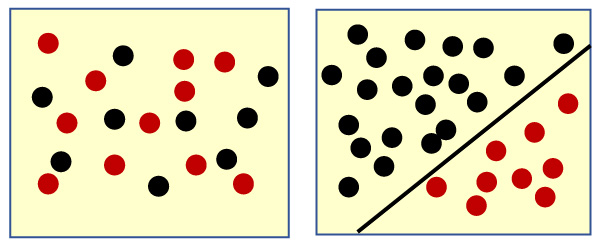}}
\caption{The schematic for atomic (left) and grain mixing (right). In the atomic mixing the particles of different components coexist in any mesoscopic control volume and the parameters of different components relax via particle collisions.  For the grain mixing, in any mesoscopic control volume the components are separated by interfaces that may or may not resolved by the computational grid. The parameters of components on each side of the interface relax by wave propagation and heat conduction in continuum mechanics.}
\label{fig:atomic_grain_mix}
\end{figure}

In our previous works we have extended Kapila's five-equation model to incorporate some transport phenomena such as electron heat conduction, ion viscosity and inter-diffusion \citep[][]{ZhangC_PhysRevE2023,zhang2022a}. These models still  assume electron-ion temperature equilibrium and allow us to consider relatively low-temperature HED plasma flows. 
However, as temperature increases the materials turn into plasma sate, which is typically an electrically quasi-neutral medium of unbound positive ion and negative electron particles (i.e. the overall charge of a plasma is roughly zero). The applicability of these models are questionable since the temperature relaxation time between ions and electrons is large enough to impact hydrodynamics, thermodynamics and nuclear reaction in ICF. The models proposed in this paper will get rid of this assumption.

The present work also investigates the formally unified description of heterogeneous mixing in ICF or other multi-component  HED flows. The concept of heterogeneous mixing consists in the  coexistence of the mixing at the atomic scale as well as the macroscopic/mesoscopic grain scale (Fig. \ref{fig:atomic_grain_mix}).
The atomic mixing consists in the following aspects: first, the plasma itself can be viewed as an atomic mixture of ions and electrons; second, the mass diffusion between different ion species is driven by concentration, temperature, and pressure gradients, especially as temperature significantly increases in the ICF deceleration stage.

The laser/X-ray ablation in ICF or specialized HED experiments (e.g., the MARBLE campaign \citep[][]{haines2020nc,albright2022experimental}) creates huge spatial temperature non-uniformity. The low-temperature region, where the materials (especially high-Z ones) are not or only weakly ionized, allows the mesoscopic grain mixing with materials separated by unresolved interfaces on the computational grid. The dynamics of these unresolved interfaces may have major impact on the mean flow field.


The unified description of these mixing patterns is within the scope of the current work. To derive governing equations for such heterogeneous mixing of plasmas, we introduce the method of multiphase flow to the multi-component plasma flows, i.e., the spatial averaging procedure with the spatial characteristic function as a kernel function \citep[][]{drew1983mathematical,ABGRALL2003361,CHINNAYYA2004490}. We apply the spatial averaging procedure to a multi-component BGK model \citep[][]{Haack2017ACE}, which leads to a fully-disequilibrium multi-material hydrodynamic plasma model with relaxations at the grain interfaces. The derived model can be viewed as two sets of the two-phase Baer-Nunziato model \citep{BAER1986861} for ions and electrons in multi-component plasma flows coupled through the ion-electron relaxations.

The obtained average multi-component plasma model consists of 14 equations (among which 7 for ions and 7 for electrons) in the case of two components. It is formidable to directly solve the fully disequilibrium model from perspective of computational efficiency and algorithm complexity, therefore, reduction is necessary. The reduction is performed on the basis of the evaluation of the characteristic time scales of mechanical/thermal relaxation. The relaxation time scales are evaluated at both the atomic scale and the mesoscopic grain scale under ICF conditions. Our evaluations have shown that  the temperature relaxation times at grain scale maybe comparable to the characteristic times of HED processes. Meanwhile, the mechanical relaxation times are at least 2 orders smaller than temperature relaxation time with the characteristic length 1$\mu$m and temperture less than 1keV. The model formulation is then generalized to the scenario of atomic mixing by defining different relaxation mechanism.

On the basis of the above evaluations, we derive a series of reduced models with different levels of atomic/grain disequilibrium being retained.   Furthermore, thanks to the smallness of the electron mass, the models are further reduced by neglecting its contribution to the plasma mixture mass and momentum  along with the quasi-neutral property of plasma. 

Among these reduced models, the nine-equation model can be reduced to the Baer-Nunziato model in the absence of plasma effect, or to the Zeldovich model\citep[][]{zeldovich1967} in the absence of the multi-component effect. Therefore, this nine-equation model is termed as the Baer-Nunziato-Zeldovich model (or the BNZ model) and is capable of dealing with mechanical and thermal disequilibria. It allows four temperatures including two ion temperatures and two electron temperatures in the case of two components. The BNZ model is then reduced to give birth to a series of simpler models consisting of 8, 6, and 5 equations, respectively. The reduced 6-equation model is reduced to Kapila's five-equation model in non-HED state and to Zeldovich model for one pure component, thus termed as the Kapila-Zeldovich model (KZ model). The simplest one is equipped  with only two equilibrium temperatures.

Apart from that, the inter-penetration effect due to the velocity disequilibrium has also shown to be an important effect for ICF, especially at the deceleration stage \citep[][]{vold2021plasma}. 
At the atomic scale the inter-penetration is supported by various field gradients similar to the Fick's law.
By utilizing the smallness of the Knudsen number $\textrm{Kn}<<1$, we provide a set of one-velocity governing equations with inter-penetration effect.

All these derived models are compatible with existing classical ones. The entropy dissipation of the derived models has been proven to be satisfied with properly defined constitution laws.


For a quick reference, we list the main results as follows
\begin{enumerate}[label=\roman*.]
    \item The 9-equation BNZ model (eq. \ref{eq:nine_eqn_plasma_model})
    \item The 8-equation model (eq. \ref{eq:eight_eq_model})
    \item The 6-equation KZ model (eq. \ref{eq:multi_plasma_six_eqn})
    \item The 5-eqaution model (eq. \ref{eq:multi_plasma_five_eqn})
    \item The 8-equation model with mass diffusion (eq. \ref{eq:reduced_avbn})
    \item The 5-equation model with mass diffusion (eq. \ref{eq:reduced_five_diffusion})
\end{enumerate}

The proposed models are formally a set of hyperbolic-parabolic-relaxation partial differential equations (PDEs). We adopt the fractional step method for its numerical resolution.  With the aid of such method, the model is split into three basic subsystems, i.e., the hyperbolic subsystem, the parabolic subsystem and the relaxation subsystem.  For the numerical resolution of the hyperbolic, we used the finite volume method (FVM) with a specially designed a path-conservative HLLEM Riemann solver\citep[][]{Dumbser2016ANE}{}{}. High-order extensions are also performed with the SSP RK for time integration and the MUSCL scheme for spatial reconstruction. 
The stiff relaxation subsystem is solved with the a stiff ODE solver. We have verified the numerical methods against some benchmark problems. The proposed model together with the numerical methods are then applied to simulate ICF relevant problems.

The rest of the article is organized as follows. In Section \ref{sec:two_fluid_plasma} we re-derive the classical two-fluid plasma model from a multi-component BGK model. In Section \ref{sec:multi_plasma} we derive the multi-component plasma model for grain mixing from the two-fluid plasma model.  The relaxation time scales are then evaluated in Section \ref{sec:relax_time}. On the basis of the evaluations, a series of reduced models is derived in Section \ref{sec:Reduced_model}. The numerical methods for the solution of the proposed model are described in Section 6. Some numerical results are displayed in Section 7. Finally, a concise conclusion is drawn.

\section{The atomic-mixing plasma model}\label{sec:two_fluid_plasma}
In this section we derive the multi-fluid plasma model from a multi-component BGK model. Typically for atomic mixing we are only concerned with three species, i.e., ions of both components and electrons. However, here the electrons of different ions are treated as different species for further extension to the grain mixing scenario in next section. 

Let us consider the multi-species Boltzmann equation in the following BGK formulation \citep[][]{Haack2017ACE}{}{}
\begin{equation}
\dudx{f_r}{t} + \vc{v}\cdot \nabla_{\vc{x}}f_r = \Omega_r[\vc{f}],
\end{equation}
where $r$ is the index of the particle species, $r\in\left\{i1, i2, e1, e2 \right\}$, $f_r$ is the density of particle $r$ in the measure $\textrm{d}\vc{v}\textrm{d}\vc{x}$, and $\vc{f}=(f_{i1},\;f_{i2},\;f_{e1},\;f_{e2})^{\textrm{T}}$. The subscripts $i1,\;i2$ represent the ions of component 1 and 2,  while  $e1,\;e2$ are their electrons.

The collision operator is approximated with the BGK operator as follows
\begin{equation}
\Omega_r[\vc{f}] =  \Omega_r^{BGK}[\vc{f}] = \sum \Omega_{r,j}^{BGK}[\vc{f}] = \sum_{j} \nu_{r,j} \left(  \mathcal{M}_{r,j} - f_{r} \right).
\end{equation}

The Maxwellians are defined as
\begin{equation}
\mathcal{M}_{r,j} = n_r \left( \frac{m_r}{2\pi \overline{T}_{r,j}} \right)^{\frac{3}{2}}  \textrm{exp} \left(  - \frac{m_r \left( \vc{v} - \overline{\vc{u}}_{rj} \right)^2}{2\overline{T}_{rj}} \right).
\end{equation}

The parameters $\overline{T}_{r,j}$ and $\overline{\vc{u}}_{r,j}$ are defined in such a way that the following constraints are satisfied
\begin{subeqnarray}\label{eq:BGK_constraints1}
&&\int \Omega_{r,j}^{BGK} \textrm{d}\vc{v} = 0, \\
&&\int \Omega_{r,j}^{BGK} m_r \vc{v} \textrm{d}\vc{v} + \int \Omega_{j,r}^{BGK} m_j \vc{v} \textrm{d}\vc{v} = 0,  \label{eq:BGK_constraints2}\\
&&\int \Omega_{r,j}^{BGK} m_r |\vc{v}|^2 \textrm{d}\vc{v} + \int \Omega_{j,r}^{BGK} m_j |\vc{v}|^2 \textrm{d}\vc{v} = 0. \label{eq:BGK_constraints3}
\end{subeqnarray}

By performing the Chapman-Enskog expansion, one can derive the multi-fluid plasma equations as follows
\begin{subeqnarray}\label{eq:multifluid_model}
  &&\dudx{n_r}{t} + \dvv{n_r \vc{u}_r} = 0, \\
  &&\dudx{n_r m_r \vc{u}_r}{t} + \dvv{n_r m_r \vc{u}_r \vc{u}_r + {\tensor{P}}_r} = \vc{R}_r, \label{eq:model0:mom}\\
   &&\dudx{n_r m_r E_r}{t} +  \dvv{ n_r m_r \left( E_r + \frac{T_r}{m_r} \right)\vc{u}_r + \overline{\overline{\Pi}}_r \cdot \vc{u}_r  + \vc{q}_r} = Q_r , \label{eq:model0:en}
\end{subeqnarray}
where the variables $n_r,\;\vc{u}_r,\;p_r,\;T_r,\;\overline{\overline{P}}_r,\;\vc{q}_r$ are the number density, the bulk velocity, the pressure, the temperature (in energy unit in this section), the stress tensor and the heat flux of $r$ particles. 

\[\overline{\overline{P}}_r = p_r \tensor{I} + \tensor{\Pi}_r,\] where $\tensor{I}$ , $p_r$ and $\tensor{\Pi}_r$  are the unit tensor, the pressure and the viscous stress tensor, respectively. 

The component total energy $E_r$ and internal energy $\varepsilon_r$ are
\[E_r = \varepsilon_r  + \frac{1}{2} |\vc{u}_r|^2, \quad \varepsilon_r = \frac{3}{2}\frac{T_r}{m_r}.\]


The term $\vc{R}_r$ is the friction term between different particle species.

\begin{eqnarray}\label{eq:Rr}
  \vc{R}_r = \sum_j \int m_r  \nu_{r,j} \vc{v} \left( \mathcal{M}_{r,j} - f_r \right) \textrm{d}\vc{v} = \sum_j {\nu}_{r,j} n_r m_r \left( \overline{\vc{u}}_{r,j} - \vc{u}_r\right).
\end{eqnarray}

The heat exchange term ${Q}_r$ is
\begin{eqnarray}\label{eq:Qr}
  {Q}_r &=& \sum_j \int  \nu_{r,j} m_r  \frac{|\vc{v}|^2}{2}  \left( \mathcal{M}_{r,j} - f_r \right) \textrm{d}\vc{v}, \nonumber\\
  &=& \frac{3}{2} \sum_j {\nu}_{r,j} n_r \left( \overline{{T}}_{r,j} - T_r\right) + \frac{1}{2}\sum_j \nu_{r,j} n_r m_r \left( \overline{\vc{u}}_{r,j} + \vc{u}_r \right) \left( \overline{\vc{u}}_{r,j} - \vc{u}_r \right),
\end{eqnarray}
where $\overline{\vc{u}}_{r,j}$ is a mass-weighted average of ${\vc{u}}_{r}$ and ${\vc{u}}_{j}$, while $\overline{T}_{r,j}$ is a special average of ${T}_{r}$ and ${T}_{j}$ such that the constraints (\ref{eq:BGK_constraints1}) are satisfied.

They are given in \cite{Haack2017ACE} as follows
\begin{subeqnarray}
\overline{\vc{u}}_{r,j} &=& \frac{n_{r}m_r\nu_{r,j}\vc{u}_r + n_{j}m_j\nu_{j,r}\vc{u}_j}{n_{r}m_r\nu_{r,j} + n_{j}m_j\nu_{j,r}}, \\
\overline{T}_{r,j} &=& \frac{n_{r}\nu_{r,j}T_r + n_{j}\nu_{j,r}T_j}{n_{r}\nu_{r,j} + n_{j}\nu_{j,r}} \nonumber\\ &+& \frac{n_{r}m_r\nu_{r,j} (|\vc{u}_r|^2 - |\overline{\vc{u}}_{r,j}|^2) + n_{j}m_j\nu_{j,r} (|\vc{u}_j|^2 - |\overline{\vc{u}}_{r,j}|^2)}{3 (n_r\nu_{r,j} + n_j\nu_{j,r})}.
\end{subeqnarray}

%
%

The following variables, i.e., mass density, pressure, and specific internal energy, can be defined as
\begin{equation}\label{vars}
  \widehat{\rho}_r = m_r n_r, \;\; \widehat{p}_r = n_r T_r, \;\; \widehat{\varepsilon}_r = \frac{3}{2} n_r T_r ,
\end{equation}
where the over-hat is used to show the difference of variables from the grain mixing case that follows below.

With these definitions, one can further write the following familiar hydrodynamic equations
\begin{equation}\label{eq:dUdFS}
\dudx{\vc{U}}{t} + \dvv{\vc{F}} = \vc{S},
\end{equation}
where the conservative, flux and source vectors are defined as follows
\[\vc{U} = [\widehat{\rho}_r, \; \widehat{\rho}_r\hat{\vc{u}}_r, \; \widehat{\rho}_r \hat{E}_r]^{\textrm{T}},\quad\vc{F} = [\widehat{\rho}_r\widehat{\vc{u}}_r, \; \widehat{\rho}_r \widehat{\vc{u}}_r \widehat{\vc{u}}_r + \widehat{\tensor{P}}_r, \; \widehat{\rho}_r \widehat{E}_r \widehat{\vc{u}}_r + \widehat{\tensor{P}}_r\cdot\widehat{\vc{u}}_r + \widehat{\vc{q}}_r]^{\textrm{T}},\]
\[\vc{S} = [0, \; \widehat{\vc{R}}_r, \;\widehat{Q}_r]\]
and the stress tensor $\widehat{\tensor{P}}_r = \widehat{p}_r\tensor{I} + \widehat{\tensor{\Pi}}_r.$

The equation for internal energy can be obtained by using eq. (\ref{eq:multifluid_model}), which reads
\begin{eqnarray}\label{eq:multifluid_inten}
   \dudx{\widehat{\rho}_r \widehat{\varepsilon}_r}{t} +  \dvv{\widehat{\rho}_r \widehat{\varepsilon}_r \widehat{\vc{u}}_r} + \left( \widehat{\tensor{P}}_r \cdot \nabla \right) \cdot {\widehat{\vc{u}}_r} + \dvvo{\widehat{\vc{q}}_r} &=& \widehat{Q}_r^{\varepsilon},
\end{eqnarray}
where

\begin{eqnarray}
\widehat{Q}_r^{\varepsilon} = \widehat{Q}_r - \widehat{\vc{u}}_r \cdot \widehat{\vc{R}}_r = \frac{3}{2}\sum_{j} \widehat{\mu}^{T}_{r,j} \left( T_j -T_r \right) + \widehat{Q}_r^f,
\end{eqnarray}
where
\begin{eqnarray}
&&\widehat{\mu}^{T}_{r,j} = \frac{n_r\nu_{r,j}n_j\nu_{j,r}}{n_r \nu_{r,j} + n_j \nu_{j,r}}, \label{eq:nu_T_rj} \\
&&\widehat{Q}_r^f = \frac{1}{2}\sum_j \frac{n_j m_j \nu_{j,r} \; n_r m_r \nu_{r,j}}{n_j m_j \nu_{jr} + n_r m_r \nu_{r,j}}\left( \frac{n_r \nu_{r,j}}{n_r \nu_{r,j} + n_j \nu_{j,r}} + \frac{n_r m_r \nu_{r,j}}{n_r m_r \nu_{r,j} + n_j m_j \nu_{j,r}} \right) |\widehat{\vc{u}}_j - \widehat{\vc{u}}_r|^2. \nonumber \\\label{eq:def_Qrf}  
\end{eqnarray}

The non-negative term $\widehat{Q}_r^f$ represents the heat generation due to frictions between particles.  

\section{The grain-mixing plasma model}\label{sec:multi_plasma}
\subsection{The grain-size mixing model}
\begin{figure}
\centering
{\includegraphics[width=0.6\textwidth]{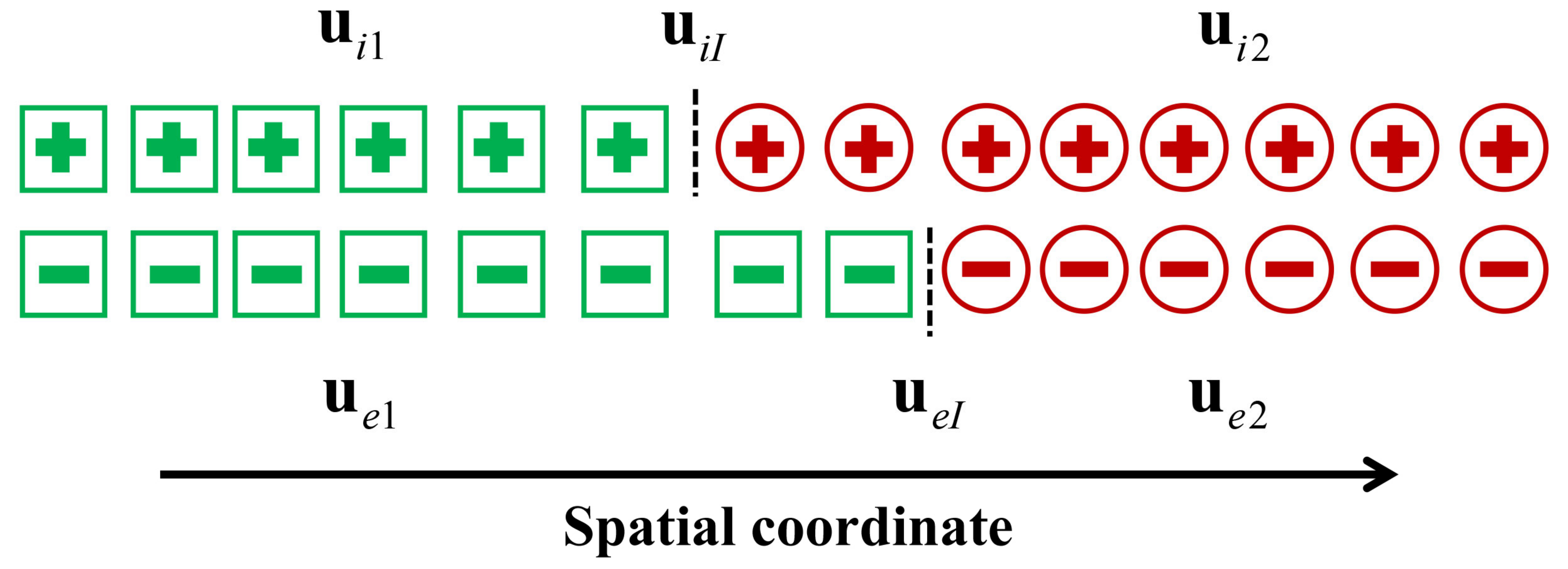}}
\caption{The grain mixing between plasmas of different materials. The ions and electrons are displayed on the top and bottom, respectively. Interfaces are represented by dashed lines.}
\label{fig:mix_on_grid}
\end{figure}

In this section we the consider the grain mixing of two plasmas. This type of mixing may appear in the early stage of ICF. The mixing topology may take various forms, e.g., stratified flow or dilute/dense dispersed flow. Whatever the topology, there exist macroscopic or mesoscopic material interfaces that may or may not be resolved by the computational grid.
In such scenario the distribution function $f_r$ is discontinuous in space, as displayed in Figure \ref{fig:mix_on_grid}. To describe the discontinuous flows in the whole domain, we introduce the characteristic function, which is a Heaviside function as follows
\[X_r(\vc{x},t) = 1,  \textrm{ if the point } \vc{x} \textrm{ is in species } r,\]
\[X_r(\vc{x},t) = 0,  \textrm{ otherwise. }\]
The characteristic function obeys
\begin{equation}
\dudx{X_r}{t} + \vc{u}_{rI}^{\sigma} \cdot \grado{X_r} = 0, 
\end{equation}
where $\vc{u}_{rI}^{\sigma}$ is the local interface velocity. It depends on the mean velocities $\vc{u}_{r}$ and    $\vc{u}_{r^{\prime}}$ with the species $r^{\prime}$ on the other side of the interface. 

 Then we can integrate the BGK equation in the following form
\begin{equation}\label{eq:int_int_BGK}
\int X_r \int \phi(\vc{v}) \left( \dudx{f_r}{t} + \vc{v}\cdot \nabla_{\vc{x}}f_r \right)\textrm{d}{\vc{v}} \textrm{d}{\vc{x}} = \int X_r \int \phi(\vc{v}) \Omega_r \textrm{d}{\vc{v}}\textrm{d}{\vc{x}},
\end{equation}
where we have omitted the superscript $BGK$ over the operator $\Omega_r$ without causing confusion. Such spatial averaging procedure is usually applied to the Euler equations to derive the multi-phase flow models \citep[][]{drew1983mathematical,CHINNAYYA2004490,ABGRALL2003361}. However, here we apply this procedure to the more fundamental BGK model since it directly gives the formulation of relaxation coefficients in grain/atomic mixing.

We assume that the electrons and ions coexist at the same spatial point and both of them occupy the whole spatial domain, thus, we have
\begin{equation}
X_{i1} + X_{i2} = 1, \quad X_{e1} + X_{e2} = 1.
\end{equation}

Therefore, only two independent characteristic functions $X_{e1}$ and $X_{i1}$ are needed. Moreover, only one local interface velocity exists for ions (or for electrons), i.e., $\vc{u}_{l1,I}^{\sigma}=\vc{u}_{l2,I}^{\sigma}=\vc{u}_{lI}^{\sigma}$ for the sake of the maximum principle.

For convenience in the separate treatment of ions and electrons,   we use double subscript $lk$ to denote the ions($l=i$)/electrons($l=e$) of the first component ($k=1$) or the second ($k=2$) in the following. We also use conjugate subscripts $k$ or $k^{\prime}$, i.e., $k,k^{\prime}\in \{ 1,2\}$ and $k\neq k^{\prime}$. Similar conjugate subscripts are defined as $l,l^{\prime}\in \{ i,e\}$ and $l\neq l^{\prime}$.

Here, the collision operator is also spatially discontinuous. The particle $i1$ may collide  with the particles $e1$, $e2$ and its own species inside the domain it occupies and with $i2$ at the interface. The collision operator $\Omega_{lk}$ is a function of $f_r$ and $X_r$. It can be expressed as 


\begin{eqnarray}\label{eq:omega_lk}
\Omega_{lk} = \Omega_{lk,lk} 
+ \underbrace{\sum_{j=k,k^{\prime}}X_{l^{\prime}j} \Omega_{lj,l^{\prime}j}}_{\textrm{(1)}i-e \textrm{ volume collision}}
+ \underbrace{X_{lk^{\prime}} \Omega_{lk,lk^{\prime}}}_{\textrm{(2)}i-i/e-e \textrm{ volume collision}}
+  \underbrace{\vc{n}\cdot\nabla X_{lk^{\prime}}  \Omega_{lk,lk^{\prime}}^{A}}_{\textrm{(3)} i-i/e-e \textrm{ interface collision}}
\end{eqnarray}

The term $X_{lk^{\prime}} \Omega_{lk,lk^{\prime}}$ always vanishes when substituting into  eq. (\ref{eq:int_int_BGK}) since $X_{lk}X_{lk^{\prime}} = 0$ for grain mixing. 
The term $\Omega_{lk,lk^{\prime}}^{A}$ represents the change rate of $f_{lk}$ due to collision on a unit area of the interface. The  term $\vc{n} \cdot \nabla X_{lk^{\prime}}  \Omega_{lk,lk^{\prime}}^{A}$ is the effective collision term per volume in the spirit of the continuum surface model \citep[][]{BRACKBILL1992335}. The vector $\vc{n}$ is the average local normal to the interface.

We first perform the Chapman-Enskog expansion to obtain a set of hydrodynamic equations for each component. Then by applying the spatial averaging procedure \citep[][]{drew1983mathematical,CHINNAYYA2004490}, one can obtain the following equations

\begingroup
\allowdisplaybreaks
\begin{subeqnarray}
\label{eq:14_eqn_plasma_model}
&&\dudx{\alpha_{lk}\rho_{lk}}{t} +  \dvv{\alpha_{lk}\rho_{lk}\vc{u}_{lk}} = 0, \\
&&\dudx{\alpha_{lk}\rho_{lk}\vc{u}_{lk}}{t} + \dvv{\alpha_{lk}\rho_{lk} \vc{u}_{lk} \vc{u}_{lk}}  + \dvv{\alpha_k \overline{\overline{P}}_{lk}} =  \underbrace{{\color{black}\alpha_{lk}\vc{R}_{lk}}}_{(1)} +\underbrace{{\color{black}{\overline{\overline{P}}}_{lI}\cdot\grado{\alpha_{lk}}} + {\color{black}\mathbf{M}_{lk}}}_{(3)}, \\
&&\dudx{\alpha_{lk}\rho_{lk} E_{lk}}{t} + \dvv{\alpha_{lk} \rho_{lk} E_{lk} \vc{u}_{lk}} + \dvv{\alpha_{lk}\overline{\overline{P}}_{lk}\cdot\vc{u}_{lk}} + \dvv{\alpha_{kl}\vc{q}_{lk}} = \nonumber\\ &&\underbrace{{\color{black}\alpha_{lk}Q_{lk}}}_{(1)}
  + \underbrace{{\color{black} {\color{black}\vc{u}_{lI}} \cdot\left({\overline{\overline{P}}}_{lI}\cdot\grado{\alpha_{lk}}\right)} + {\color{black}\vc{q}_{lI}\cdot \nabla\alpha_k} - {\color{black}\tilde{p}_{lI}\mathcal{P}_{lk}} + {\color{black}\tilde{\vc{u}}_{lI}\cdot\mathbf{M}_{lk}} + {\color{black}\mathcal{Z}_{lk}}}_{(3)}, \\
&&\dudx{\alpha_{lk}}{t} +\underbrace{{\color{black}\vc{u}_{lI} \grado{\alpha_{lk}}}}_{(3)} = \underbrace{{\color{black}\mathcal{P}_{lk}}}_{(3)},
\end{subeqnarray}
\endgroup
where we have denoted the correspondence between right-hand-side terms with the collision operator in eq. (\ref{eq:omega_lk}). 

The volume fraction is defined as the volume integration of the characteristic function $X_{lk}$,
\begin{equation}\label{eq:def_alpha}
\alpha_{lk} = \frac{1}{V} \int_{V} X_{lk} \textrm{d}V.
\end{equation}

The phase variables are defined as
\begin{equation}\label{eq:def_phaseV}
\Phi_{lk} = \frac{1}{V}\int_{V} X_{lk} \Phi_{l} \textrm{d}V, \quad \Phi = \rho, \rho\vc{u}, \rho E, \tensor{P}.
\end{equation}

The subscript ``I'' denotes the interfacial variable.  A fundamental problem consists in the definition of these interfacial variables that depend on the material properties, flow field and phase connectivity. 
The average interface velocity is 
\[\vc{u}_{lI} = \frac{1}{V}\int_{V} \vc{u}_{lI}^{\sigma} \textrm{d}V. \]
The simplest definition is 
\begin{equation}\label{eq:BN_pI_uI}
    \mathbf{u}_{lI} = \tilde{\mathbf{u}}_{lI} = \mathbf{u}_{l2}, \; p_{lI} = \tilde{p}_{lI}= p_{l1}, 
\end{equation}
where material 1 and 2 correspond to the more compressible and less compressible material. This definition leads to the Baer-Nunziato (BN) model \citep[][]{BAER1986861} for each species without the (1) terms and viscous terms. 

A more complicated definition is derived by approximating the Riemann problem at the sub-cell scale with the acoustic  Riemann solver, which gives \citep[][]{CHINNAYYA2004490,saurel2018diffuse}
\begin{equation}\label{eq:int_vel}
\mathbf{u}_{lI} =  \tilde{\mathbf{u}}_{lI} + \textrm{sgn}\left(\grado{\alpha_{l1}}\right)\frac{p_{l2}-p_{l1}}{Z_{l1} + Z_{l2}}, \quad \tilde{\mathbf{u}}_{lI} = \frac{Z_{l1} \mathbf{u}_{l1} + Z_{l2} \mathbf{u}_{l2}}{Z_{l1} + Z_{l2}}.
\end{equation}

\begin{equation}\label{eq:int_pres}
{p}_{lI} = \tilde{p}_{lI}  + \textrm{sgn}\left(\grado{\alpha_{l1}}\right) \cdot \frac{Z_{l1}Z_{l2} \left( {\mathbf{u}}_{l2}- {\mathbf{u}}_{l1} \right)}{Z_{l1} + Z_{l2}}, \quad \tilde{p}_{lI} = \frac{Z_{l1} p_{l2} + Z_{l2} p_{l1}}{Z_{l1} + Z_{l2}}.
\end{equation}

Both of these definitions ensure the entropy production of the hydrodynamic part. Moreover, the BN definition ensures the linear degeneracy of the material interface, and thus yields a unique jump condition. For the study on the interface variable definitions, also see \cite{gallouet2004numerical}. 

The velocity, pressure, temperature interface relaxation terms
\begin{subeqnarray}\label{eq:int_vel_relax}
\mathbf{M}_{lk} = \mu^{u}_{lk,lk^{\prime}} \left( \vc{u}_{lk^{\prime}} - \vc{u}_{lk} \right),
\\\label{eq:int_pres_relax}
\mathcal{P}_{lk} = \mu^{p}_{lk,lk^{\prime}} \left( {p}_{lk} - {p}_{lk^{\prime}} \right),
\\\label{eq:int_temp_relax}
\mathcal{Z}_{lk} = \mu^{T}_{lk,lk^{\prime}} \left( {T}_{lk^{\prime}} - {T}_{lk} \right),
\end{subeqnarray}
where the coefficients $\mu^{u}_{lk,lk^{\prime}}$, $\mu^{p}_{lk,lk^{\prime}}$, $\mu^{T}_{lk,lk^{\prime}}$ are the velocity, pressure, temperature relaxation rates at the grain-scale interface. Moreover, it has been shown that the interface heat flow ${\color{black}\vc{q}_{lI}\cdot \nabla\alpha_k}$ can be incorporated into the $\mathcal{Z}_{lk}$ with a modified $\mu^{T}_{lk,lk^{\prime}}$ \citep[][]{petitpas2014discrete}.

We have the following constraints
\begin{equation}\label{eq:constraints}
  \sum_{k} \alpha_{lk} = 1,\;\; \sum_{k} \mathbf{M}_{lk} = 0,\;\; \sum_{k} \mathcal{P}_{lk} = 0,\;\; \sum_{k} \mathcal{Z}_{lk} = 0.
\end{equation}

The (1) collision relaxation terms are
\begin{eqnarray}\label{eq:Rlk}
\vc{R}_{lk} &=&  \sum_{j={k,k^{\prime}}} \alpha_{l^{\prime}j} \int m_{lk}  \nu_{lk,l^{\prime}j} \vc{v} \left( \mathcal{M}_{lk,l^{\prime}j} - f_{lk} \right) \textrm{d}\vc{v}
\nonumber\\&=& \sum_{j={k,k^{\prime}}} \alpha_{l^{\prime}j} \nu_{lk,l^{\prime}j} n_{lk} m_{lk} \left( \overline{\vc{u}}_{lk,l^{\prime}j} - \vc{u}_{lk}\right).
\end{eqnarray}
\begin{eqnarray}\label{eq:Qlk}
  {Q}_{lk} &=& \sum_{j={k,k^{\prime}}} {\alpha_{l^{\prime}j}} \int  \nu_{lk,l^{\prime}j} m_{lk}  \frac{|\vc{v}|^2}{2}  \left( \mathcal{M}_{lk,l^{\prime}j} - f_{lk} \right) \textrm{d}\vc{v} \nonumber\\
 & =& \frac{3}{2} \sum_{j={k,k^{\prime}}} {\alpha_{l^{\prime}j}} {\nu}_{lk,l^{\prime}j} n_{lk} \left( \overline{{T}}_{lk,l^{\prime}j} - T_{lk}\right) \nonumber\\
 &+& \frac{1}{2}\sum_j {\alpha_{l^{\prime}j}} \nu_{lk,l^{\prime}j} n_{lk} m_{lk} \left( \overline{\vc{u}}_{lk,l^{\prime}j} + \vc{u}_{lk} \right) \left( \overline{\vc{u}}_{lk,l^{\prime}j} - \vc{u}_{lk} \right).
\end{eqnarray}

From the total energy equation and the momentum equation, one can derive the internal energy equation
\begin{eqnarray}\label{eq:alpha_rho_ep_lk}
 \dudx{\alpha_{lk}\rho_{lk} {\varepsilon}_{lk}}{t} + \dvv{\alpha_{lk} \rho_{lk} {\varepsilon}_{lk} \vc{u}_{lk}}
 + \left( \alpha_{lk}\overline{\overline{P}}_{lk} \cdot \nabla \right) \cdot\vc{u}_{lk} + \dvv{\alpha_{kl}\vc{q}_{lk}} = \nonumber\\ {\color{black}{\alpha_{lk}}{Q}_{lk}^{\varepsilon}} + {\color{black}p_{lI}\left(\vc{u}_{lI} - \vc{u}_{lk}\right)\cdot\grado{\alpha_{lk}}+ {\color{black}\left(\vc{u}_{lI} - \vc{u}_{lk} \right)} \cdot\left({\overline{\overline{\Pi}}}_{lI}\cdot\grado{\alpha_{lk}}\right)} + {\color{black}\vc{q}_{lI}\cdot \nabla\alpha_k} \nonumber\\ - {\color{black}\tilde{p}_{lI}\mathcal{P}_{lk}} + {\color{black}\left(\tilde{\vc{u}}_{lI} - \vc{u}_{lk}\right)\cdot\mathbf{M}_{lk}} + {\color{black}\mathcal{Z}_{lk}},
\end{eqnarray}
{\color{black}where
\begin{eqnarray}
{Q}_{lk}^{\varepsilon} = Q_{lk} - \vc{R}_{lk}\cdot\vc{u}_{lk}\nonumber\\
= \sum_{j={k,k^{\prime}}} \left[\alpha_{l^{\prime}j} \widehat{\mu}^{T}_{lk,l^{\prime}j} \left( {T}_{l^{\prime}j} - {T}_{lk}\right) + \alpha_{l^{\prime}j}{Q}_{lk}^{f}\right].
\end{eqnarray}
}

For further analysis we need the equations with respect to the primitive variables.
The material derivative of a variable $\Phi$ related to the velocity $\vc{u}_m \;(m=lk,lI)$ is
\[ \frac{\textrm{D}_m \Phi }{\textrm{D} t} = \dudx{\Phi}{t} + \vc{u}_m \cdot \nabla \Phi . \]

The multi-component plasma model can be recast into the following form with respect to the primitive variables $\left[ s_{lk}, \vc{u}_{lk}, p_{lk}, \alpha_{lk} \right]$
\begin{subeqnarray}\label{eq:grainModel}
  \alpha_{lk} \rho_{lk} T_{lk} \frac{\textrm{D}_{lk} s_{lk}}{\textrm{D} t} &=& \left(\tilde{\vc{u}}_{lI}-\vc{u}_{lk}\right) \cdot \mathbf{M}_{lk} + \left(p_{lk}-\tilde{p}_{lI}\right) {\mathcal{P}}_{lk} \nonumber\\
  &+&\left(p_{lI}-p_{lk}\right)\left(\vc{u}_{lI}-\vc{u}_{lk}\right) \cdot \nabla \alpha_{lk}  \nonumber\\ &+& \left( \vc{u}_{lI} -\vc{u}_{lk} \right) \cdot \left(  \overline{\overline{\Pi}}_{lI} \cdot \nabla \alpha_k \right) + \mathcal{G}_{lk}, \label{eq:sk}\\
  \alpha_{lk} \rho_{lk} \frac{\mathrm{D}_{lk} \vc{u}_{lk}}{\mathrm{D} t} &=& - \nabla\cdot \left( \alpha_{lk} \overline{\overline{P}}_{lk} \right) + \overline{\overline{P}}_{lI} \cdot \nabla \alpha_{lk} + {\mathbf{M}}_{lk} + \alpha_{lk} {\mathbf{R}}_{lk} \label{eq:uk},\\
  \alpha_{lk}\frac{\mathrm{D}_{lk} p_{lk}}{\mathrm{D} t} &=&  - \alpha_{lk} \rho_{lk} a_{lk}^2 \nabla\cdot \vc{u}_k + \Gamma_{lk}\left(\tilde{\vc{u}}_{lI}-\vc{u}_{lk}\right) \cdot \mathbf{M}_{lk} \nonumber\\ &-& \left[\rho_{lk} a_{lI}^2 + \Gamma_{lk}\left(
  p_{lI} - \tilde{p}_{lI}\right) \right] {\mathcal{P}}_{lk}   
 - \rho_{lk} a_{lI}^2 \left(\vc{u}_{lI}-\vc{u}_{lk}\right) \cdot \nabla \alpha_{lk} \nonumber\\ &+& \Gamma_{lk} \left( \vc{u}_{lI} -\vc{u}_{lk} \right) \cdot \left(  \overline{\overline{\Pi}}_{lI} \cdot \nabla \alpha_k \right) + \Gamma_{lk} \mathcal{G}_{lk}\label{eq:pk},\\
  \frac{\mathrm{D}_{lI} \alpha_{lk}}{\mathrm{D} t} &=& \mathcal{P}_{lk}. \label{eq:alpk}
\end{subeqnarray}

The Grüneisen coefficient
\[\Gamma_{lk} = \frac{1}{\rho_{lk}} \dudx{p_{lk}}{e_{lk}} \vert _{\rho_{lk}}. \]

The adiabatic exponent
\[\gamma_{lk} = \frac{\rho_{lk} a_{lk}^2}{p_{lk}}. \]

We define interface sound velocity $a_{Ik}$ as follows
\[{\rho_{lk} a_{Ik}^{2}} = \gamma_{lk} p_{lk} + \Gamma_k \left({p_{lI}-p_{lk}}\right)= {\rho_{lk} a_{lk}^{2}} + \Gamma_k \left({p_{lI}-p_{lk}}\right).\]

The entropy generation part is gathered into $\mathcal{G}_{lk}$, i.e.,
\[\mathcal{G}_{lk} = \mathcal{S}_{lk} + \mathcal{Z}_{lk} + {\alpha_{lk}}\mathcal{Q}_{lk}^{\varepsilon} + q_{lk}, \]
\[q_{lk} = -\dvv{\alpha_{kl}\vc{q}_{lk}}, \quad \mathcal{S}_{lk} = - \left( \alpha_{lk} \overline{\overline{\Pi}}_{lk} \cdot \nabla \right) \cdot \vc{u}_{lk}.\]

The mixture plasma entropy equation is
\begin{eqnarray}
&&\sum_{l,k} \left[ \dudx{\alpha_{lk} \rho_{lk} s_{lk}}{t} + \nabla\cdot\left( \alpha_{lk} \rho_{lk} s_{lk} \mathbf{u}_{lk} \right) + \nabla\cdot\left( \frac{\mathbf{q}_{lk}}{T_{lk}} \right) \right] \nonumber\\
&&= \frac{\mu^{T}_{i1,i2}\left( T_{i1} - T_{i2} \right)^2}{T_{i1} T_{i2}} + \frac{\mu^{T}_{e1,e2}\left( T_{e1} - T_{e2} \right)^2}{T_{e1} T_{e2}} \nonumber\\
&&+ \sum_k \frac{\alpha_{i1}\alpha_{ek}\widehat{\mu}^{T}_{i1,ek}\left( T_{i1} - T_{ek} \right)^2}{T_{i1} T_{ek}} + \sum_k \frac{\alpha_{i2}\alpha_{ek}\widehat{\mu}^{T}_{i2,ek}\left( T_{i2} - T_{ek} \right)^2}{T_{i2} T_{ek}} \nonumber\\
&&+ \sum_{l,k} \frac{1}{T_{lk}}\left[ \left(\tilde{\vc{u}}_{lI}-\vc{u}_{lk}\right) \cdot \mathbf{M}_{lk} + \left(p_{lk}-\tilde{p}_{lI}\right) {\mathcal{P}}_{lk} \right] \nonumber\\
&&+ \sum_{l,k} \frac{1}{T_{lk}}\left[ \left(p_{lI}-p_{lk}\right)\left(\vc{u}_{lI}-\vc{u}_{lk}\right) \cdot \nabla \alpha_{lk}  + \left( \vc{u}_{lI} -\vc{u}_{lk} \right) \cdot \left(  \overline{\overline{\Pi}}_{lI} \cdot \nabla \alpha_{lk} \right) \right] \nonumber\\
&&+ \sum_{l,k} \left[  \mathbf{q}_{lk} \cdot  \nabla \left(\frac{1}{T_{lk}} \right) +  \frac{\mathcal{S}_{lk}}{T_{lk}} + \alpha_{lk} {Q}_{lk}^{f}  \right].
\end{eqnarray}

An important criterion in building multi-phase flow models is to define the interface variables that ensure non-negativity of the right hand side terms, as mentioned above. 
The BN-type model provides definition for interface pressure $p_I$ and interface velocity $\vc{u}_I$ that ensure the non-negativity except the term with $\tensor{\Pi}_{lI}$. 

As for the definition of the interface viscous stress $\tensor{\Pi}_{lI}$, one possible definition is 
\begin{equation}\label{eq:PI_I}
    \tensor{\Pi}_I = f \left( \vc{u}_I - \vc{u}_k\right) \nabla \alpha_{lk}, 
\end{equation}
where $f$ is a positive scalar depends on the set of objective quantities. This definition makes the term $\left(\vc{u}_{lI} -\vc{u}_{lk} \right) \cdot \left(  \overline{\overline{\Pi}}_{lI} \cdot \nabla \alpha_{lk} \right)$ non-negative.  Note that the tensor $\left( \vc{u}_I - \vc{u}_k\right) \nabla \alpha_{lk}$ is a frame-indifferent quantity \citep[][]{drew1983mathematical}. This term will be cancelled out in the velocity equilibrium models.

\subsection{Reformulation for atomic mixing}
In fact in the case of atomic mixing one typically only needs three distribution functions, i.e., $f_{i1}$, $f_{i_2}$ and $f_e$. However, thanks to its additivity, here we still split $f_e = f_{e1} + f_{e2}$ in order to conform to the grain-mixing governing equations.
Since all particles occupy the whole domain in the case of atomic mixing, the collision operator should be
\begin{equation}\label{eq:omega_lk_atom}
\Omega_{lk} = \Omega_{lk,lk} +  \sum_{j=k,k^{\prime}}\Omega_{lk,l^{\prime}j} + \Omega_{lk,lk^{\prime}}.
\end{equation}

This expression can be considered as a particular case of eq. (\ref{eq:omega_lk}) when $X_{r} = 1$, which means that all particles exist in the whole space. In this case the interface collision term $\nabla X_{lk}\Omega_{lk,lk^{\prime}}^{A}$ vanishes due to zero gradient. Instead, different ions (or electrons) interact via $\Omega_{lk,lk^{\prime}}$.  

With the collision operator in eq. (\ref{eq:omega_lk_atom}), one should arrive at eq. (\ref{eq:dUdFS}) with the Chapman-Enskog expansion. However, here we aim to obtain a unified form for both types of mixing by means of re-definitions of relaxations. For this purpose, we need find a unified description for interactions between particles $lk$ and $lk^{\prime}$.

\begin{figure} 
\centering
{\includegraphics[width=0.45\textwidth]{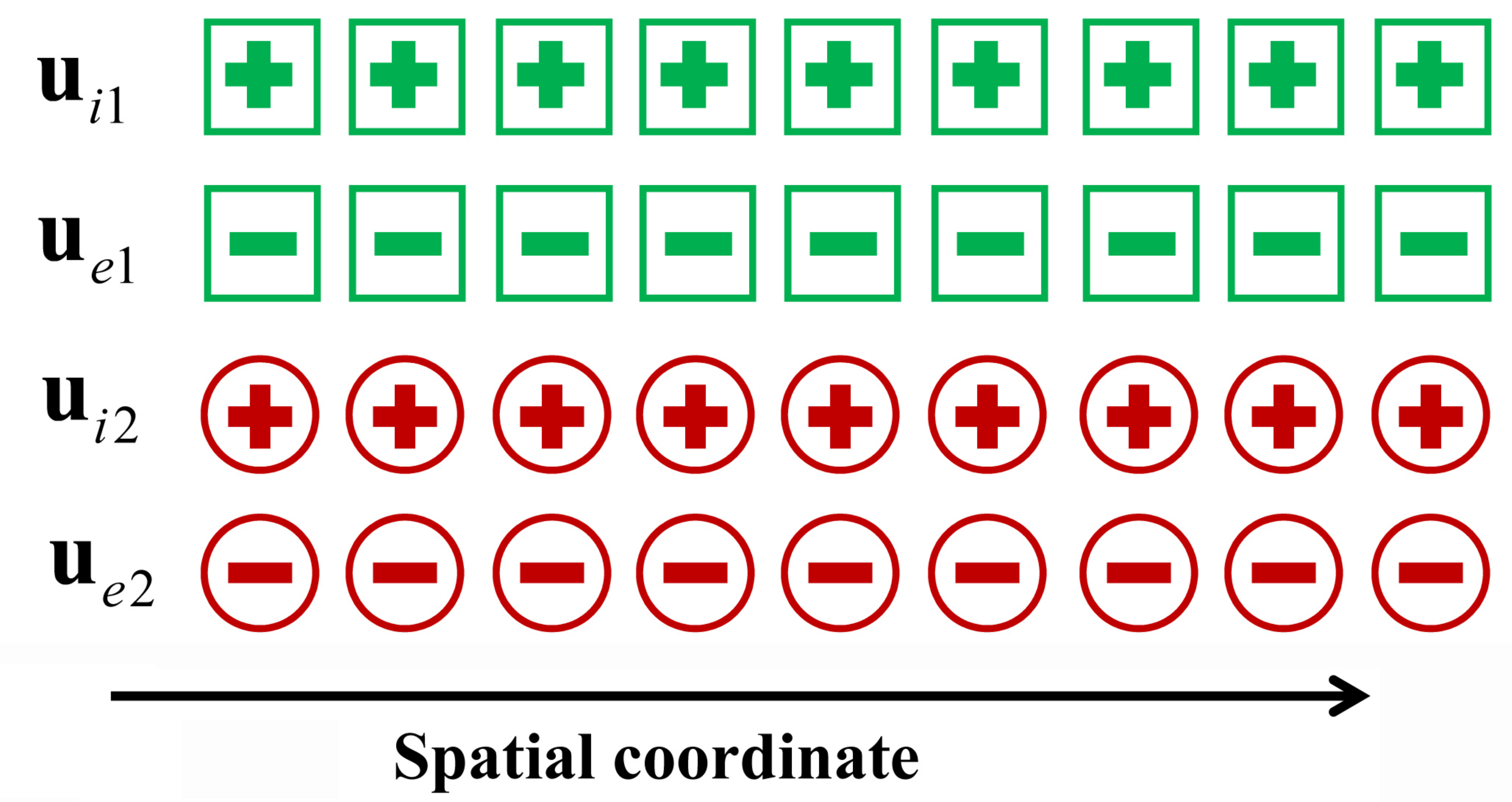}}
\caption{The atomic mixing between plasmas of different materials.}
\label{fig:atomic_mix_on_grid}
\end{figure}




First of all, due to the absence of an interface ($\nabla X_r = \vc{0}$), the terms $\vc{M}_{lk} = \vc{0}$ and $\mathcal{Z}_{lk} = {0}$ in atomic mixing.
For the grain mixing BN model primitive variables are $\alpha_k{\rho}_{lk}$, ${u}_{lk}$, ${p}_{lk}$, $\alpha_{lk}$,
while in classical two-fluid plasma description (\ref{eq:dUdFS}) are  $\widehat{\rho}_{lk}$, $\widehat{u}_{lk}$, $\widehat{p}_{lk}$. Moreover, in the latter case there is no concept of pressure relaxation, nor is the volume fraction as in the BN-type model (\ref{eq:14_eqn_plasma_model}).
Since each particle species occupies the whole space (Fig. \ref{fig:atomic_mix_on_grid}), the extension  of eq. (\ref{eq:def_alpha}) to atomic mixing is obviously not feasible. However, it is possible to formally unify the description of both cases by introducing the parameters ${{p}^{*}_{lk}}$ and ${\alpha}^{*}_{lk}$ to replace ${\widehat{p}_{lk}}$ such that
\begin{equation}\label{eq:alpha_hat}
     {\widehat{p}_{lk}} = {\alpha}^{*}_{lk} {{p}^{*}_{lk}}, \quad \sum{\alpha}^{*}_{lk} = 1.
\end{equation}

To define a unique mapping from  ${\widehat{p}_k}$ to ${\alpha}^{*}_k$ and ${{p}^{*}_k}$, we need an additional closure. A possible choice is ${{p}^{*}_k} = {{p}^{*}_{k^{\prime}}}$, or in an equivalent relaxation scheme
%
\begin{equation}\label{eq:dp_relax}
     \dudx{{{p}^{*}_{lk}}}{t} = \zeta^{*}_{lk} \left({{p}^{*}_{lk^{\prime}}} - {{p}^{*}_{lk}} \right), \quad \sum_k\zeta^{*}_{lk} \to \infty,
\end{equation}
where we only retain the pseudo-pressure relaxation mechanism for simplicity.


Then we define the relaxation rates $\zeta^{*}_{lk}$ in a way compatible with the first and second laws of thermodynamics. By invoking the mass/energy conservation, we have
\begin{subeqnarray}\label{eq:dp_relax_sys}
  \sum_k {\widehat{\rho}_{lk}} \dudx{v_{lk}^{*}}{t} &=& 0, \\
  \sum_k {\widehat{\rho}_{lk}} \dudx{{\varepsilon}_{lk}^{*}}{t} = \sum_k \widehat{\rho}_{lk}\left[ \frac{\gamma_{lk} - \Gamma_{lk}}{\Gamma_k}p_{lk}^{*} \dudx{v_{lk}^{*}}{t} + \frac{v_{lk}^{*}}{\Gamma_{lk}} \dudx{p_{lk}^{*}}{t} \right] &=& 0,
\end{subeqnarray}
where $\widehat{\rho}_{lk} = {\alpha}^{*}_{lk} {\rho}^{*}_{lk} = {\alpha}^{*}_{lk} / {v}^{*}_{lk}$.

Inserting eqs. (\ref{eq:dp_relax}) into eqs. (\ref{eq:dp_relax_sys}), one can solve for $\dudx{v_{lk}^{*}}{t}$ and $\dudx{{\alpha}_{lk}^{*}}{t} = \widehat{\rho}_{lk}\dudx{v_{lk}^{*}}{t}$.

If we define the pressure relaxation coefficient and the interfacial variables in the same way as in eq. (\ref{eq:pk}), one can recover the BN-type model, which ensures entropy production.


With such definitions, we see that the parameter ${\alpha}^{*}_{lk}$ is in fact the partial pressure fraction, which equals to the volume fraction ${\alpha}_{lk}$  under the temperature equilibrium (Dalton's law).
The new relaxation system can be obtained if we combine eq. (\ref{eq:alpha_hat}) and the original model (\ref{eq:multifluid_model}) (with $\widehat{p}_{lk}$ being replaced by $\alpha_{lk}^{*}$ and $p_{lk}^{*}$ and  $\widehat{\rho}_{lk}$  by $\alpha_{lk}^{*}$ and ${\rho}_{lk}^{*}$). This new system can be fitted into eq. (\ref{eq:grainModel}) by adjusting the interaction terms and interfacial stress tensor (see Table \ref{tab:GrainAtomicDefs}). 

In fact, to ensure that the relaxation system tend to the solution of the original model eq. (\ref{eq:grainModel}), we have to prove the so-called sub-characteristic condition:
\begin{equation}
[{\textrm{min}}(\widehat{\vc{\Lambda})}, {\textrm{max}}(\widehat{\vc{\Lambda}})] \subseteq [{\textrm{min}}(\vc{\Lambda}^{*}), {\textrm{max}}(\vc{\Lambda}^{*})],
\end{equation}
where $\widehat{\vc{\Lambda}}$ and $\vc{\Lambda}^{*}$ are the vector of characteristic values of hyperbolic parts of the original and the relaxation system, respectively. They are
\[\widehat{\vc{\Lambda}} = \left[ \widehat{u}_{l1} - \widehat{c}_{l1}, \;\widehat{u}_{l1}, \;\widehat{u}_{l1} + \widehat{c}_{l1}, \; \widehat{u}_{l2} - \widehat{c}_{l2}, \;\widehat{u}_{l2}, \;\widehat{u}_{l2} + \widehat{c}_{l2}   \right], \]
\[\vc{\Lambda}^{*} = \left[ u_{l1}^{*} - c_{l1}^{*}, \;u_{l1}^{*}, \;u_{l1}^{*} + c_{l1}^{*}, \; u_{l2}^{*} - c_{l2}^{*}, \;u_{l2}^{*}, \;u_{l2}^{*} + c_{l2}^{*}, {u}_{I}^{*}  \right], \]

The velocities are $u_{lk}^{*} = \widehat{u}_{lk}$. The sound velocity \[\widehat{c}_{lk} = \sqrt{\gamma_{lk}\frac{\widehat{p}_{lk}}{\widehat{\rho}_{lk}}} = \sqrt{\gamma_{lk}\frac{\alpha_{lk}^{*}{p}_{lk}^{*}}{\alpha_{lk}^{*}{\rho}_{lk}^{*}}} = c_{lk}^{*}. \]
If we define ${u}_{I}^{*}$ as convex combination of $\vc{u}_{lk}$, then the sub-characteristic condition is satisfied.

To sum up, the atomic mixing model can be adjusted to the form of the grain-mixing model for ions or electrons with the re-definitions in the Table \ref{tab:GrainAtomicDefs}.


\begin{table}
  \begin{center}
\def~{\hphantom{0}}
\begin{tabular}{lcc}
\hline
  Mixing types & Grain mixing & Atomic mixing \\
  \hline
 The collision between $lk$ and $lk^{\prime}$  & $\nabla X_{lk} \Omega_{lk,lk^{\prime}}^{A}$ & $\Omega_{lk,lk^{\prime}}$  \\
  The parameter $\alpha$ & Volume fraction & Pressure fraction \\
  Interface pressure relaxation & $\mathcal{P}_{lk}$ & $\mu^{p}_{lk,lk^{\prime}}\to\infty$ \\
  The velocity relaxation  & $\vc{M}_{lk}$ & $\vc{R}_{lk}$ \\
   The temperature relaxation  & $\mathcal{Z}_{lk}$ & $\mathcal{Q}_{lk}^{\varepsilon}$ \\
   \hline
\end{tabular}
  \caption{Comparison of definitions in grain and atomic mixing}
  \label{tab:GrainAtomicDefs}
  \end{center}
\end{table}

\section{Relaxation time scales}\label{sec:relax_time}
Here we perform an analysis of various time scales involved in the model. The mixing takes place at very different length scales from the atomic mixing to the grain-sized mixing. The relaxation time scales in the atomic mixing is evaluated on the basis of particle collision frequency and those in the grain-sized mixing is evaluated through the wave propagation and heat conduction in continuum mechanics.

Since this work is motivated by ICF plasma flows, in  the following evaluations the characteristic scales of ICF will be employed. The typical materials involved in ICF include Carbon(C), Hydrogen (H), Deuterium(D) and Tritium (T).

The ICF-relevant characteristic scales used are as follows:
\begin{enumerate}
  \item[(a)]  Density: 0.1g/cm$^3$, 1g/cm$^3$, 10g/cm$^3$
  \item[(b)]  Temperature: 0 - 10keV
  \item[(c)]  Length: 0.01$\mu$m, 0.1$\mu$m, 1$\mu$m
\end{enumerate}

\subsection{The relaxation time at atomic scales}
We use the Coulomb collision theory to evaluate the characteristic velocity/temperature relaxation times in the multi-component plasma, we have
\begin{subeqnarray}
  \dudx{{\vc{u}}_r}{t} &=& \nu_{r,{r^{\prime}}}^{u}\left( {\vc{u}}_{r^{\prime}} - {\vc{u}}_r \right)\label{eq:dukdt}, \\
  \dudx{T_r}{t} &=& \nu_{r,{r^{\prime}}}^{T}\left( T_{r^{\prime}} - T_r \right)\label{eq:dTkdt}.
\end{subeqnarray}

The velocity relaxation rate \citep[][]{Braginskii1965ReviewsOP}
\begin{equation}\label{eq:nu_velRelax}
    \nu_{r,{r^{\prime}}}^{u} = \frac{n_{r^{\prime}}m_{rr^{\prime}}}{m_{r}} \frac{4\sqrt{2\pi}\Lambda_{rr^{\prime}} Z_r Z_{r^{\prime}} e^4}{3\sqrt{m_{r{r^{\prime}}}} (k_B T)^{3/2}}.
\end{equation}

For temperature relaxation rate we use the formula from \cite{NRL_formulary_2004}
\begin{equation}\label{eq:nu_tempRelax}
  \nu_{r,{r^{\prime}}}^{T} = 1.8\times 10^{-19} \frac{(m_{r} m_{r^{\prime}} )^{1/2} Z_{r}^2 Z_{r^{\prime}}^2 n_{r^{\prime}} \Lambda_{rr^{\prime}} }{\left( m_{r} T_{r^{\prime}} + m_{r^{\prime}} T_{r} \right)^{3/2}},
\end{equation}
where the Coulomb logarithm $\Lambda_{rr^{\prime}}$ is evaluated differently for ion-ion and ion-electron combinations.

According to eqs. (\ref{eq:dukdt}), the velocity and temperature timescales have the following evaluations
\begin{equation}\label{eq:relax_times}
    \tau_{{r,{r^{\prime}}}}^{u} = 1/\left(\nu_{r,{r^{\prime}}}^{u} + \nu_{{r^{\prime}},r}^{u} \right), \quad \tau_{{r,{r^{\prime}}}}^{T} = 1/\left(\nu_{r,{r^{\prime}}}^{T} + \nu_{{r^{\prime}},r}^{T} \right).
\end{equation}

\begin{figure} 
\centering
\subfloat[$\tau_{i1,i2}^{T}, \; \tau_{i,e}^{T}$ for $y_{C} = 10\%$]{\includegraphics[width=0.5\textwidth]{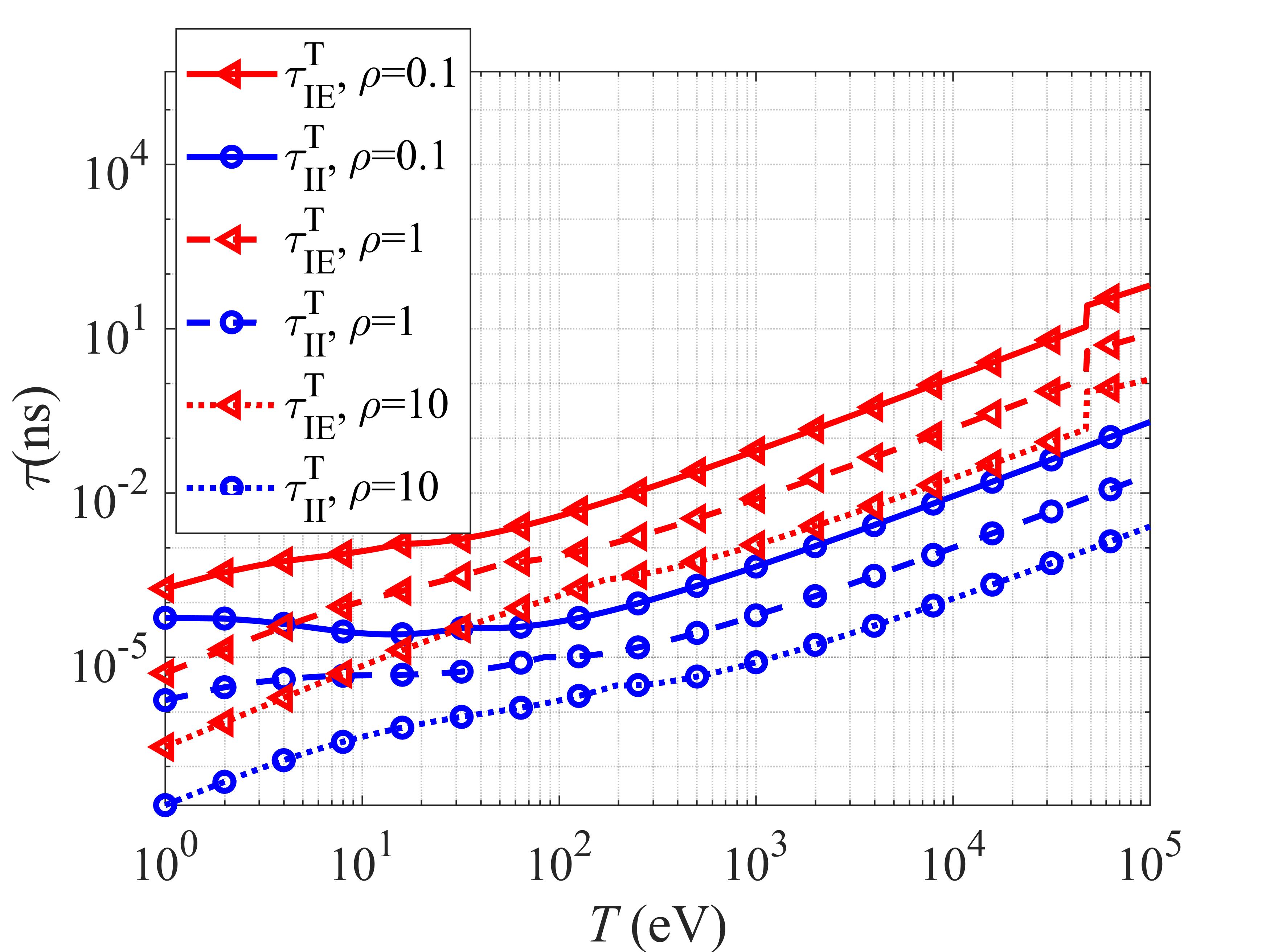}}
\subfloat[$\tau_{i1,i2}^{T}, \; \tau_{i,e}^{T}$ for $y_{C} = 50\%$]{\includegraphics[width=0.5\textwidth]{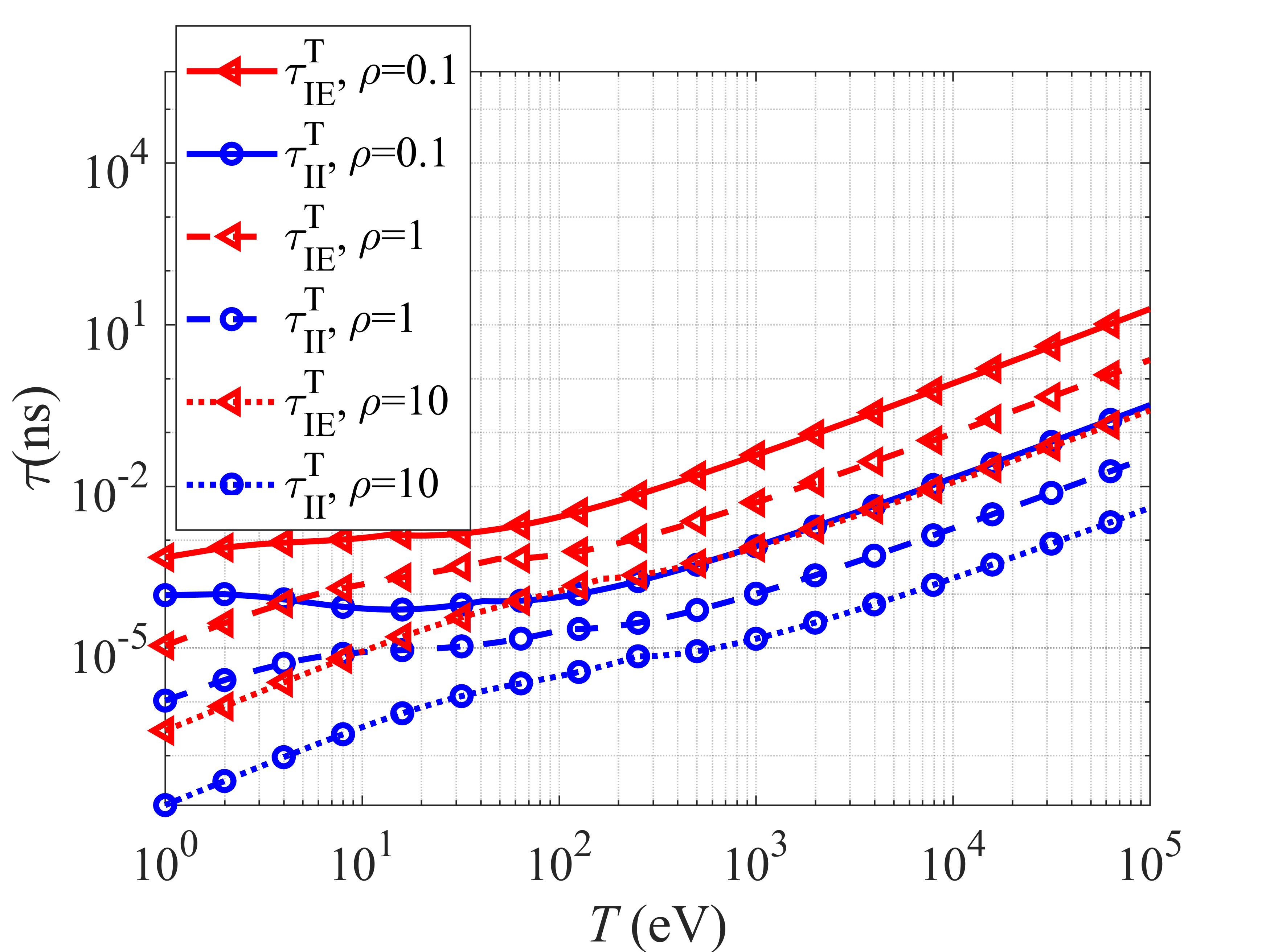}}\\
\subfloat[$\tau_{i1,i2}^{T}, \; \tau_{i,e}^{T}$ for $y_{C} = 90\%$]{\includegraphics[width=0.5\textwidth]{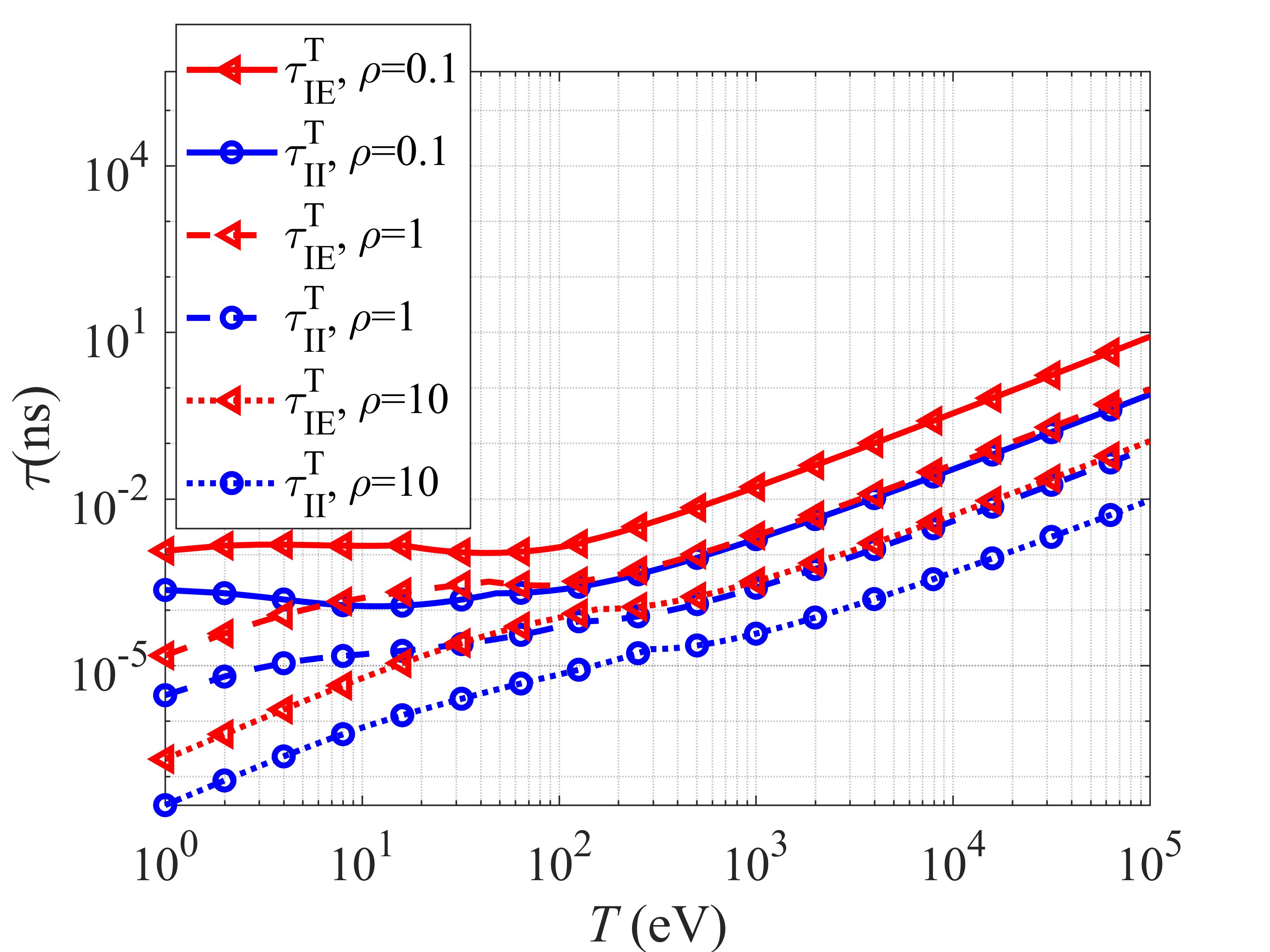}}
\subfloat[$\tau_{i1,i2}^{T}, \; \tau_{i,e}^{T}$ for $y_{C} = 10\%$]{\includegraphics[width=0.5\textwidth]{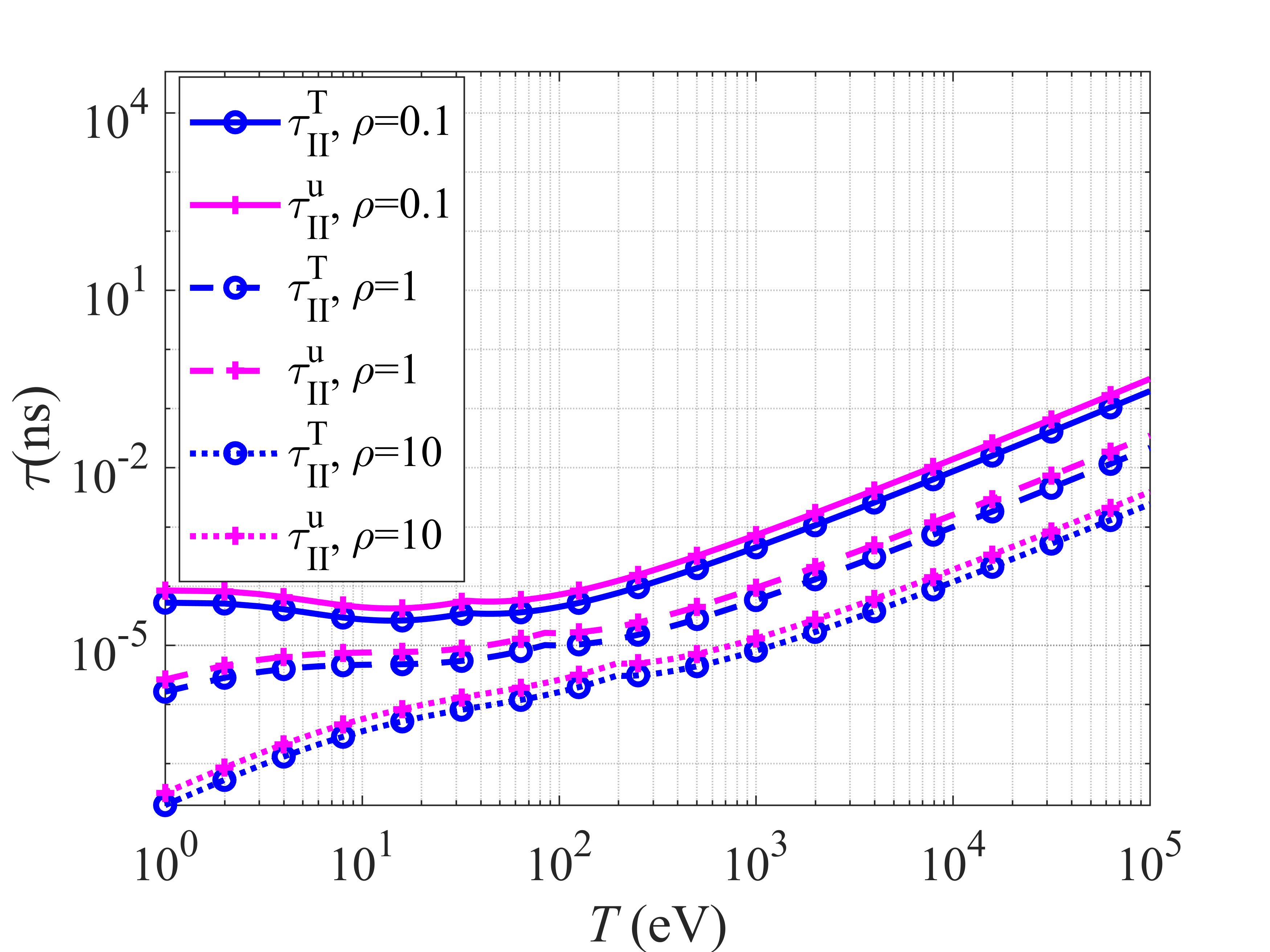}}\\
\subfloat[$\tau_{i1,i2}^{u}, \; \tau_{i1,i2}^{T}$  for $y_{C} = 50\%$]{\includegraphics[width=0.5\textwidth]{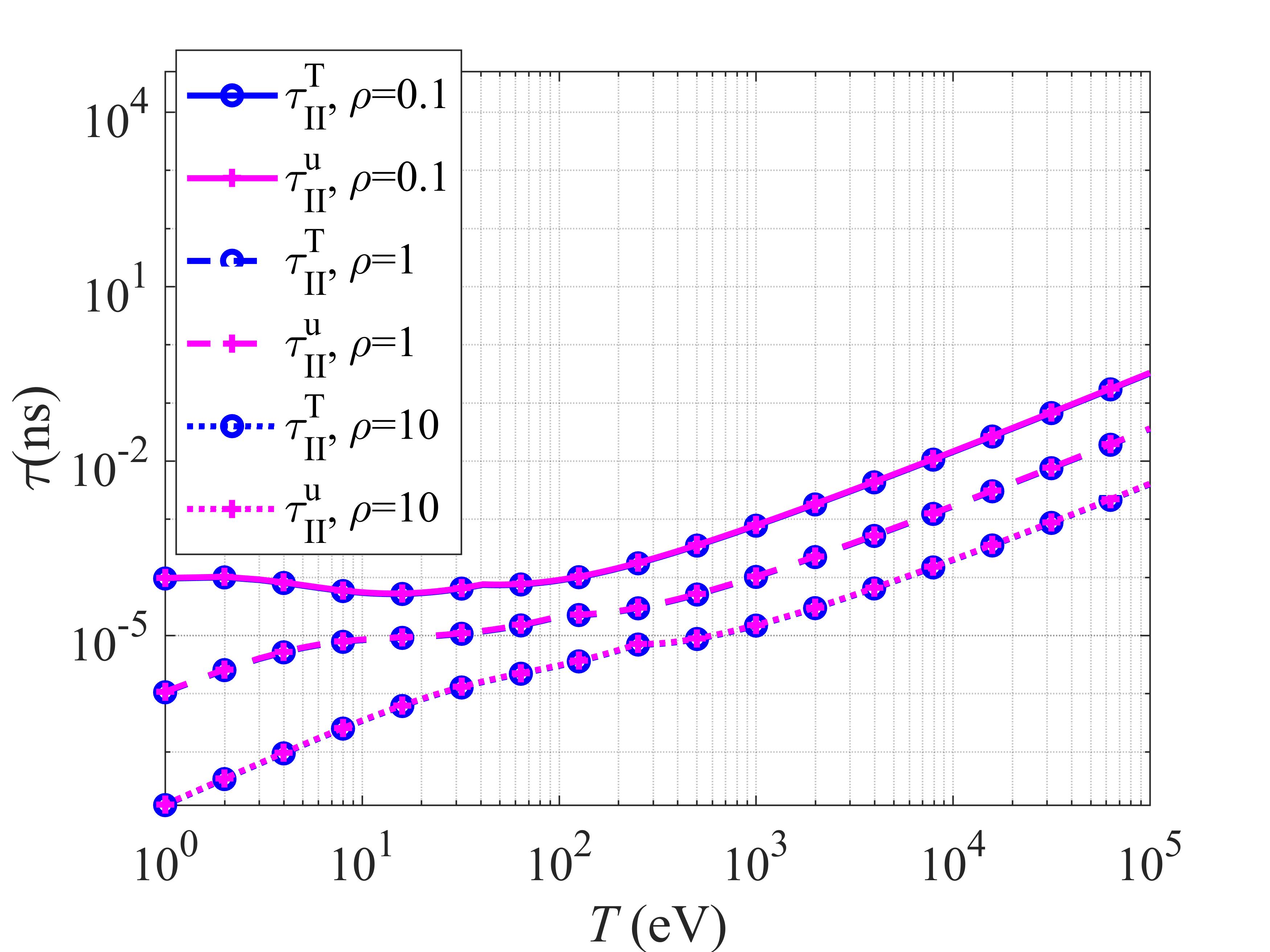}}
\subfloat[$\tau_{i1,i2}^{u}, \; \tau_{i1,i2}^{T}$  for $y_{C} = 90\%$]{\includegraphics[width=0.5\textwidth]{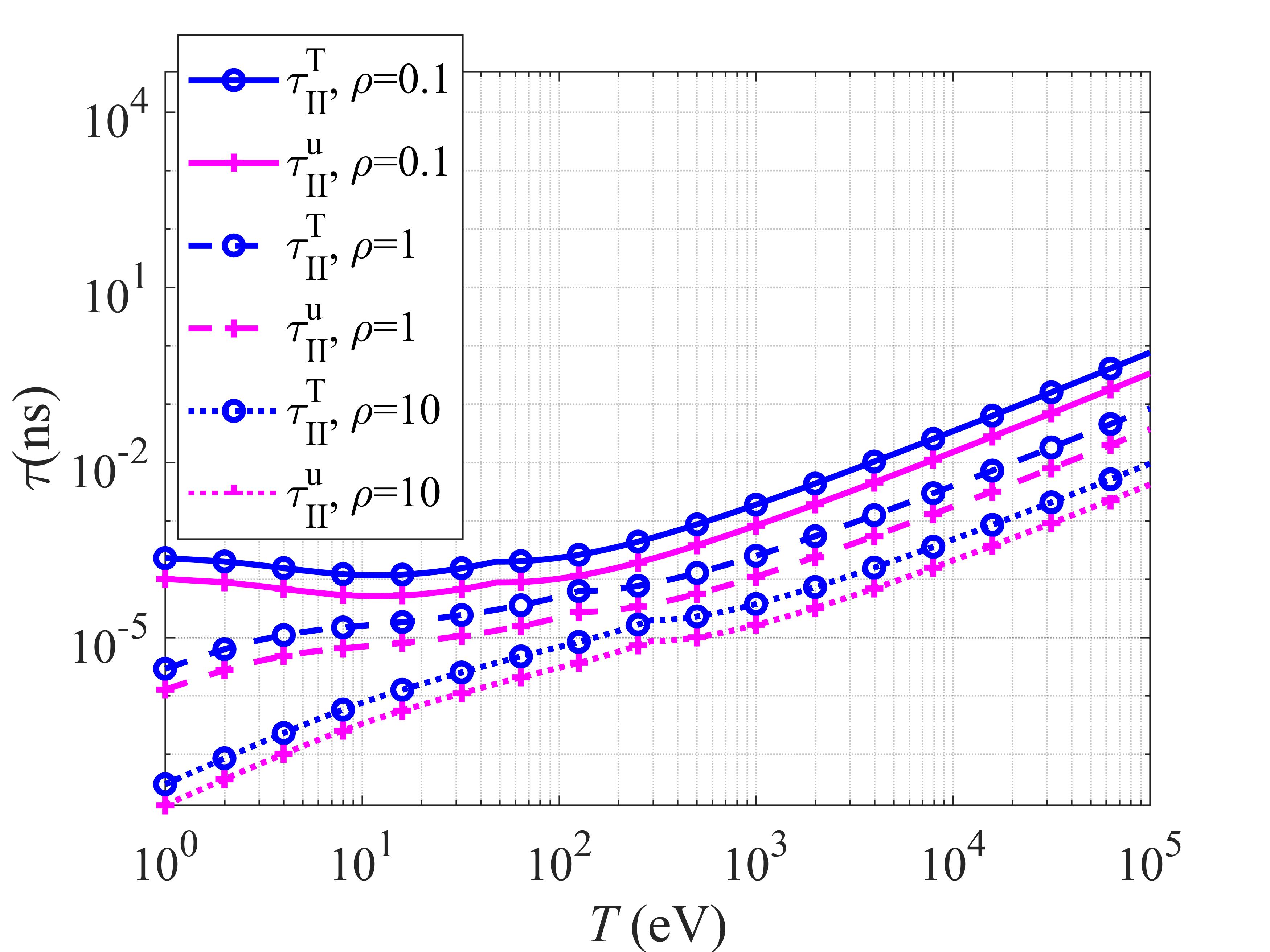}}
\caption{The temperature relaxation time scales for C-D plasma mixture under different densities, temperatures and mass fractions. The i-e relaxation times are evaluated with number-weighted atom number $Z$ and atom number $A$. The ionization is calculated with the method of \cite{bound1985pressure}.}
\label{fig:relaxation_time_atomic}
\end{figure}

In particular, for electron-ion mixture, the characteristic relaxation time is dominated by the shortest one (i.e., $\tau_{e,e}$) due to frequency additivity. Thus, we have
\begin{equation}\label{eq:ie_relaxa_times}
    \tau_{i,e}^{u} = 1 / \left( \frac{m_e n_e}{m_i n_i \tau_{e,e}^{u}} + \frac{1}{\tau_{e,e}^{u}} \right), \quad \tau_{i,e}^{T} = 1 / \left( \frac{ 2m_e n_e}{m_i n_i \tau_{e,e}^{T}} + \frac{2m_e}{m_i \tau_{e,e}^{T}} \right),
\end{equation}
where $\tau_{e,e} = 1/ \nu_{e,e}$.


Since $m_e << m_i$ and $m_en_e << m_i n_i$ the velocity relaxation rate can be considered to be infinite in comparison with the temperature relaxation rate.
This evaluation supports the following assumption of a unified characteristic function for ions and electrons, i.e., $X_{ik} = X_{ek}$ or $\vc{u}_{iI} = \vc{u}_{eI}$.

Figure \ref{fig:relaxation_time_atomic} displays relaxation times for different carbon-deuterium (C-D) mixture, which is representative in ICF. Here in the evaluation of electron/ion temperature relaxation time, we use the number-weighted atomic mass and number. 
We see that the differences between $\tau_{i,e}^{T}$, $\tau_{i1,i2}^{T}$, $\tau_{i1,i2}^{u}$ depend on the composition of the plasma mixture.  In the case of $y_c = 10\%$ and $T>10^2$eV, the ion-electron temperature relaxation time is much greater than that of ion-ion with  ${\tau_{i,e}^{T}}/{\tau_{i1,i2}^{T}} \approx 10^2$. As $y_c$ increases to $90\%$, $\tau_{ie}^{T}$ and $\tau_{i1,i2}^{T}$ maybe of the same order with ${\tau_{i,e}^{T}}/{\tau_{i1,i2}^{T}} < 10$. The ion-ion temperature and velocity relaxation times are of the same order despite of densities, temperatures and mass fractions.  These evaluations indicates that the ion-ion temperature separation is more prominent on the high-Z component (carbon in this particular case) side in mixing region. Moreover, ion-ion velocity separation is as significant as temperature separation.


\subsection{The ion relaxation times at mesoscopic scales}
\begin{figure} 
\centering
{\includegraphics[width=0.38\textwidth]{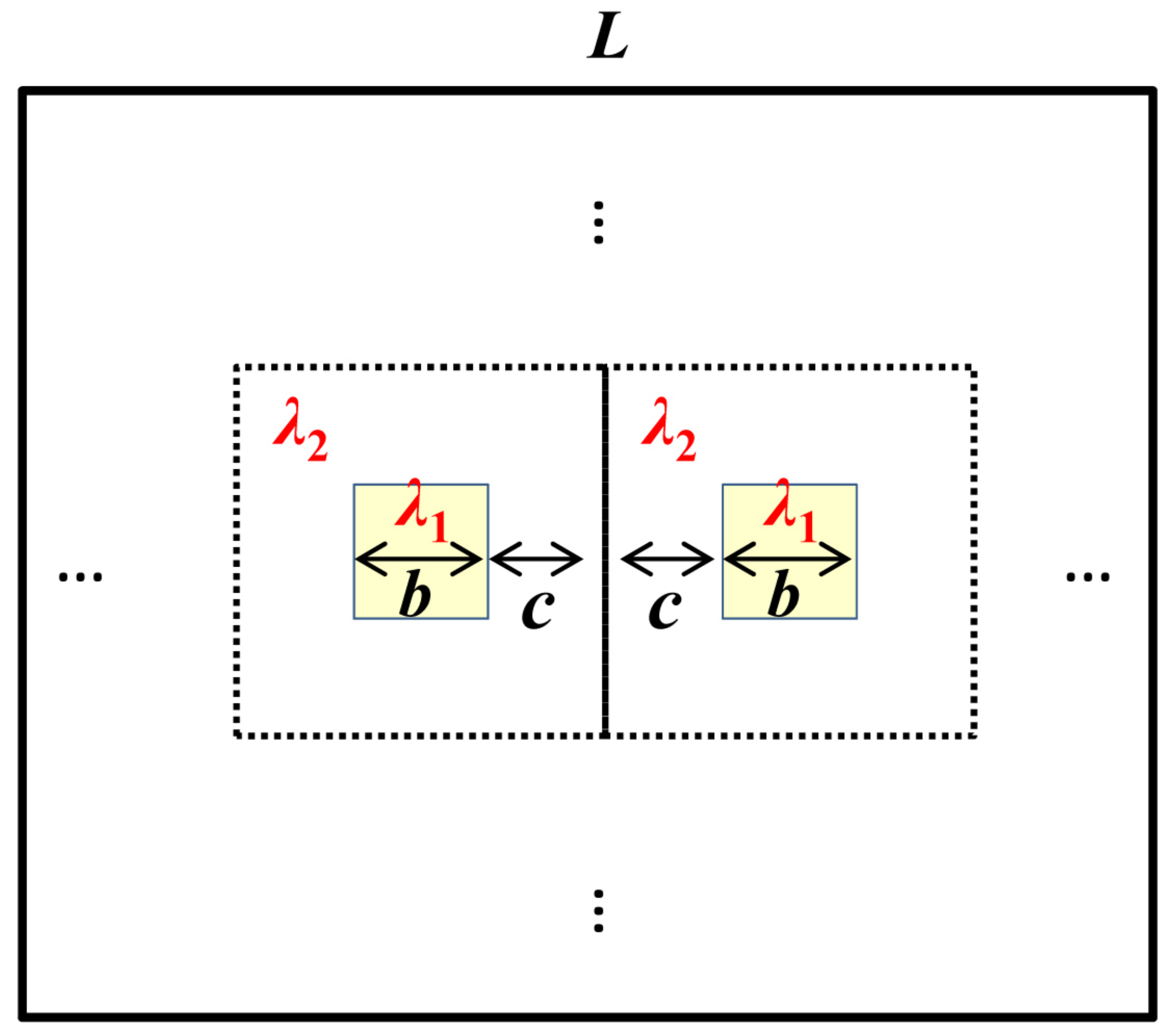}}
\caption{The schematic for deriving the grain temperature relaxation timescale. $L, \; d, \;b$ are the characteristic length of the physical domain, sub-domain, grain, respectively. $n$ is the number of grains.}
\label{fig:heatConductionDemo}
\end{figure}
We assume that $n$ grains of size $b$ are uniformly distributed in the concerned physical domain of characteristic length $L$ (see Figure \ref{fig:heatConductionDemo}). In 2D scenario we define $V = L^2$, $V_1 = n b^2$, $V_2 = V - V_1$ and  $\alpha_k = V_k / V$.

\cite{CHINNAYYA2004490} has derived the mechanical relaxation rates as follows
\begin{eqnarray}\label{eq:presRelax}
  \mu_{lk,lk^{\prime}}^{p} = \frac{2 nb}{\left( Z_{l1} + Z_{l2} \right)V}, \label{eq:mu_presRelax}\\
   \mu_{lk,lk^{\prime}}^{u} = Z_{l1} Z_{l2} \mu_{lk,lk^{\prime}}^{p}, \label{eq:mu_velRelax}
\end{eqnarray}
where $nb$ is the contact area, which depends on the grain size.

By using the primitive equations (\ref{eq:uk}), one can obtain the following evaluations:

(a) The pressure relaxation time scale
\begin{equation}\label{eq:tauP}
  \tau_{lk,lk^{\prime}}^{p} = \frac{1}{\mu_{lk,lk^{\prime}}^{p} \left( {\rho_{l1} a_{l1I}^2}/{\alpha_{l1}} + {\rho_{l2} a_{l2I}^2}/{\alpha_{l2}} \right)},
\end{equation}

(b) The velocity relaxation time scale
\begin{equation}\label{eq:tauV}
  \tau_{lk,lk^{\prime}}^{u} = \frac{1}{\mu_{lk,lk^{\prime}}^{u} \left(  1/\left( \alpha_{l1} \rho_{l1} \right) + 1/\left( \alpha_{l2} \rho_{l2} \right) \right)},
\end{equation}

(c) The temperature relaxation time scale is
\begin{equation}\label{eq:tauT}
  \tau_{lk,lk^{\prime}}^{T} = \frac{1}{\mu_{lk,lk^{\prime}}^{T} \left[ 1 / (\alpha_{l1} \rho_{l1} C_{vl1}) +  1 / (\alpha_{l2} \rho_{l2} C_{vl2}) \right]},
\end{equation}
where
\begin{equation}\label{eq:mu_mechTempRelax}
\mu_{lk,lk^{\prime}}^{T} = \frac{4nb\lambda_f}{l_T L^2}.
\end{equation}

Here we temporarily neglect the relative motion between the grains and the background fluid. Let us assume that the temperature relaxation in grain scale is governed by the heat conduction. The evaluation (\ref{eq:tauT}) can be derived from the following conservation equations
\begin{subeqnarray}
 \rho_{l1} V_1 C_{v1} \dudx{T_{l1}}{t} &=& 4b\lambda_{lf} \frac{\left( T_{l2} - T_{l1} \right)}{l_{T}}, \\
 \rho_{l2} V_2 C_{v2} \dudx{T_{l2}}{t} &=& 4b\lambda_{lf} \frac{\left( T_{l1} - T_{l2} \right)}{l_{T}},
\end{subeqnarray}
where the plasma heat conductivity is calculated with the model of \cite{Spitzer1953TRANSPORTPI}. $\lambda_{lf}$ is the heat conductivity on the interface.

The temperature gradient length scale $l_T$ can be determined by dividing the physical domain into $n$ uniform sub-domains of size $d = \sqrt{L^2/n}$. The characteristic length scale $l_T = (b+c)/2, \; c = (d-b)/2$ . For flux continuity, the heat conductivity at the cell face is averaged by
\[\frac{d}{\lambda_{lf}} = \frac{b}{2\lambda_{l1}} + \frac{c}{2\lambda_{l2}}.\]

\begin{figure} 
\centering
\subfloat[$l$ = 1$\mu$m]{\includegraphics[width=0.5\textwidth]{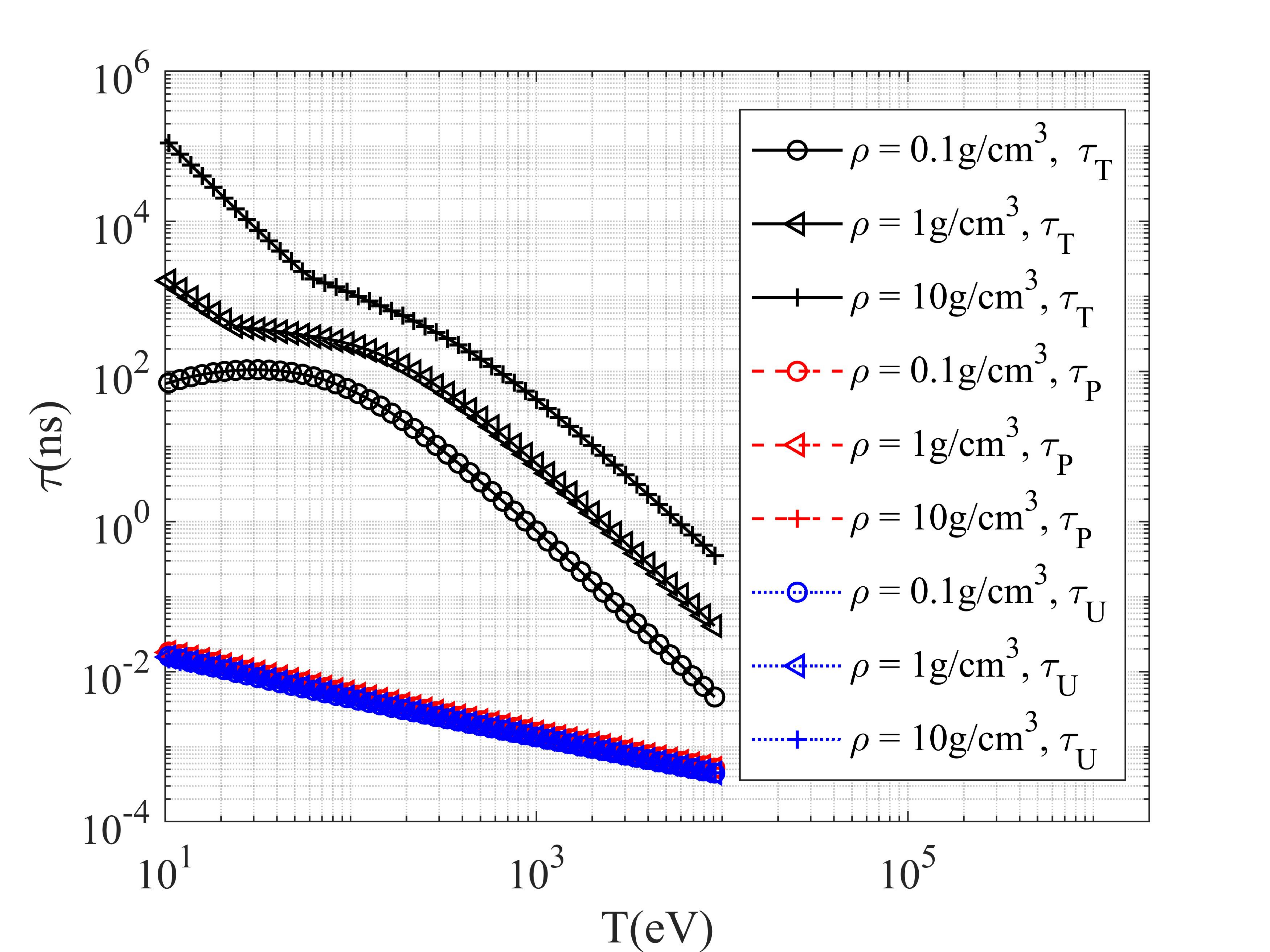}}
\subfloat[$l$ = 0.1$\mu$m]{\includegraphics[width=0.5\textwidth]{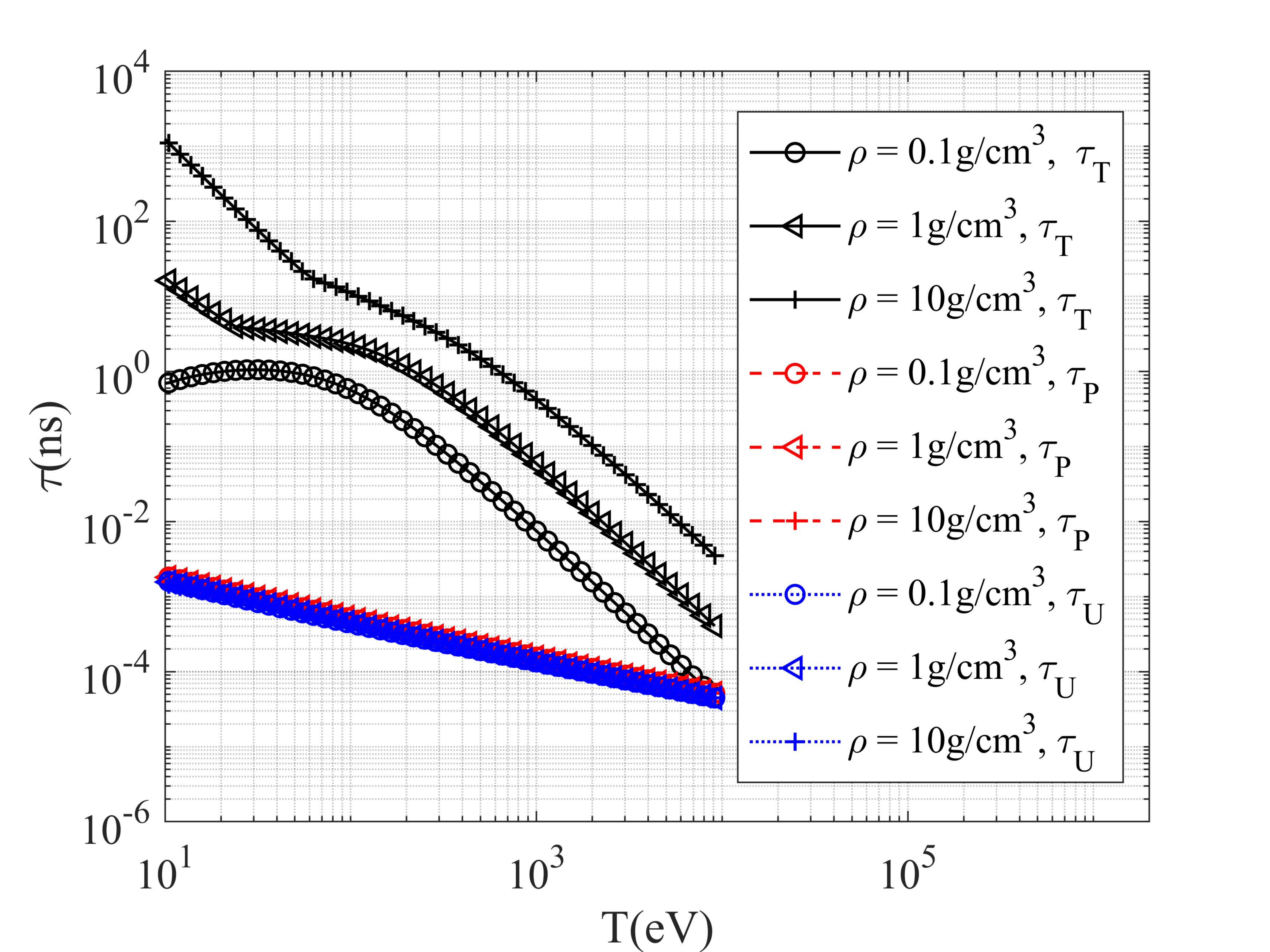}}\\
\subfloat[$l$ = 0.01$\mu$m]{\includegraphics[width=0.5\textwidth]{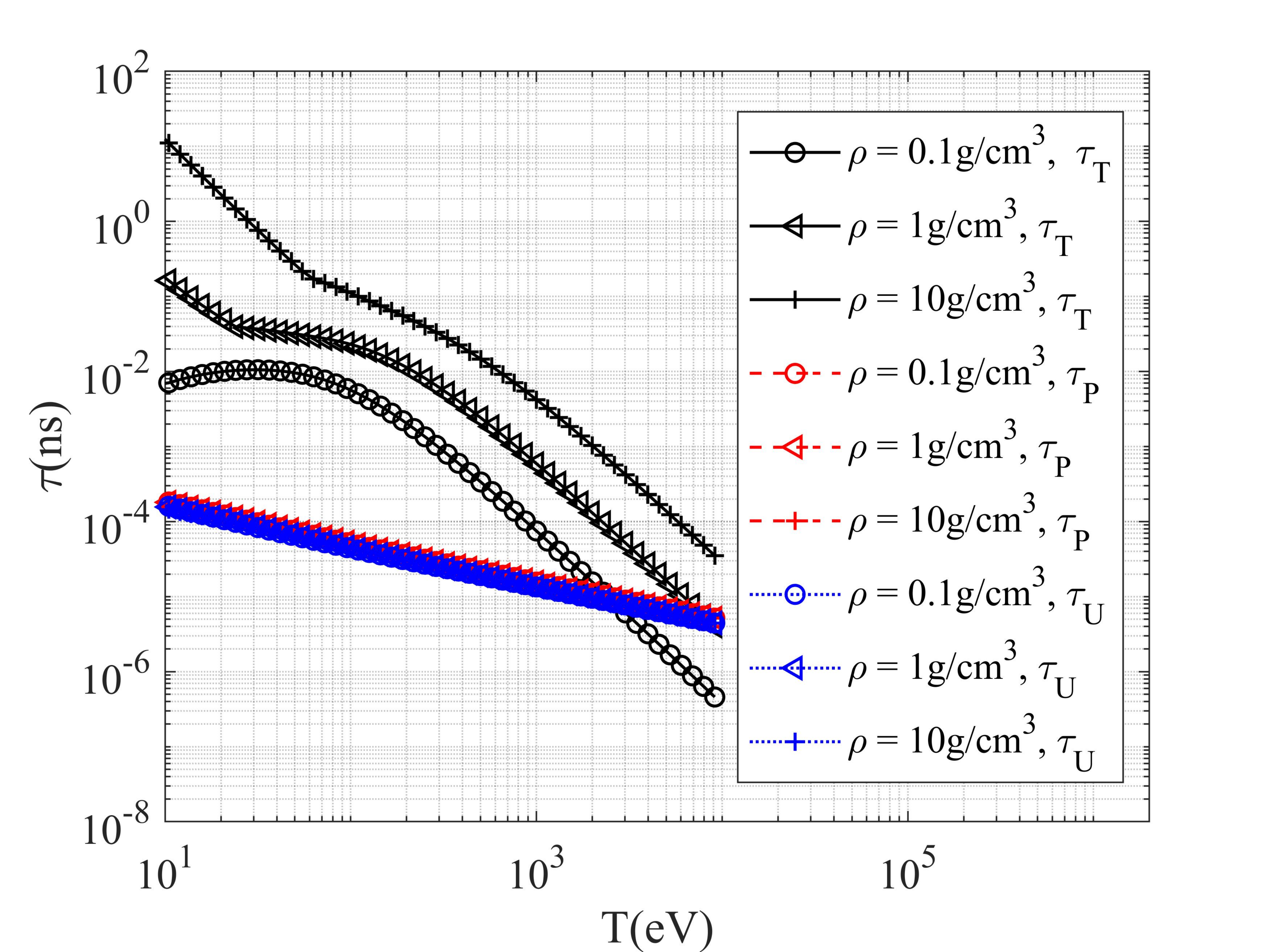}}
\caption{The velocity relaxation time $\tau^u$, the pressure relaxation time $\tau^p$ and the temperature relaxation time $\tau^T$ between carbon and deuterium under different densities. The characteristic length scales are  1$\mu$m, 0.1$\mu$m and 0.01$\mu$m, respectively. The volume fraction of carbon in the mixture is assumed to be 0.1. The phase densities are $\rho_{C} = 3\rho_{0}$ and $\rho_{D} = 0.25\rho_{0}$, respectively, with $\rho_{0} = 0.1,\;1,\;10$g/cm$^3$. }
\label{fig:relaxation_time_grain}
\end{figure}

With the above characteristic parameters, one can evaluate the mesoscopic pressure, velocity and temperature relaxation time scales under the ICF relevant condition. Our evaluations for ion-ion relaxations in C-D plasma are displayed in Fig. \ref{fig:relaxation_time_grain}. It can be seen that the mechanical relaxation times are less than the temperature relaxation scale by at least 2 orders under $T \sim$ 10$^2$eV and $l \sim$ 10$^{-2}$-10$^0\mu$m, which indicates that one can make the assumption of negligible mechanical relaxation time under such conditions.

\section{Reduction of the model}\label{sec:Reduced_model}
The model (\ref{eq:14_eqn_plasma_model}) consists of 14 equations and is in fully disequilibrium in grain-size mixing at the interface. Ions and electrons are only coupled through the collision terms $\vc{R}$ and $Q$. Thus, the established numerical methods for Baer-Nunziato type model can be extended to the solution of the proposed model. Although the algorism is straightforward, the computation is complicated and extremely computationally expensive. Reductions of the fully-disequilibrium model are entailed.

\begin{figure} 
\centering
{\includegraphics[width=0.9\textwidth]{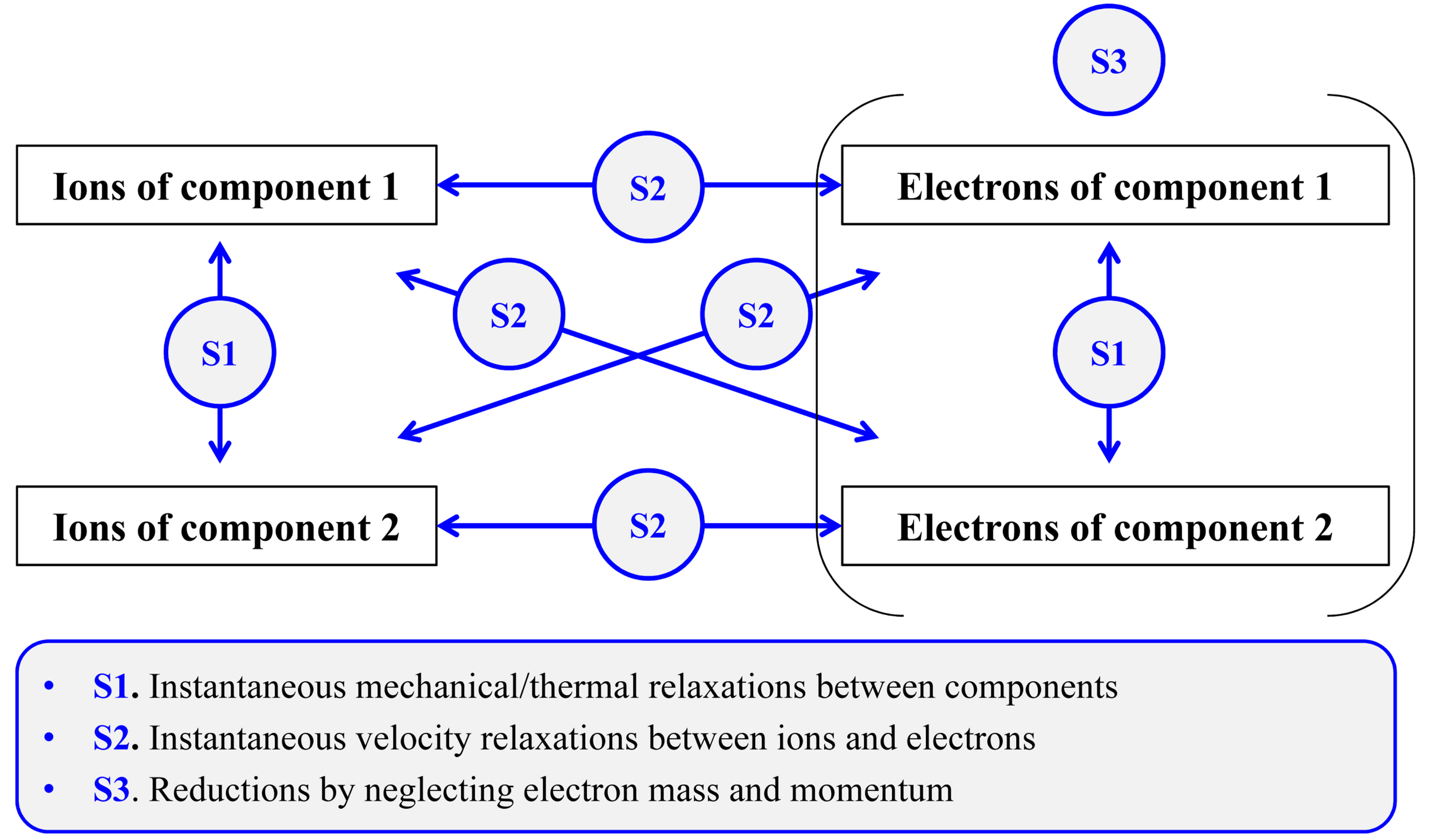}}
\caption{The map for reducing the BN-type multi-component plasmas model.}
\label{fig:reduction_map}
\end{figure}

On the basis of the evaluations of various relaxation time scales, we derive some reduced versions of the fully disequilibrium model (\ref{eq:model0:en}). 
We present the reduction map in Figure \ref{fig:reduction_map} for clarity. The reductions are performed in the following steps:

\begin{description}
  \item[Step S1.] We first consider reductions within each particle species (ions or electrons) in grain mixing. This step can be performed in a totally similar manner to that of the non-HED case, just with additional ion-electron coupling term as a source term. The derivations are presented in Section \ref{sec:A1}.
  \item[Step S2.] We consider the simplification of the coupling between ions and electrons. This step is realized with the elementary assumptions that the ion-electron velocity relaxation time approaches zero in comparison with that of the temperature relaxation. For corresponding details see Section \ref{sec:A2}.
  \item[Step S3.] We further reduce the model by assuming the quasi-neutral condition of plasma and that the electron mass is negligibly small in comparison with that of ions. The corresponding reductions are described in Section \ref{sec:A3}.
\end{description}

On the basis of the above reductions, we write some typical reduced multi-plasma models including the nine-equation model (Section \ref{sec:multi_plasma_eight_eqn}), the eight-equation model (Section \ref{sec:multi_plasma_seven_eqn}), the six-equation model (Section \ref{sec:multi_plasma_six_eqn}), the five-equation model (Section \ref{sec:multi_plasma_five_eqn}), and the simplified inter-penetration model (Section \ref{sec:multi_plasma_inter_mix}).

\subsection{Reductions within the particle species}\label{sec:A1}
In this section we perform the S1 step in Fig. \ref{fig:reduction_map}. For each particle species (ions or electrons) we derive a series of reduced models, including
\begin{enumerate}[label=\roman*.]
  \item Model in the limit of instantaneous velocity relaxation (i.e., the u-eq model),
  \item Model in the limit of instantaneous velocity and pressure relaxation (i.e., the pu-eq model),
  \item Model in the limit of instantaneous velocity, pressure and temperature relaxation (i.e., the puT-eq model),
  \item Model in the limit of instantaneous velocity, pressure, temperature and compressibility relaxation (i.e., the puT$\gamma$-eq model).
\end{enumerate}

\subsubsection{The u-eq model}\label{sec:6eqn}
As has been evaluated above, the mechanical (pressure and velocity) relaxation time is far smaller than that of thermal relaxation in grain mixing (see Fig. \ref{fig:relaxation_time_grain}). However, in this section we only drive the velocity into equilibrium while pressure disequilibrium is retained. Such model has certain advantages. Firstly, it is well-posed and hyperbolic. It is well known that the pressure-equilibrium, velocity-disequilibrium  model has long been plagued by the loss of hyperbolicity \citep[][]{butler1982modeling,nigmatulin1990dynamics,saurel2008modelling}. Note a recent progress in resolving the former issue has been made \citep[][]{Hantke2021}. Secondly, it demonstrates more numerical robustness \citep[][]{saurel2009simple}.

To obtain a velocity-equilibrium model we consider the reduced model in the limit $\tau_{lk,lk^{\prime}}^{u}\to 0$. In this case we assume the asymptotic expansion in the form of $\vc{Z} = \vc{Z}^{(0)} + \tau_{lk,lk^{\prime}}^{u}\vc{Z}^{(1)} + \mathcal{O}\left( \left(\tau_{lk,lk^{\prime}}^{u}\right) ^2\right)$, where $\mathbf{Z} = \left[ s_{lk}, \; \vc{u}_{lk},\; p_{lk}, \;\alpha_{lk} \right]$.  Inserting such expansion into the multi-component plasma model, one obtains $\vc{u}_{l1}^{(0)} = \vc{u}_{l2}^{(0)} = \vc{u}_{l}^{(0)}$ in the order $\mathcal{O}\left( 1 / \tau_{lk,lk^{\prime}}^{u} \right)$.

In the order $\mathcal{O}(1)$, we have
\begin{eqnarray}
\rho_l\left(\vc{u}_{l1}^{(1)} - \vc{u}_{l2}^{(1)}\right) = -\alpha_{l2}\rho_{l2}\nabla \cdot \left( \alpha_{l1} \tensor{P}_{l1} \right) + \alpha_{l1}\rho_{l1}\nabla \cdot \left( \alpha_{l2} \tensor{P}_{l2} \right)  + \rho_l \tensor{P}_{lI} \cdot \nabla\alpha_{l1} \nonumber\\+ \left( \alpha_{l2}\rho_{l2} \vc{R}_{l1} - \alpha_{l1}\rho_{l1} \vc{R}_{l2} \right).
\end{eqnarray}

The drift velocity can be therefore determined as
\begin{equation}\label{eq:grain_drfit_vel}
\vc{w}_{lk} = \vc{u}_{lk} - \vc{u}_{l} = y_{lk^{\prime}} \tau_{lk,lk^{\prime}}^{u} \left( \vc{u}_{lk}^{(1)} - \vc{u}_{lk^{\prime}}^{(1)} \right),
\end{equation}
where the mass-averaged velocity $\vc{u}_{l} = \sum y_{lk} \vc{u}_{lk}$.

In the order $\mathcal{O}\left( 1\right)$, the model is reduced to
\begin{subeqnarray}
  \alpha_{lk} \rho_{lk} T_{lk} \frac{\textrm{D}_{l} s_{lk}}{\textrm{D} t} = \left(p_{lk}-p_{lI}\right) {\mathcal{P}}_{lk}+ \mathcal{G}_{lk}, \label{eq:six_sk}\\
   \rho_{l} \frac{\mathrm{D}_{l} \vc{u}_{l}}{\mathrm{D} t} = - \nabla\cdot \tensor{P}_l  + {\mathbf{R}}_{l},\label{eq:six_uk}\\
  \alpha_{lk}\frac{\mathrm{D}_{l} p_{lk}}{\mathrm{D} t}=  - \alpha_{lk} \rho_{lk} a_{lk}^2 \nabla\cdot \vc{u}_{l} - \left[\rho_{lk} a_{lI}^2 + \Gamma_{lk}\left(
  p_{lI} - \tilde{p}_{lI}\right) \right] {\mathcal{P}}_{lk} + \Gamma_{lk} \mathcal{G}_{lk},\label{eq:six_pk}\\
  \frac{\mathrm{D}_{l} \alpha_{lk}}{\mathrm{D} t} = \mathcal{P}_{lk} \label{eq:six_alpk},
\end{subeqnarray}
where \[\tensor{P}_l = \sum_k \alpha_{lk} \overline{\overline{P}}_{lk}, \quad {\mathbf{R}}_l = \sum_k \alpha_{lk} {\mathbf{R}}_{lk}.\]

With respect to variables $[\alpha_{lk}\rho_{lk}, \; \alpha_{lk}\rho_{lk}\vc{u}_l, \; \alpha_{lk}\rho_{lk} \varepsilon_{lk}, \; \alpha_{lk}]$,
\begin{subeqnarray}\label{eq:six_eq_model}
  \dudx{\alpha_{lk}\rho_{lk}}{t} +  \dvv{\alpha_{lk}\rho_{lk}\vc{u}_{l}} = 0, \\
  \dudx{\rho_{l}\vc{u}_{l}}{t} + \dvv{\rho_{l} \vc{u}_{l} \vc{u}_{l}} + \dvvo{ \overline{\overline{P}}_{l}} = \vc{R}_{l},  \\
    \dudx{\alpha_{lk}\rho_{lk} {\varepsilon}_{lk}}{t} + \dvv{\alpha_{lk} \rho_{lk} {\varepsilon}_{lk} \vc{u}_{l}} +
   \alpha_{lk}{{p}}_{lk} \nabla   \cdot\vc{u}_{l}  =  {\color{black}\mathcal{G}_{lk}} - {\color{black}\tilde{p}_{lI}\mathcal{P}_{lk}},   \\
  \dudx{\alpha_{lk}}{t} + {\color{black}\vc{u}_{l} \cdot \grado{\alpha_{lk}}} = {\color{black}\mathcal{P}_{lk}}.
\end{subeqnarray}

\subsubsection{The pu-eq model}
The model can be further simplified by assuming instantaneous mechanical (pressure and velocity) relaxations within each type of particle ($l=i$ or $l=e$). The derivation procedure is totally similarly to that of \cite{kapila2001two}, thus details are omitted.
In the limit $\tau_{lk,lk^{\prime}}^{p} = \tau_{lk,lk^{\prime}}^{u} \to 0$, one can obtain the following equations
\begin{subeqnarray}
  \DuDx{l}{s_{lk}}{t} &=& \frac{\mathcal{G}_{lk}}{\alpha_{lk}\rho_{lk}T_{lk}},\label{eq:reduced_5ion_entropy} \\
  \DuDx{l}{\vc{u}_l}{t} &=& - \frac{1}{\rho_l}\dvvo{\overline{\overline{P}}_l} + \frac{\vc{R}_{l}}{\rho_l}, \label{eq:reduced_5ion_mom}\\
  \DuDx{l}{p_l}{t} &=& -A_l\dvvo{\vc{u}_l} + A_l\sum_k \frac{\Gamma_{lk}\mathcal{G}_{lk}}{A_{lk}},\label{eq:reduced_5ion_pres} \\
  \DuDx{l}{\alpha_{lk}}{t} &=&  \mathcal{P}_{lk}^{(1)}. \label{eq:reduced_5ion_vol}
\end{subeqnarray}
The right hand side of eq. (\ref{eq:reduced_5ion_vol}) represents the motion of the interface under the first order pressure disequilibrium
\begin{equation}\label{eq:Flk1}
\mathcal{P}_{lk}^{(1)} = p_{lk^{\prime}}^{(1)} - p_{lk}^{(1)} = \alpha_{lk} \frac{A_l-A_{lk}}{A_{lk}}\dvvo{\vc{u}_l} + A_l\sum_{m\neq k} \frac{\mathcal{G}_{lk}\alpha_{lm} - \mathcal{G}_{lm}\alpha_{lk}}{A_{lm}A_{lk}},
\end{equation}
where \[A_{lk} = {\rho}_{lk}a_{lk}^2, \;\; 1/A_l = \sum_k {\alpha}_{lk}/A_{lk}. \]

By using the relation
\[\dero{p} = \rho \Gamma T \dero{s} + \gamma p / \rho \dero{\rho}, \]
and equations in (\ref{eq:reduced_5ion_entropy}), one can obtain
\begin{equation}\label{eq:reduced_5ion_mass}
  \DuDx{l}{\alpha_{lk}\rho_{lk}}{t} = - \alpha_{lk}\rho_{lk} \dvvo{\vc{u}_l}.
\end{equation}

From equation (\ref{eq:reduced_5ion_mass}) follows
\begin{equation}\label{eq:DykDt}
  \DuDx{l}{y_{lk}}{t} = 0, \quad y_{lk} = \frac{\alpha_{lk}\rho_{lk}}{\rho_l}.
\end{equation}

The mixture ion internal energy is
\begin{equation}\label{eq:mix_ion_inten}
 \DuDx{l}{\varepsilon_l}{t} = \DuDx{l}{\left(\sum{y_{lk}\varepsilon_{lk}}\right)}{t} = \sum y_{lk} \DuDx{l}{\varepsilon_{lk}}{t}.
\end{equation}

By using the Gibbs relation, one can deduce
\begin{equation}\label{eq:reduced_5ion_inten}
  \dudx{ \rho_{l} \varepsilon_l}{t} + \dvv{ \rho_{l} \varepsilon_l \vc{u}} + p_l \dvvo{\vc{u}} = \mathcal{G}_{l},
\end{equation}
where $\mathcal{G}_{l} = \sum_k \mathcal{G}_{lk} = \sum_k \left(\mathcal{S}_{lk} + {\alpha_{lk}}\mathcal{Q}_{lk} + q_{lk} \right)$, $\mathcal{Z}_{lk}$ is canceled out since $\sum_k \mathcal{Z}_{lk} = 0$.

Thus, the model can be rewritten in the following form:
\begin{subeqnarray}
  \dudx{\alpha_{lk}\rho_{lk}}{t} +   \dvv{\alpha_{lk}\rho_{lk}\vc{u}_l} = 0, \\
   \dudx{\rho_l\vc{u}_l}{t} + \dvv{\rho_l \vc{u}_l \vc{u}_l} + \dvvo{\overline{\overline{P}}_l} = 0,\\
  \dudx{ \rho_l \varepsilon_l}{t} + \dvv{ \rho_l \varepsilon_l \vc{u}_l} + p_l \dvvo{\vc{u}_l} = \mathcal{G}_{l}, \\
  \dudx{\alpha_{lk}}{t} + \vc{u}_l \cdot \grado{\alpha_{lk}} =  \mathcal{P}_{lk}^{(1)},\label{eq:fiveEqn:vol}
\end{subeqnarray}
where the last equation for volume fraction is only written for volume fraction of one species.

The total energy equation can be obtained by manipulating mass, momentum and internal energy equation,
\begin{equation}\label{eq:totEl}
  \dudx{\rho_l E_l}{t} + \dvv{\rho_l E_l \vc{u}_l + \tensor{P}_{l} \cdot \vc{u}_l + \sum{\alpha_{lk}\vc{q}_{lk}}} = \sum_k{{\alpha_{lk}}Q_{lk}}.
\end{equation}

An equivalent form of this model can be derived by invoking the Gibbs relation
\begin{subeqnarray}
  \dudx{\alpha_{lk}\rho_{lk}}{t} +   \dvv{\alpha_{lk}\rho_{lk}\vc{u}_l} = 0, \\
   \dudx{\rho_l\vc{u}_l}{t} + \dvv{\rho_l \vc{u}_l \vc{u}_l} + \dvvo{\overline{\overline{P}}_l} = 0,\\
  \dudx{ \alpha_{lk}\rho_{lk} \varepsilon_{lk}}{t} + \dvv{ \alpha_{lk} \rho_{lk} \varepsilon_{lk} \vc{u}} + \alpha_{lk} p_{l} \dvvo{\vc{u}_l} = \mathcal{G}_{lk} - p_{l} \mathcal{P}_{lk}^{(1)}.
\end{subeqnarray}

Moreover, the last equation can be replaced by the following  total energy equation for species $lk$
\begin{eqnarray}\label{eq:totAlphalEl}
  \dudx{\alpha_{lk}\rho_{lk}E_{lk}}{t} + \dvv{\alpha_{lk}\rho_{lk}E_{lk}\vc{u}_l +  \alpha_{lk}\tensor{P}_{lk}\cdot \vc{u}_l + \alpha_{lk} \vc{q}_{lk}}  \nonumber\\ = \left( \alpha_{lk} - y_{lk} \right) \vc{u}_l \cdot \grado{p_l} + p_l \vc{u}_l \cdot \grado{\alpha_{lk}} - p_l \mathcal{P}_{lk}^{(1)} + \mathcal{G}_{lk}.
\end{eqnarray}

\subsubsection{The puT-eq model and the puT$\gamma$-eq model}
On the basis of the above reduced model in the limit of $\tau_{lk,lk^{\prime}}^{u} = \tau_{lk,lk^{\prime}}^{p} \to 0$, we further assume that  $\tau_{lk,lk^{\prime}}^{T} \to 0$, then we obtain the following model in both mechanical and thermal equilibrium
\begin{subeqnarray}
  \dudx{y_{lk}\rho_{l}}{t} +   \dvv{y_{lk}\rho_{l}\vc{u}_l} = 0, \\
   \dudx{\rho_l\vc{u}_l}{t} + \dvv{\rho_l \vc{u}_l \vc{u}_l} + \dvvo{\overline{\overline{P}}_l} = 0,\\
  \dudx{ \rho_l \varepsilon_l}{t} + \dvv{ \rho_l \varepsilon_l \vc{u}} + p_l \dvvo{\vc{u}_l} = \mathcal{G}_{l}.
\end{subeqnarray}

If iso-compressibility is additionally assumed, then we arrive at
\begin{subeqnarray}
  \dudx{\rho_{l}}{t} +   \dvv{\rho_{l}\vc{u}_l} = 0, \\
   \dudx{\rho_l\vc{u}_l}{t} + \dvv{\rho_l \vc{u}_l \vc{u}_l} + \dvvo{\overline{\overline{P}}_l} = 0,\\
  \dudx{ \rho_l \varepsilon_l}{t} + \dvv{ \rho_l \varepsilon_l \vc{u}} + p_l \dvvo{\vc{u}_l} = \mathcal{G}_{l}.\label{eq:reduced_ele3_inten}
\end{subeqnarray}

In general, it is sufficient to describe the electron gas with the simplest model in mechanical, thermal and compressibility equilibrium (puT$\gamma$-eq model). However, discontinuity in electron temperature may arise due to ion-election relaxations.
As for ions, the disequilibrium between species is maintained to different levels according to the physical context.

\subsection{Reductions of the interactions between ions and electrons}
\subsubsection{Instantaneous velocity relaxation between ions and electrons}\label{sec:A2}
The mixture electron momentum equation
\begin{equation}
\dudx{\rho_e\vc{u}_e}{t} + \dvv{\rho_e \vc{u}_e \vc{u}_e} + \dvvo{\overline{\overline{P}}_e} = \mathbf{R}_e, \label{eq:reduced_ele3_mom}
\end{equation}
where \[\rho_e \vc{u}_e = \sum \alpha_{ek} \rho_{ek} \vc{u}_{ek}, \quad \overline{\overline{P}}_e = \sum_k \alpha_{ek} \overline{\overline{P}}_{ek}, \quad \mathbf{R}_e = \sum_k \alpha_{ek} \mathbf{R}_{ek}.\]

The momentum equation for electrons that coexist in space with the ions ``$ik$'' can be obtained by
\[\int X_{ik} \left[ \dudx{\rho_e\vc{u}_e}{t} + \dvv{\rho_e \vc{u}_e \vc{u}_e} + \dvvo{\overline{\overline{P}}_e} \right] \textrm{d}\vc{x} = \int X_{ik} \mathbf{R}_e \textrm{d}\vc{x},\]
which leads to
\begin{equation}
\dudx{{\alpha_{ik}}\widetilde{\rho}_{ek} \widetilde{\vc{u}}_{ek}}{t} + \dvv{\alpha_{ik}\widetilde{\rho}_{ek} \widetilde{\vc{u}}_{ek} \widetilde{\vc{u}}_{ek}} + \dvv{\alpha_{ik}\widetilde{\overline{\overline{P}}}_{ek}} = {\alpha_{ik}} \widetilde{\mathbf{R}}_{ek}.
\end{equation}

Here one should note the difference between the variables with/without the overhead tilde. The latter represent mixture electron quantities averaged in the volume occupied by $k$-th ion, i.e.,
\[ \widetilde{\Phi}_{ek} = \frac{1}{V}\int_{V} X_{ik} \Phi_e \textrm{d}V, \quad \Phi_e = \sum \alpha_{ek}\Phi_{ek}, \quad \Phi = \rho, \rho\vc{u}, \tensor{P}. \]

The ion-electron coupling reduction is based on the fact that the ion/electron velocities equilibrate much faster than their temperatures. This means that $\tau^{u}_{i,e} = 1/\mu^{u}_{i,e} \to 0$.
The asymptotic analysis is performed in the similar manner to Kapila's work with $\tau^{u}_{i,e}$ as the small parameter. Taking the difference of the momentum equations, we find that

In the order of $\mathcal{O}(1/\tau^{u}_{i,e})$, we have
\begin{equation}\label{eq:u_ie_eq}
\vc{u}_{ik}^{(0)} = \widetilde{\vc{u}}_{ek}^{(0)} = \vc{u}_{k}^{(0)} = \frac{\widetilde{\rho}_{ek}\widetilde{\vc{u}}_{ek} + \rho_{ik}\vc{u}_{ik}}{\widetilde{\rho}_{ek} + \rho_{ik}}.
\end{equation}

In the order of $\mathcal{O}(1)$, we have
\begin{eqnarray}\label{eq:Re1}
{\alpha_{ik}}\vc{R}_{ik} = - {\alpha_{ik}}\widetilde{\vc{R}}_{ek} = \frac{\widetilde{\rho}_{ek}\left(\dvv{\alpha_{ik}\tensor{P}_{ik}}  - \tensor{P}_{iI} \cdot  \grado{\alpha_{ik}} - \mathbf{M}_{ik} \right) }{\widetilde{\rho}_{ek} + \rho_{ik}} \nonumber\\
- \frac{ \rho_{ik}\left( \dvv{\alpha_{ik}\widetilde{\tensor{P}}_{ek}} - \widetilde{\tensor{P}}_{eI} \cdot  \grad{\alpha_{ik}} - \widetilde{\mathbf{M}}_{ek}\right)}{\widetilde{\rho}_{ek} + \rho_{ik}},
\end{eqnarray}
where the superscript ``(0)'' for right-hand side variables are omitted for simplicity.


If given initial distribution $X_{ek}|_{t=0} = X_{ik}|_{t=0}$ and $\alpha_{ek}|_{t=0} = \alpha_{ik}|_{t=0}$, the ion and electron characteristic function equations (and volume fraction equations) are reduced to be one equation in order O(1) thanks to the instantaneous ion-electron velocity relaxation.

In order O(1), we have
\begin{equation}
\dudx{X_k}{t} + \vc{u}_{I}^{\sigma} \cdot \grado{X_k} = 0, \quad \vc{u}_I^{\sigma} = \frac{\rho_{ik}\vc{u}_{iI}^{\sigma} + \widetilde{\rho}_{ek}\vc{u}_{eI}^{\sigma}}{\rho_{ik} + \widetilde{\rho}_{ek}}.
\end{equation}

Following \cite{CHINNAYYA2004490}, the pressure relaxation $\mathcal{P}_{lk}$ comes from the fluctuations of the interface velocity, thus
\begin{equation}\label{eq:Pek_Pik}
\mathcal{P}_{ik} = \mathcal{P}_{ek} = \mathcal{P}_{k} =  \frac{\rho_{ik}\mathcal{P}_{ik} + \widetilde{\rho}_{ek}\widetilde{\mathcal{P}}_{ek}}{\rho_{ik} + \widetilde{\rho}_{ek}} = (\vc{u}_I - \vc{u}_I^{\sigma})\nabla X_k.    
\end{equation}

Thus, the ion-electron velocity relaxation leads to the following ion-electron density weighted volume fraction equation
\begin{eqnarray}\label{eq:vof_eqns}
\dudx{\alpha_{k}}{t} + \vc{u}_{I} \cdot \grado{\alpha_{k}} =  \mathcal{P}_{k}.
\end{eqnarray}

By using eq. (\ref{eq:presRelax}) and assuming \[O\left(\frac{p_{e1}-p_{e2}}{p_{i1} - p_{i2}}\right) \sim O\left(1\right),\] it can be estimated
\begin{equation}\label{eq:neglect_pek_on_vol}
 O\left(\frac{\widetilde{\rho}_{ek}\widetilde{\mathcal{P}}_{ek}}{\rho_{ik}\mathcal{P}_{ik}}\right) \sim O\left(\sqrt{\frac{\rho_{ek}}{\rho_{ik}}}\right),
\end{equation}
and
\begin{equation}\label{eq:mu_p_ratio}
 O\left(\frac{\mu^{p}_{e1,e2}}{\mu^{p}_{i1,i2}}\right) \sim O\left(\sqrt{\frac{\rho_{ik}}{\rho_{ek}}}\right).
\end{equation}


\subsubsection{Neglecting electron mass and momentum}\label{sec:A3}
The mass of electron is negligibly small in comparison with that of ions. For example, for Deuterium, $m_i/m_e \approx 3600$. Moreover, due to plasma neutrality  $n_e = Z n_i$, the number densities of electron and ion are comparable in plasmas consisting of carbon and hydrogen. This means that $\rho_e << \rho_i$ and $\rho_{ek} << \rho_{ik}$. Therefore, we can further simplify the model by neglecting the electron mass and momentum. By doing this, we have the following approximation
\begin{subeqnarray}
  \rho_k = \rho_{ik} + \rho_{ek} \approx \rho_{ik}, \label{eq:rhoek_neg}\\
   {\alpha_k}\vc{R}_{ik} \approx - \dvv{\alpha_{ik}\tensor{P}_{ek}} + \tensor{P}_{eI} \cdot  \grad{\alpha_{ik}} + \mathbf{M}_{ek}, \label{eq:Rik}\\
   {p}_{ek} \approx {{p}}_{ek^{\prime}}.  \label{eq:Pe_eq}
\end{subeqnarray}

Pressure equilibrium in eqs. (\ref{eq:Pe_eq}) follows from the estimation in eq. (\ref{eq:mu_p_ratio}), which means the electron pressure relaxation  happens instantaneously in comparison with those of ions.

To omit the dependence on electron density $\rho_{ek}$, we have to reformulate the electron EOS:
\begin{equation}\label{eq:sim_eEOS}
  p_{ek} = \bracein{\gamma_e - 1} \rho_{ek} \varepsilon_{ek} = \bracein{\gamma_e - 1} \rho_{ik} \varepsilon_{ek}^{\prime}, \quad \varepsilon_{ek}^{\prime} =  C_{v{ek}}^{\prime} T_{ek}
\end{equation}
\begin{equation}\label{eq:Cvp}
  C_{v{ek}}^{\prime} = \frac{Z_{k}R}{A_{ik} \left( \gamma_e -1 \right)}, \quad R = k_B N_A,
\end{equation}
where $Z_{k}$ is the ionization degree, $R$ is the universal gas constant.

The sound speed $a_{{ek}}^{\prime}$ is defined as
\begin{equation}\label{eq:e_soundSpeed}
  \rho_{ek} a_{ek}^2 = \gamma_e p_{ek}  = \rho_{ik} \left(a_{{ek}}^{\prime}\right)^2
\end{equation}
\!

With such definitions and the reduced model equation (\ref{eq:alpha_rho_ep_lk}) (with $l=e$), we obtain
\begin{eqnarray}\label{eq:reduced_3ele_intenp}
 \dudx{\alpha_{k}\rho_{ik} \varepsilon_{ek}^{\prime}}{t} + \dvv{\alpha_{k}\rho_{ik} \varepsilon_{ek}^{\prime} \vc{u}_{k}}
 + \left( \alpha_{k}\overline{\overline{P}}_{ek} \cdot \nabla \right) \cdot\vc{u}_{k} + \dvv{\alpha_{k}\vc{q}_{ek}} = \nonumber\\ {\color{black}\alpha_{k}{Q}_{ek}^{\varepsilon}} + {\color{black}p_{eI}\left(\vc{u}_{I} - \vc{u}_{k}\right)\cdot\grado{\alpha_{k}}+ {\color{black}\left(\vc{u}_{I} - \vc{u}_{k} \right)} \cdot\left({\overline{\overline{\Pi}}}_{eI}\cdot\grado{\alpha_{k}}\right)} + {\color{black}\vc{q}_{eI}\cdot \nabla\alpha_{k}} \nonumber\\ - {\color{black}\tilde{p}_{eI}\mathcal{P}_{ek}} + {\color{black}\left(\tilde{\vc{u}}_{I} - \vc{u}_{k}\right)\cdot\mathbf{M}_{ek}} + {\color{black}\mathcal{Z}_{ek}}.
\end{eqnarray}

The electron internal energy equation can also be recast into a conservative formulation  with the electron  entropy $s_{ek}^{\prime}$
\begin{eqnarray}\label{eq:sep}
T_{ek} \left[ \dudx{\alpha_{k} \rho_{ik} s_{ek}^{\prime}}{t} + \dvv{\alpha_{k} \rho_{ik}  s_{ek}^{\prime} \vc{u}_{k}} \right] 
 + \left( \alpha_{k}\overline{\overline{\Pi}}_{ek} \cdot \nabla \right) \cdot\vc{u}_{k} + \dvv{\alpha_{k}\vc{q}_{ek}} = \nonumber\\ {\color{black}\alpha_{k}{Q}_{ek}^{\varepsilon}} + {\color{black}\left(p_{eI}-p_{ek}\right)\left(\vc{u}_{I} - \vc{u}_{k}\right)\cdot\grado{\alpha_{k}}+ {\color{black}\left(\vc{u}_{I} - \vc{u}_{k} \right)} \cdot\left({\overline{\overline{\Pi}}}_{eI}\cdot\grado{\alpha_{k}}\right)} + {\color{black}\vc{q}_{eI}\cdot \nabla\alpha_{k}} \nonumber\\ + \left(p_{ek}  - \tilde{p}_{eI} \right){\color{black}  \mathcal{P}_{ek}} + {\color{black}\left(\tilde{\vc{u}}_{I} - \vc{u}_{k}\right)\cdot\mathbf{M}_{ek}} + {\color{black}\mathcal{Z}_{ek}}, \label{eq:sek_prime}
\end{eqnarray}
where \[\textrm{exp}\left( \frac{s_{ek}^{\prime}}{C_{vek}} \right) = \frac{\gamma_e - 1}{\rho_{ik}^{\gamma_e-1}} \varepsilon_{ek}^{\prime}.\]

\subsection{Some reduced multi-plasma models}\label{sec:red_models}
In the above sections we have reduced the models for each particle species and their interaction. Combining these results we can obtain the following typical reduced multi-plasma models that are approximations  of the parent model in different levels:
\begin{enumerate}[label=\roman*.]
  \item Ion fully disequilibrium model + electron energy equations = the multiple-plasma nine-equation model (the BNZ model),
  \item Ion u-eq model + electron energy equations = the multiple-plasma eight-equation model,
  \item Ion pu-eq model + mixture electron energy equation = the multiple-plasma six-equation model (the KZ model),
  \item Ion puT-eq model + mixture electron energy equation = the multiple-plasma five-equation model,
  \item Ion fully disequilibrium model + electron energy equations + diffusion velocity order analysis = the multiple-plasma eight-equation model for inter-penetration mixing.
\end{enumerate}

All these models are compatible with the second law of thermodynamics, as shown in \ref{appA}.
We present these reduced models in the following.

\subsubsection{Nine-equation model for multiple-plasma flows in full disequilibrium}\label{sec:multi_plasma_eight_eqn}

Keeping the ions of different species in full disequilibrium,  and using the approximations (\ref{eq:rhoek_neg}) and eq. (\ref{eq:reduced_3ele_intenp}), we obtain a nine-equation model as follows
\begingroup
\allowdisplaybreaks
\begin{subeqnarray}\label{eq:nine_eqn_plasma_model}
  && \dudx{\alpha_{k}\rho_{ik}}{t} +  \dvv{\alpha_{k}\rho_{ik}\vc{u}_{k}} = 0, \\
  && \dudx{\alpha_{k}\rho_{ik}\vc{u}_{k}}{t} + \dvv{\alpha_{k}\rho_{ik} \vc{u}_{k} \vc{u}_{k}} + \dvv{ \alpha_{k} \overline{\overline{P}}_{k} } = {\color{black}{\overline{\overline{P}}}_{I}\cdot\grado{\alpha_{k}}} + {\color{black}\mathbf{M}_{k}}, \label{eq:nine_eqn_plasma_model:momk}\\
  && \dudx{\alpha_{k}\rho_{ik} E_{ik}}{t} + \dvv{\alpha_{k} \rho_{ik} E_{ik} \vc{u}_{k}} + \dvv{\alpha_{k}\tensor{{P}}_{ik}  \cdot\vc{u}_{k}} + \dvv{\alpha_{k}\vc{q}_{ik}} = \nonumber\\
 &&  {\color{black}\alpha_{k} Q_{ik}^{\varepsilon}  -    \left[\dvv{\alpha_{k}\tensor{{P}}_{ek}} - \tensor{{P}}_{eI}\cdot \grado{\alpha_k} - \mathbf{M}_{ek} \right] \cdot \vc{u}_k } \nonumber\\
 &&  + {\color{black} {\color{black}\vc{u}_{I}} \cdot\left({\overline{\overline{P}}}_{iI}\cdot\grado{\alpha_{k}}\right)} + {\color{black}\vc{q}_{iI}\cdot \nabla\alpha_{k}}
  - {\color{black}\tilde{p}_{iI}\mathcal{P}_{ik}} + {\color{black}\tilde{\vc{u}}_{I}\cdot\mathbf{M}_{ik}} + {\color{black}\mathcal{Z}_{ik}} \label{eq:nine_eqn_plasma_model:Eik},\\
  &&  \dudx{\alpha_{k}\rho_{ik} {\varepsilon}_{ek}^{\prime}}{t} + \dvv{\alpha_{k} \rho_{ik} {\varepsilon}_{ek}^{\prime} \vc{u}_{k}} +
  \left( \alpha_{k}\tensor{{P}}_{ek} \cdot \nabla \right)  \cdot\vc{u}_{k} + \dvv{\alpha_{k}\vc{q}_{ek}} = {\color{black}\alpha_{k} Q_{ek}^{\varepsilon}} \nonumber\\ 
  && + {\color{black} {\color{black}\left(\vc{u}_{I}-\vc{u}_{k}\right)}
  \cdot\left({\overline{\overline{P}}}_{eI}\cdot\grado{\alpha_{k}}\right)} + {\color{black}\vc{q}_{eI}\cdot \nabla\alpha_{k}}
  - {\color{black}\tilde{p}_{eI}\mathcal{P}_{ek}} + {\color{black}\left(\tilde{\vc{u}}_{I} -\vc{u}_{k}\right)\cdot\mathbf{M}_{ek}} + {\color{black}\mathcal{Z}_{ek}} \label{eq:nine_eqn_plasma_model:epek},\\
  && \dudx{\alpha_{k}}{t} + {\color{black}\vc{u}_{I} \cdot \grado{\alpha_{k}}} = {\color{black}\mathcal{P}_{k}}. \label{eq:nine_eqn_plasma_model:vol}
\end{subeqnarray}
\endgroup

By summing energy equations for ion and electron and the kinetic energy, one obtains the total energy equation for the $k$-th component
\begin{eqnarray}
\dudx{\alpha_{k}\rho_{ik} E_{k}}{t} + \dvv{\alpha_{k} \rho_{ik} E_{k} \vc{u}_{k}} + \dvv{\alpha_{k}\tensor{{P}}_{k}  \cdot\vc{u}_{k}} + \dvv{\alpha_{k}\vc{q}_{k}} = \nonumber\\
  + {\color{black} {\color{black}\vc{u}_{I}} \cdot\left({\overline{\overline{P}}}_{I}\cdot\grado{\alpha_{k}}\right)} + {\color{black}\vc{q}_{I}\cdot \nabla\alpha_{k}}
  - {\color{black}\tilde{p}_{iI}\mathcal{P}_{ik}} - {\color{black}\tilde{p}_{eI}\mathcal{P}_{ek}} + {\color{black}\tilde{\vc{u}}_{I}\cdot\mathbf{M}_{k}} + {\color{black}\mathcal{Z}_{k}}, \label{eq:nine_eqn_plasma_model:Ek}
\end{eqnarray}
where \[E_k = \varepsilon_{ik} + \varepsilon_{ek}^{\prime} + |\vc{u}_k|^2 / 2 , \;\; \tensor{{P}}_{I} = \tensor{{P}}_{iI} + \tensor{{P}}_{eI},\]
\[\Phi_{k} = \Phi_{ik} + \Phi_{ek}, \;\; \Phi = \tensor{{P}}, \; \vc{{q}}, \; \mathbf{M}, \; \mathcal{Z}.\]

One can check that the above nine-equation model is a combination of the classical Baer-Nunziato model and the Zeldovich model in a thermodynamically compatible way, and thereby termed as Baer-Nunziato-Zeldovich (BNZ) model.

\subsubsection{Eight-equation model for multi-plasma flows in velocity equilibrium}\label{sec:multi_plasma_seven_eqn}
The combination of the ion u-eq model (\ref{eq:six_eq_model}) and the electron energy equations leads to
\begin{subeqnarray}\label{eq:eight_eq_model}
 && \dudx{\alpha_{k}\rho_{ik}}{t} +  \dvv{\alpha_{k}\rho_{ik}\vc{u}} = 0, \\
  &&  \dudx{\rho_{i}\vc{u}}{t} + \dvv{\rho_{i} \vc{u} \vc{u}} + \dvvo{\tensor{P}}   =  0, \\
  &&  \dudx{\alpha_{k}\rho_{ik} {\varepsilon}_{ik}}{t} + \dvv{\alpha_{k} \rho_{ik} {\varepsilon}_{ik} \vc{u}} +
  \left( \alpha_{k}\tensor{{P}}_{ik} \cdot \nabla \right)  \cdot\vc{u} + \dvv{\alpha_{k}\vc{q}_{ik}} = \nonumber\\ 
 &&  {\color{black}\alpha_k Q_{ik}^{\varepsilon}} + {\color{black}\vc{q}_{iI}\cdot \nabla\alpha_k} - {\color{black}\tilde{p}_{iI}\mathcal{P}_{ik}} + {\color{black}\mathcal{Z}_{ik}}, \\
 &&   \dudx{\alpha_{k}\rho_{ik} {\varepsilon}_{ek}^{\prime}}{t} + \dvv{\alpha_{k} \rho_{ik} {\varepsilon}_{ek}^{\prime} \vc{u}} +
  \left( \alpha_{k}\tensor{{P}}_{ek} \cdot \nabla \right)  \cdot\vc{u} + \dvv{\alpha_{k}\vc{q}_{ek}} = \nonumber\\ 
  && {\color{black}\alpha_{k}Q_{ek}^{\varepsilon}}
   + {\color{black}\vc{q}_{eI}\cdot \nabla\alpha_{k}}
  - {\color{black}\tilde{p}_{eI}\mathcal{P}_{ik}}  + {\color{black}\mathcal{Z}_{ek}}, \\
  && \dudx{\alpha_{k}}{t} + {\color{black}\vc{u} \cdot \grado{\alpha_{k}}} = {\color{black}\mathcal{P}_{k}}.
\end{subeqnarray}

The mixture total stress tensor
\[\tensor{P} = \sum_k \tensor{P}_k = \sum_k \left( \tensor{P}_{ik} + \tensor{P}_{ek}  \right).\]

This model can be viewed as the reduced equations of (\ref{eq:nine_eqn_plasma_model}) in the limit of instantaneous velocity relaxation. 

\subsubsection{Six-equation model for multi-plasma flows in mechanical equilibrium}\label{sec:multi_plasma_six_eqn}
For the above eight-equation model, if we further assume
\[\mu_{i,i}^{p} = \mu_{e,e}^{p} = \mu^{p} \to \infty, \]
we obtain the total pressure equilibrium in order $O(1/\mu^p)$, i.e., \[p_{i1} = p_{i2} = p_i, \; p_{e1} = p_{e2} = p_e\]

In order $O(1)$ one can similarly derive a reduced model in mechanical equilibrium as follows
\begin{subeqnarray}\label{eq:multi_plasma_six_eqn}
 && \dudx{\alpha_{k}\rho_{ik}}{t} +   \dvv{\alpha_{k}\rho_{ik}\vc{u}} = 0, \\
 &&  \dudx{\rho\vc{u}}{t} + \dvv{\rho \vc{u} \vc{u}} + \dvvo{\overline{\overline{P}}} = 0,\\
 && \dudx{\rho E}{t} + \dvv{\rho E\vc{u} +  {\overline{\overline{P}}}\cdot \vc{u} + \vc{q} }= 0, \\
  && \dudx{\rho {\varepsilon}_{e}^{\prime}}{t} + \dvv{\rho {\varepsilon}_{e}^{\prime} \vc{u}} +
  \left(\tensor{{P}}_{e} \cdot \nabla \right)  \cdot\vc{u} + \dvvo{\vc{q}_{e}} =  Q_{e}^{\varepsilon},\\
  && \dudx{\alpha_{ik}}{t} + \vc{u} \cdot \grado{\alpha_{ik}} = \mathcal{P}_{k}^{(1)},
\end{subeqnarray}
where we have used the following definitions
\[\rho E = \sum_k {\alpha_{k}\rho_{ik} E_{k}}, \;\; \rho {\varepsilon}_e = \sum_k {\alpha_{k}\rho_{ik} {\varepsilon}_{ek}}, 
\]
\[\vc{q} = \sum_k \alpha_{k} \vc{q}_{k}, \;\; \vc{q}_e = \sum_k \alpha_{k} \vc{q}_{ek}, \;\; Q_{e}^{\varepsilon} = {\color{black}\sum \alpha_k Q_{ek}^{\varepsilon}}\]
\[\mathcal{P}_{k}^{(1)} = \alpha_{k} \frac{A_i-A_{ik}}{A_{ik}}\dvvo{\vc{u}} + A_i\sum_{m\neq k} \frac{\mathcal{G}_{ik}\alpha_{m} - \mathcal{G}_{im}\alpha_{ik}}{A_{im}A_{ik}},\]

This model is termed as the Kapila-Zeldovich model (or the KZ model) since it is reduced to the Kapila's five-equation model \citep[][]{kapila2001two} in the non-HED condition. 
It contains two independent pressures, i.e., the ion pressure $p_i$ and electron pressure $p_{e}$.
The electron pressures $p_{e}$ can be calculated directly from the mixture electron internal energy $\rho {\varepsilon}_{e}^{\prime}$,
\[\sum_k \frac{\alpha_k p_e}{\gamma_{ek} - 1} = \rho {\varepsilon}_{e}^{\prime}.\]
The ion pressures $p_{i}$ can be solved with the aid of the total pressure equilibrium
\[\sum_k \frac{\alpha_k p_i}{\gamma_{ik} - 1} = \rho {\varepsilon}_{i} = \rho E - \rho {\varepsilon}_{e}^{\prime} - \rho \frac{|\vc{u}|^2}{2}.\]

The ion and electron temperatures are calculated in the following way
\[
T_{lk} = T_{lk}\left( \rho_k, p_l \right), \;\; l = i,e.
\]

%
%

\subsubsection{Five-equation model for multi-plasma flows in mechanical and ion-temperature equilibrium}\label{sec:multi_plasma_five_eqn}
We further assume the ion temperatures of different species reach equilibrium instantaneously, so do the electron temperatures. Then one can derive a closed model in both mechanical equilibrium, ion-ion temperature equilibrium, and electron-electron temperature equilibrium. It reads as follows
\begin{subeqnarray}\label{eq:multi_plasma_five_eqn}
 && \dudx{y_{ik}\rho}{t} +   \dvv{y_{ik}\rho\vc{u}_i} = 0, \\
 &&  \dudx{\rho\vc{u}}{t} + \dvv{\rho \vc{u} \vc{u}} + \dvvo{\tensor{P}} = 0,\\
 && \dudx{ \rho \varepsilon_i}{t} + \dvv{ \rho \varepsilon_i \vc{u}} + p_i \dvvo{\vc{u}} = \mathcal{G}_{i}, \\
  &&\dudx{\rho \varepsilon_e^{\prime}}{t} + \dvv{\rho \varepsilon_e^{\prime} \vc{u}} + p_e \dvvo{\vc{u}} = \mathcal{G}_e.
\end{subeqnarray}
A conservative form can be obtained by replacing the last two equations by entropy equations.

\subsection{Model for interpenetration mixing}\label{sec:multi_plasma_inter_mix}

To consider the interpenetration mixing effect, one can directly solve the two-velocity nine-equation model. However, for simplicity and computation efficiency one may need a one-velocity model including ion interpenetration.
The derivation of such a model is similar to our previous work in \cite{ZhangC_PhysRevE2023}. It starts from the nine-equation model (\ref{eq:nine_eqn_plasma_model}). We replace $\vc{u}_k$ by $\vc{u} + \vc{w}_k$ and gather the diffusion-related terms containing $\vc{w}_k$. The obtained formulation reads
\begingroup
\allowdisplaybreaks
\begin{subeqnarray}
&&\dudx{\alpha_k\rho_{ik}}{t} + \nabla\cdot(\alpha_k\rho_{ik} {\vc{u}}) = - \nabla\cdot \vc{J}_{ik},\label{eq:avbn:mass}\\
&&\dudx{\alpha_k\rho_{ik} \vc{u}}{t} + \nabla\cdot\left( \alpha_k \rho_{ik} \vc{u} \vc{u} + \alpha_k {\tensor{P}}_{ak} \right) =  {\tensor{P}}_{iI} \cdot \nabla \alpha_k + \mathcal{M}_{ik}  \nonumber\\
&&+ \dudx{\alpha_k \rho_{ik} \vc{w}_k}{t} + \nabla \cdot \left( 2\alpha_k \rho_{ik} \vc{u} \vc{w}_k +  \alpha_k \rho_{ik} \vc{w}_k \vc{w}_k\right) - \nabla \cdot \left( \alpha_k {\tensor{P}}_{wk} \right), \label{eq:avbn:mom}
\\
&&\dudx{\alpha_k \rho_{ik} {\varepsilon}_{ik}}{t} + \nabla \cdot \left( \alpha_k \rho_{ik} {\varepsilon}_{ik} \vc{u} \right) = - \alpha_k p_{ik} \nabla \cdot \vc{u} - p_{iI} \mathcal{P}_k  + \vc{w}_k \cdot \mathcal{M}_{ik} \nonumber\\ &&+ \vc{w}_k \cdot \left( \tensor{P}_{iI} \cdot \nabla \alpha_k \right) - \nabla \cdot \left( \alpha_k \rho_{ik} {\varepsilon}_{ik} \vc{w}_k \right)
- \alpha_k p_{ik} \nabla \cdot \vc{w}_k + \mathcal{G}_{ik},
\\
&&\dudx{\alpha_k \rho_{ek} {\varepsilon}_{ek}^{\prime}}{t} + \nabla \cdot \left( \alpha_k \rho_{ek} {\varepsilon}_{ek}^{\prime} \vc{u} \right) = - \alpha_k p_{ek} \nabla \cdot \vc{u} - p_{eI} \mathcal{P}_{ek}  + \vc{w}_k \cdot \mathcal{M}_{ek} \nonumber\\ &&+ \vc{w}_k \cdot \left( \tensor{P}_{eI} \cdot \nabla \alpha_k \right) - \nabla \cdot \left( \alpha_k \rho_{ik} {\varepsilon}_{ek}^{\prime} \vc{w}_k \right)
- \alpha_k p_{ek} \nabla \cdot \vc{w}_k + \mathcal{G}_{ek},
\\&&\dudx{\alpha_k }{t} + \vc{u} \cdot \nabla\alpha_k = \mathcal{P}_k + (\vc{u} - \vc{u}_I)\cdot\nabla\alpha_k \label{eq:avbn:vol},
\end{subeqnarray}
\endgroup
where the diffusion flux \[ \vc{J}_{ik} = \alpha_k \rho_{ik} \vc{w}_{k}.\]

The total energy equation can be reformulated as
 \begin{eqnarray}\label{eq:bn:int_en1}
&&\dudx{\alpha_k \rho_{ik} {{E}}_{ak}}{t} +
  \nabla\cdot\left(
\alpha_k \rho_{ik} {{E}}_{ak} \vc{u}   + \alpha_k {\tensor{P}}_{ak} \cdot \vc{u}
  \right)
  =   \vc{u}_{I} \cdot \left( {\tensor{P}}_I \cdot \nabla {\alpha_k} \right) \nonumber\\&&+ \vc{q}_{I}\cdot\grado{\alpha_k} - p_{iI} \mathcal{P}_{ik} - p_{eI} \mathcal{P}_{ek} + \vc{u} \cdot \mathcal{M}_k - \vc{u} \cdot \mathcal{M}_{ek} + q_k + \mathcal{Z}_{k}
  \nonumber\\
   &&{\color{black}-} \dudx{\alpha_k \rho_{ik} {E}_{wk} }{t} - \nabla \cdot \left( \alpha_k \rho_{ik}  \left(  \vc{u} E_{wk} + {E}_{ak} \vc{w}_k + E_{wk} \vc{w}_k \right) \right) \nonumber
\\&&- \nabla \cdot \left( \alpha_k {\tensor{P}}_{ak} \cdot \vc{w}_k + \alpha_k {\tensor{P}}_{wk} \cdot \vc{u} + \alpha_k {\tensor{P}}_{wk} \cdot \vc{w}_k \right) - \vc{w}_k \cdot \mathcal{M}_{ek} \label{eq:avbn:en},
\end{eqnarray}

A closure law is needed for the diffusion velocity $\vc{w}_k$. In the case of grain mixing we have shown that the ion velocities relax much faster than the temperature relaxation, by means of asymptotic expansion a pressure driven Darcy law can be obtained as in eq. (\ref{eq:grain_drfit_vel}) (see also \cite[][]{kapila2001two,GUILLARD2007288,saurel2012modelling}). 

In atomic mixing, the commonly used closure law is the Fick's law $\vc{w}_k = -D \nabla y_k / y_k$.
For moderate Mach number and collision of ions with comparable masses, it can be shown \citep[][]{kagan2014}
\begin{equation}\label{eq:w_k_kn}
\frac{|\vc{w}_k|}{u_{shock}} \approx  \frac{\lambda}{\Delta} = Kn \leq 0.001.
\end{equation}

As has been estimated in Figure \ref{fig:relaxation_time_atomic}, in atomic mixing the velocity relaxation time may be comparable with that of temperature. Therefore, we concentrate on the atomic diffusion in plasma flows.

The last inequality in (\ref{eq:w_k_kn}) is a basic assumption in continuum mechanics. With such scale estimation, in the following we will drop the terms of order  $\mathcal{O}(|\vc{w_k}^2|)$. The velocity relaxation is finite under the concerned scenario and is estimated to be $\mathcal{O}\left( 1 \right)$. The phase densities are assumed to be of the same order, thus, $O\left(\vc{w}_k\right)\sim O\left(\nabla y_k\right) \sim O\left(\nabla \alpha_k\right)$.
Moreover, for simplicity we choose $\vc{u}_I = \vc{u}$ for derivation of the following model. With other $\vc{u}_I$ definitions similar derivations can be performed. 

The general idea is to reduce the nine-equation model via order analysis. More concretely, we will drop $\mathcal{O}(|\vc{w_k}^2|)$ terms and close $\mathcal{O}(|\vc{w_k}|)$ terms with the established diffusion laws.
In the reduced model we only retain the mass-weighted velocity with the diffusion velocity determined by the gradients of the average field.
The reduced model takes the following form
\begingroup
\allowdisplaybreaks
\begin{subeqnarray}\label{eq:reduced_avbn}
&&\dudx{\alpha_k\rho_{ik}}{t} + \nabla\cdot(\alpha_k\rho_{ik} {\vc{u}}) = {\color{black} - \nabla\cdot \left( \alpha_k \rho_{ik} \vc{w}_{k} \right) },  \label{eq:reduced_avbn:mass}\\
&&\dudx{\rho \vc{u}}{t} + \nabla\cdot\left( \rho \vc{u} \vc{u} + \sum \alpha_k {\tensor{P}}_{ak} \right) = {\color{black} - \nabla \cdot \left( \sum \alpha_k {\tensor{P}}_{wk} \right)}, \label{eq:reduced_avbn:mom} 
\\
&&\dudx{\alpha_k \rho_{ik} {\varepsilon}_{ik}}{t} + \nabla \cdot \left( \alpha_k \rho_{ik} {\varepsilon}_{ik} \vc{u} \right) + \alpha_k p_{ik} \nabla \cdot \vc{u}\nonumber \\ &&=  - p_{iI} \mathcal{P}_{ik}  {\color{black}- \nabla \cdot \left( \alpha_k \rho_{ik} {\varepsilon}_{ik} \vc{w}_k \right) - \alpha_k p_{ik} \nabla \cdot \vc{w}_k } + \mathcal{G}_{ik},   \label{eq:reduced_avbn:inten}
\\
&&\dudx{\alpha_k \rho_{ek} {\varepsilon}_{ek}^{\prime}}{t} + \nabla \cdot \left( \alpha_k \rho_{ik} {\varepsilon}_{ek}^{\prime} \vc{u} \right) + \alpha_k p_{ek} \nabla \cdot \vc{u} \nonumber\\  &&=  - p_{eI} \mathcal{P}_{ek}  {\color{black} - \nabla \cdot \left( \alpha_k \rho_{ik} {\varepsilon}_{ek}^{\prime} \vc{w}_k \right) - \alpha_k p_{ek} \nabla \cdot \vc{w}_k } + \mathcal{G}_{ek}, 
\\ &&\dudx{\alpha_k }{t} + \vc{u} \cdot \nabla\alpha_k = \mathcal{P}_k \label{eq:reduced_avbn:vol}. 
\end{subeqnarray}
\endgroup



If we further assume the ion temperature/pressure equilibrium across an effective interface, we then obtain
\begin{subeqnarray}\label{eq:reduced_five_diffusion}
&&\dudx{\alpha_k\rho_{ik}}{t} + \nabla\cdot(\alpha_k\rho_{ik} {\vc{u}}) = {\color{black}- \nabla\cdot \left( \alpha_k \rho_{ik} \vc{w}_{k}  \right)},\label{eq:reduced_avbnPT:mass}\\
&&\dudx{\rho \vc{u}}{t} + \nabla\cdot\left( \rho \vc{u} \vc{u} + \sum \alpha_k {\tensor{P}}_{ak} \right) =  {\color{black}\nabla \cdot \left( \sum \alpha_k {\tensor{P}}_{wk} \right)}, \label{eq:reduced_avbnPT:mom}
\\
&&\dudx{\rho {\varepsilon}_i}{t} + \nabla \cdot \left(\rho {\varepsilon}_i \vc{u} \right) + p_i \nabla \cdot \vc{u} \nonumber\\ &&= {\color{black} - \nabla \cdot \left( \sum\alpha_k \rho_{ik} {\varepsilon}_{ik} \vc{w}_k \right) - \sum\alpha_k p_{ik} \nabla \cdot \vc{w}_k} + \mathcal{G}_{i},  \label{eq:reduced_avbnPT:inten_i}
\\
&&\dudx{\rho {\varepsilon}_e^{\prime}}{t} + \nabla \cdot \left(\rho {\varepsilon}_e^{\prime} \vc{u} \right) + p_e \nabla \cdot \vc{u} \nonumber\\&&= {\color{black} - \nabla \cdot \left( \sum\alpha_k \rho_{ik} {\varepsilon}_{ek}^{\prime} \vc{w}_k \right) - \sum\alpha_k p_{ek} \nabla \cdot \vc{w}_k} + \mathcal{G}_{e}. \label{eq:reduced_avbnPT:inten_e}
\end{subeqnarray}


\section{Mathematical analysis of the derived models}
In this section we perform analysis on the characteristic properties of the derived models, including the nine-equation BNZ model, the eight-equation model, the six-equation KZ model, the five-equation model and the four equation model. The characteristic structures of the hydrodynamic part (without relaxations and dissipations)  are summarized in Table \ref{tab:modelChar}, which demonstrates their hyperbolicity. 
For more details see  \ref{appA}.
With the hyperbolicity property, Godunov  finite volume methods can be implemented for solving these models.

\begin{table}
\begin{center}
\begin{tabular}{lcccc}
\hline
  Model & Ion eqns & Elec. eqns & Assumptions & Wave speeds \\
  \hline
  Nine-eqn & 7 & 2 & Ions in full diseq. & \makecell[c]{$u_1, u_1, u_2, u_2, u_I$, \\ $u_1\pm a_{1}$, \\ $u_2\pm a_2$, \\$a_k^2 = a_{ik}^2 + (a_{ek}^{\prime})^2$} \\
  Eight-eqn & 6 & 2 & + Ion velocities eq. & \makecell[c]{$u (6 \; \textrm{times}),$ \\ $ u \pm \sqrt{y_1 a_{1}^2 + y_2 a_{2}^2}$} \\
  Six-eqn & 5 & 1 & + Ion/elec. mechanical eq. & \makecell[c]{$u (4 \; \textrm{times}),$ \\ $ u \pm \sqrt{ \frac{1}{ \rho \left({\alpha_1} / A_1 + {\alpha_2} / A_2 \right)}},$ \\ $A_k = \rho_k  \left( a_{ik}^2 + (a_{ek}^{\prime})^2\right)$} \\
  Five-eqn & 4 & 1 & \makecell[c]{+ Ion/elec. mechanical\\ and \\ thermal eq.} & \makecell[c]{$u (3 \; \textrm{times}),$ \\ $ u \pm \sqrt{\gamma_{i} p_i/\rho + a_e^2}$, \\$\gamma_i= \frac{\sum{y_k C_{vik} (\gamma_{ik} - 1)}}{\sum{y_k C_{vik}}} + 1$}  \\
  Four-eqn & 3 & 1 & \makecell[c]{+ Ion/elec. mechanical, \\ thermal and \\ compressibility eq.} & $u,u,u\pm \sqrt{a_i^2 + (a_{e}^{\prime})^2}$ \\
  \hline
\end{tabular}
\caption{characteristic structures of the derived models}
\label{tab:modelChar}
\end{center}
\end{table}

\section{Numerical approximation}\label{sec:numer_method}
The derived models takes the general form of a quasi-conservative hyperbolic-parabolic-relaxation PDE. We use the fractional step method to split the model into several subsystems including the hydrodynamic equations, the pressure relaxation equations, the velocity relaxation equations, and the temperature relaxation equations. Below we concentrate on the numerical methods for the hydrodynamic and relaxation subsystems of the nine-equation BNZ model. The parabolic parts including heat conduction, viscous dissipation and mass diffusion can be solved with the methodology in literature \citep[][]{ZhangC_PhysRevE2023,zhang2022a,petitpas2014discrete}.

\subsection{Hydrodynamic subsystem}
The hyperbolic hydrodynamic subsystem reads
\begingroup
\allowdisplaybreaks
\begin{subeqnarray}\label{eq:nine_eqn_plasma_model_hyd}
&&  \dudx{\alpha_{k}\rho_{ik}}{t} +  \dvv{\alpha_{k}\rho_{ik}\vc{u}_{k}} = 0, \label{eq:nine_eqn_plasma_model_hyd:massk}\\
&&  \dudx{\alpha_{k}\rho_{ik}\vc{u}_{k}}{t} + \dvv{\alpha_{k}\rho_{ik} \vc{u}_{k} \vc{u}_{k}} + \grad{ \alpha_{k} p_{k} } = 0, \label{eq:nine_eqn_plasma_model_hyd:momk}\\
&&  \dudx{\alpha_{k}\rho_{ik} E_{k}}{t} + \dvv{\alpha_{k} \rho_{ik} E_{k} \vc{u}_{k}} + \dvv{\alpha_{k}p_{k}\vc{u}_{k}}  = 0,\label{eq:nine_eqn_plasma_model_hyd:Ek}\\
 &&\dudx{\alpha_{k} \rho_{ik} s_{ek}^{\prime}}{t} + \dvv{\alpha_{k} \rho_{ik}  s_{ek}^{\prime} \vc{u}_{k}} =  0 ,\\
 && \dudx{\alpha_{k}}{t} + {\color{black}\vc{u}_{I} \cdot \grado{\alpha_{k}}} = 0, \label{eq:nine_eqn_plasma_model_hyd:vol}
\end{subeqnarray}
\endgroup
Note that here the equation for the total energy of $k$-th ion $E_{ik}$ is replaced by that for  the total energy of $k$-th component $E_k$ (see eq. (\ref{eq:nine_eqn_plasma_model:Ek})). Moreover, the equation for the internal energy of $k$-th electron $\varepsilon_{ek}^{\prime}$ is replaced by that for the electron entropy $s_{ek}^{\prime}$ (see eq. (\ref{eq:sek_prime})). Such equivalent reformulation allows a quasi-conservative form that provides more numerical convenience. For its numerical solution we use the finite volume method with the HLLEM approximate Riemann solver\citep[][]{Dumbser2016ANE}. Second-order accuracy is achieved with the MUSCL-Hancock scheme\citep[][]{toro2009}{}{}.

\subsection{Pressure relaxation subsystem}
The pressure relaxation subsystem is a stiff ODE system
\begin{subeqnarray}
 && \dudx{\alpha_{k}\rho_{ik}}{t}  = 0, \label{eq:nine_eqn_plasma_model_pr:massk}\\
 && \dudx{\alpha_{k}\rho_{ik}\vc{u}_{k}}{t} = 0, \label{eq:nine_eqn_plasma_model_pr:momk}\\
 && \dudx{\alpha_{k}\rho_{ik} \varepsilon_{ik}}{t}  = - \tilde{p}_{iI}\mathcal{P}_{ik},\label{eq:nine_eqn_plasma_model_pr:epi}\\
 && \dudx{\alpha_{k} \rho_{ik} \varepsilon_{ek}^{\prime}}{t}  =  - \tilde{p}_{eI}\mathcal{P}_{ek} ,\label{eq:nine_eqn_plasma_model_pr:epe}\\
 && \dudx{\alpha_{k}}{t}  = \mathcal{P}_{ik}. \label{eq:nine_eqn_plasma_model_pr:vol}
\end{subeqnarray}
Note that on the basis of the above estimation (eq. (\ref{eq:neglect_pek_on_vol})), the effect of the electron pressure relaxation on volume fraction is negligible, therefore, eq. (\ref{eq:nine_eqn_plasma_model_pr:epe}) is decoupled from the others and its sum
\begin{equation}
\dudx{\rho_{i} \varepsilon_{e}}{t}  = 0,\label{eq:nine_eqn_plasma_model_pr:sum_epe}
\end{equation}

Moreover according to estimation (\ref{eq:mu_p_ratio}), we assume electron pressure equilibrium. 
Then one can deduce
\begin{eqnarray}
&&{\left(\rho_{i} \varepsilon_{e}^{\prime}\right)}^{(0)}  = {\left(\rho_{i} \varepsilon_{e}^{\prime}\right)}^{(1)} = \sum_k \left( \alpha_{k} \rho_{ik} \varepsilon_{ek}^{\prime} \right)^{(1)} = \sum\frac{ \left(\alpha_k p_{ek}\right)^{(1)}}{(\gamma_{ek} - 1)},\\
&&p_{e1}^{(1)} = p_{e2}^{(1)} = p_{e}^{(1)},
\end{eqnarray}
where the superscript ``(0)'' and ``(1)'' represent the variables before and after the pressure relaxation, respectively.

Thus the relaxed electron pressure is  \[p_{e}^{(1)} = \frac{{\left(\rho_{i} \varepsilon_{e}^{\prime}\right)}^{(0)}}{\sum\frac{\alpha_k}{\gamma_{ek}-1}}.\]

The subsystem for the ion pressure relaxation reads
\begin{eqnarray}\label{eq:presIonRelax}
 && \dudx{p_{ik}}{t} = \zeta_{ik}\mu_{ik,ik^{\prime}}^p \left( p_{ik} - p_{ik^{\prime}} \right), \quad \zeta_{ik} =  \frac{ \rho_{ik}a_{iI}^2 + \Gamma_{ik}\left(p_{iI} - \tilde{p}_{iI}\right)}{\alpha_k}  \label{eq:presIonRelax:pk}\\
 && \dudx{\alpha_{k}}{t} = \mu_{ik,ik^{\prime}}^p \left( p_{ik} - p_{ik^{\prime}} \right). \label{eq:presIonRelax:alpk}
\end{eqnarray}


These nonlinear ODEs are solved with the simple iteration method. In $s$-th iteration the two ODEs (\ref{eq:presIonRelax:pk}) ($k=1,2$) are linearized by taking $\zeta_{ik} = \left(\zeta_{ik}\right)^{(1),s-1}$ and $\mu_{ik,ik^{\prime}}^{p} = \left(\mu_{ik,ik^{\prime}}^{p}\right)^{(1),s-1}$. These linearized ODEs  are solved implicitly to obtain $p_{ik}^{(1),s}$, then the volume fraction are updated according to
\[
\frac{\alpha_1^{(1),s} - \alpha_1^{(0)}}{p_{i1}^{(1),s} - p_{i1}^{(0)}} = \frac{1}{\left( \zeta_{ik} \right)^{(1),s}}  .
\]

Iterations  are performed until $|p_{i1}^{(1),s+1} - p_{i1}^{(1),s}| / {p_{i1}^{(0)}}  < 10^{-6}$.

\subsection{Velocity relaxation subsystem}
The velocity relaxation subsystem reads
\begin{subeqnarray}\label{eq:velRelax}
  &&\dudx{\alpha_{k}\rho_{ik}}{t} = 0, \label{eq:velRelax:mass}\\
  &&\dudx{\vc{u}_{k}}{t} = \nu_{ik,ik^{\prime}}^{u} \left( \vc{u}_{k^{\prime}} - \vc{u}_{k} \right), \quad \nu_{ik,ik^{\prime}} = \frac{\mu_{ik,ik^{\prime}}^u}{\alpha_{k}\rho_{ik}}  \label{eq:velRelax:vel},\\
  &&\dudx{\alpha_{k}\rho_{ik} E_{ik}}{t} = {\color{black}\nu_{ik,ik^{\prime}}^{u} \tilde{\vc{u}}_{iI} \left( \vc{u}_{k^{\prime}} - \vc{u}_{k} \right)}\label{eq:velRelax:en}, \\
  &&\dudx{\alpha_{k}}{t} = 0.
\end{subeqnarray}

Note that here the electron velocity relaxation terms are neglected due to \[ O\left(\frac{|\vc{M}_{ek}|}{|\vc{M}_{ik}|}\right)  = O\left(\sqrt{\frac{\rho_{ek}}{\rho_{ik}}}\right) << O\left(1\right).\]

This nonlinear ODE system is similar to the ion pressure relaxation subsystem and so are the numerical methods.  In $s$-th iteration the two ODEs for velocity ($k=1,2$) are linearized assuming $\nu_{ik,ik^{\prime}} = \left(\nu_{ik,ik^{\prime}}\right)^{(1),s-1}$  and then implicitly solved to obtain $u_{k}^{(1),s}$. The updated total energy are determined by taking right-hand-side velocities $\vc{u}_{k} = \vc{u}_{k}^{(1),s}$.
Iterations  are performed until $|u_{1}^{(1),s+1} - u_{1}^{(1),s}| / {u_{1}^{(0)}}  < 10^{-6}$.


\subsection{Temperature relaxation subsystem}

In the case of two components the nine-equation model allows four temperatures, i.e., $T_{i1}, \; T_{i2}, \; T_{e1}, \; T_{e2}$. The temperature relaxation subsystem reads
\begingroup
\allowdisplaybreaks
\begin{subeqnarray}\label{eq:tempRelax}
&&\dudx{\alpha_k \rho_{ik}}{t} = 0, \\
&&\dudx{\alpha_k \rho_{ik} \vc{u}_k}{t} = \mathbf{0}, \\
&&\alpha_{1}\rho_{i1} C_{v,e1} \dudx{T_{e1}}{t} = \mu^{T}_{e1,e2} \left( T_{e2} - T_{e1} \right) + \alpha_1 \sum_k \alpha_k \widehat{\mu}^{T}_{e1,ik} \left( T_{ik} - T_{e1} \right),\\
&&\alpha_{2}\rho_{i2} C_{v,e2} \dudx{T_{e2}}{t} = \mu^{T}_{e2,e1} \left( T_{e1} - T_{e2} \right) + \alpha_2 \sum_k  \alpha_k \widehat{\mu}^{T}_{e2,ik} \left( T_{ik} - T_{e2} \right), \\
&&\alpha_{1}\rho_{i1} C_{v,i1} \dudx{T_{i1}}{t} = \mu^{T}_{i1,i2} \left( T_{i2} - T_{i1} \right) + \alpha_1 \sum_k \alpha_k \widehat{\mu}^{T}_{i1,ek} \left( T_{ek} - T_{i1} \right), \\
&&\alpha_{2}\rho_{i2} C_{v,i2} \dudx{T_{i2}}{t} = \mu^{T}_{i2,i1} \left( T_{i1} - T_{i2} \right) + \alpha_2 \sum_k  \alpha_k \widehat{\mu}^{T}_{i2,ek} \left( T_{ek} - T_{i2} \right), \\
&&\dudx{\alpha_k}{t} = 0.
\end{subeqnarray}
\endgroup

The first RHS term of the temperature equations represents the temperature relaxation at grain scale between different materials at the interface, while the second represents the atomic relaxation between ions and electrons inside each material. Note that when determining the temperature relaxation rates $\widehat{\nu}^T$ with eqs. (\ref{eq:nu_T_rj})(\ref{eq:nu_tempRelax}), one should use the the phase number density $n_k = \frac{\rho_k}{m_{k}}$.
Again, these nonlinear ODEs are solved with the simple iteration method. Pressures and internal energies are then modified according to updated temperatures afterwards. In the temperature relaxation we adopt procedure to maintain pressure variation equilibrium \citep[][]{ZhangC_PhysRevE2023,ZEIN20102964}.

\section{Numerical results}
In this section we first verify the model and numerical results by comparing with existing results in literature, then consider the grain scale disequilibria in the course of an ablation shock passage. The following numerical results are obtained with only the hydrodynamics and atomic/grain relaxations, which is enough for demonstrating the capability of the proposed models in dealing with mechanical and thermal disequilibria.

\subsection{The multi-component Riemann problem}
The hydrodynamic subsystem is reduced to that of the classical  BN model if we sum the equations for the ion internal energy $\alpha_k \rho_k \varepsilon_{ik}$ and electron internal energy $\alpha_k \rho_k \varepsilon_{ek}^{\prime}$ of $k$-th component. The component total pressure $p_k = p_{ik} + p_{ek}$ plays the same role as the phase pressure in BN model. If we prescribe the same initial data for density $\rho_k$, $u_k$, $p_k$ and $\alpha_k$, the proposed nine-equation model has  the same solution as the BN model. Besides, the solutions for densities, total pressures, velocities are independent of the partition between $p_{ik}$ and $p_{ek}$. Therefore, we can compare our numerical solutions to those of the BN model in literature.

The initial data is given as follows.
Within $x \in [0, 0.5]$
\[\rho_1 = 800, \; u_1 = 0, \; p_1 = 500, \; \rho_2 = 1.5, \; u_2 = 0, \; p_2 = 2, \; \alpha_1 = 0.4, \]

Within $x \in [0.5, 1.0]$
\[\rho_1 = 1000, \; u_1 = 0, \; p_1 = 600, \; \rho_2 = 1.0, \; u_2 = 0, \; p_2 = 1, \; \alpha_1 = 0.3. \]

\begin{figure} 
\centering
\subfloat[$\rho_1$]{\includegraphics[width=0.5\textwidth]{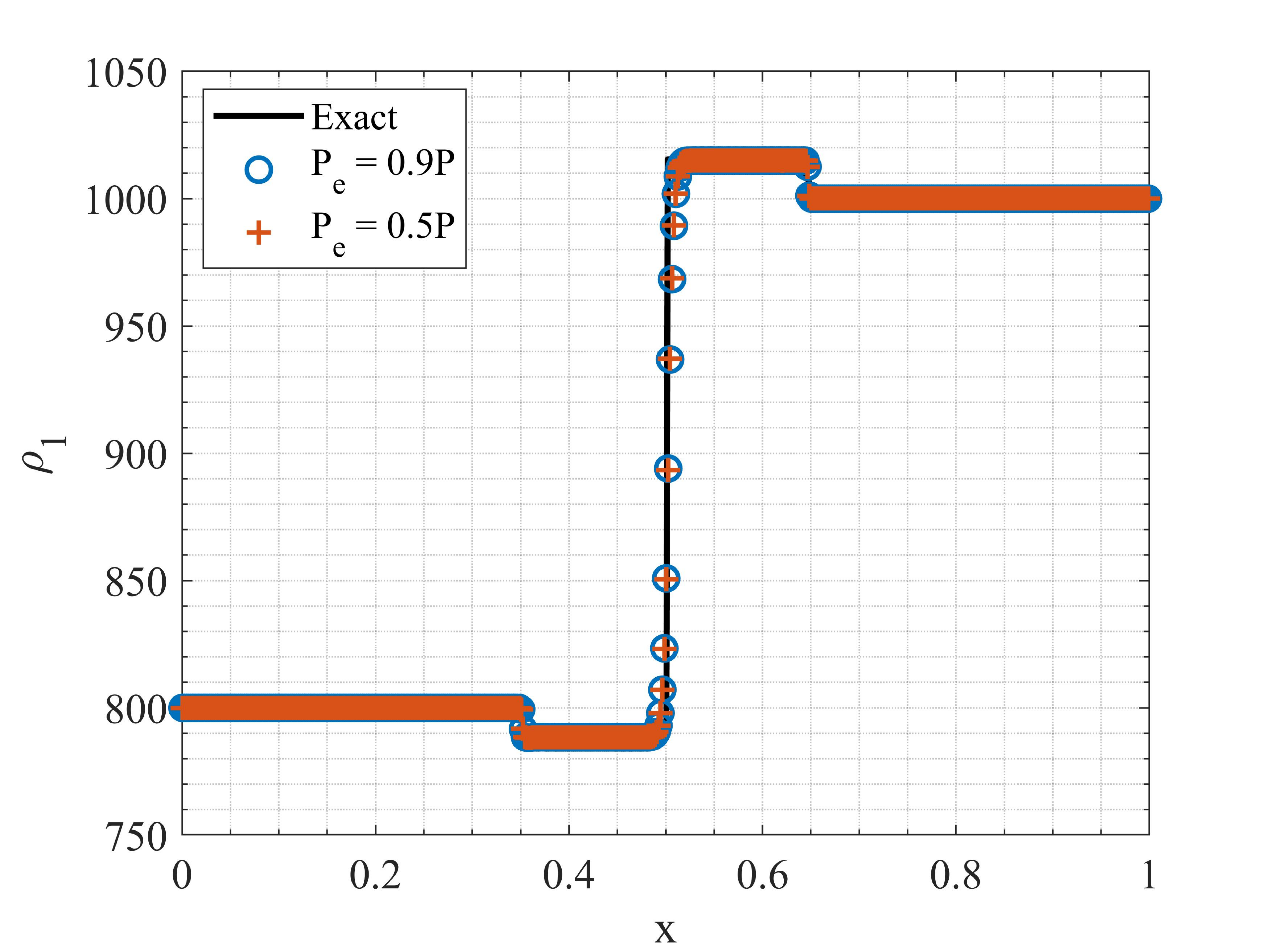}}
\subfloat[$\rho_2$]{\includegraphics[width=0.5\textwidth]{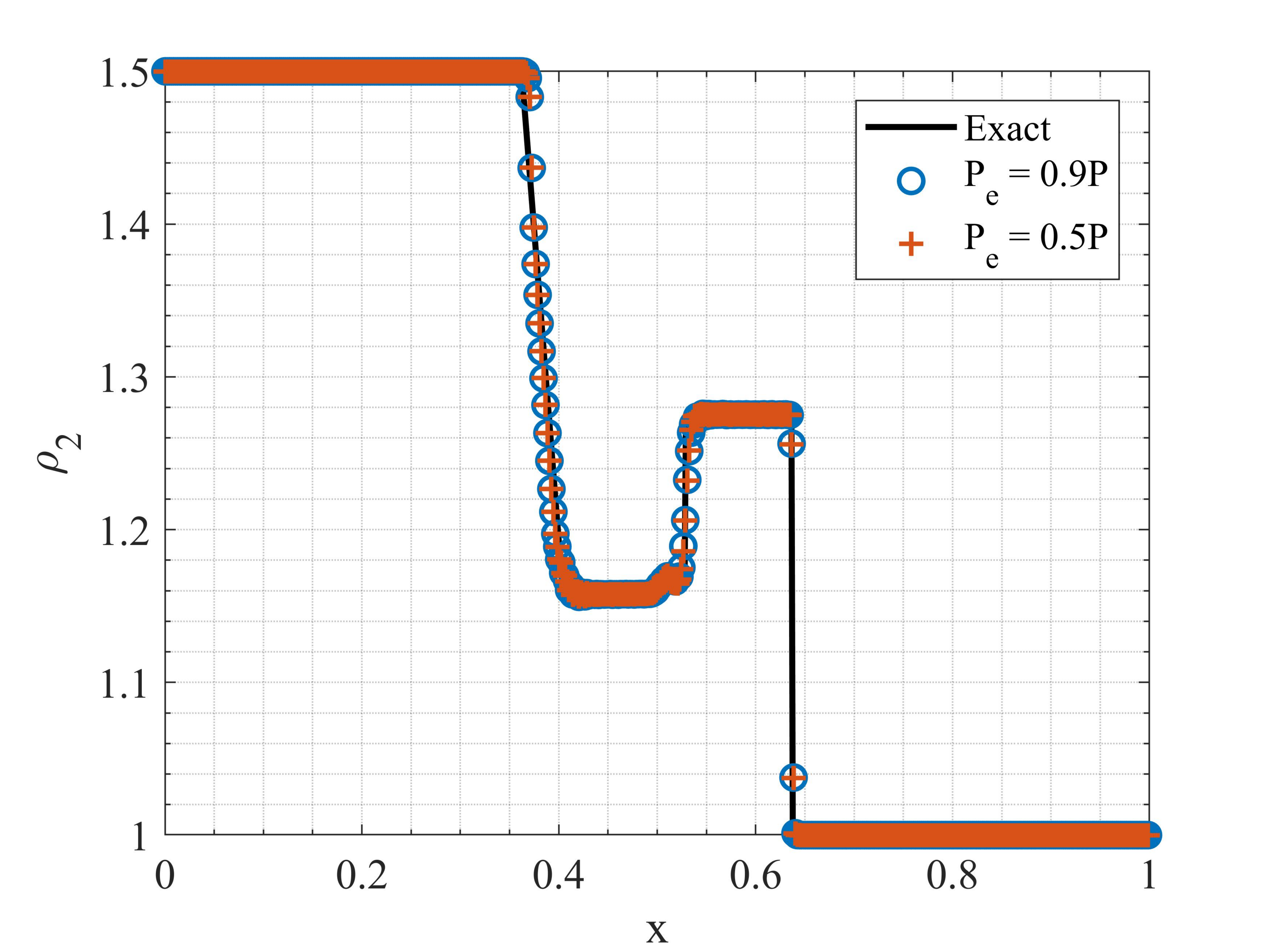}}\\
\subfloat[$S_{e1}$]{\includegraphics[width=0.5\textwidth]{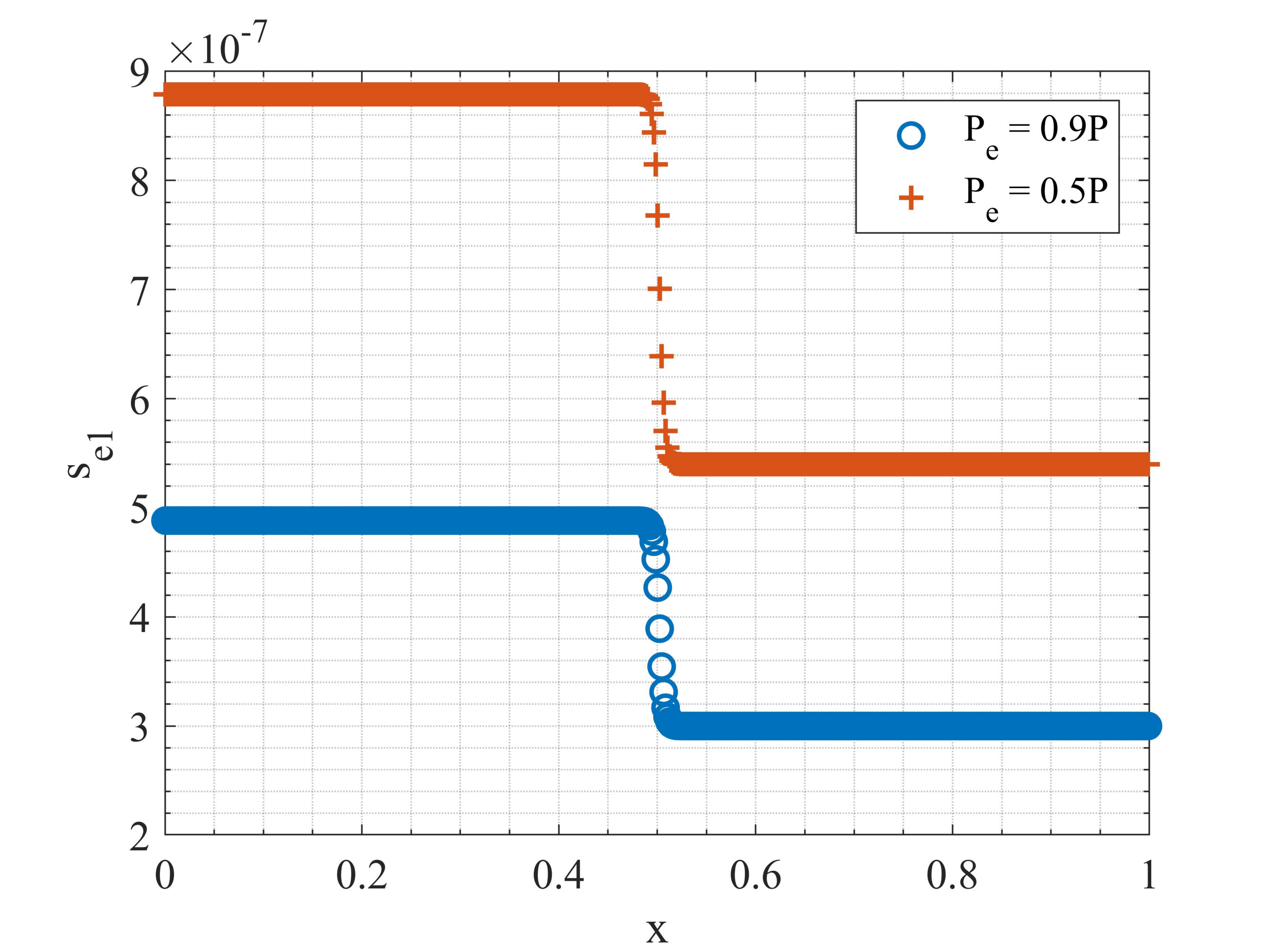}}
\subfloat[$S_{e2}$]{\includegraphics[width=0.5\textwidth]{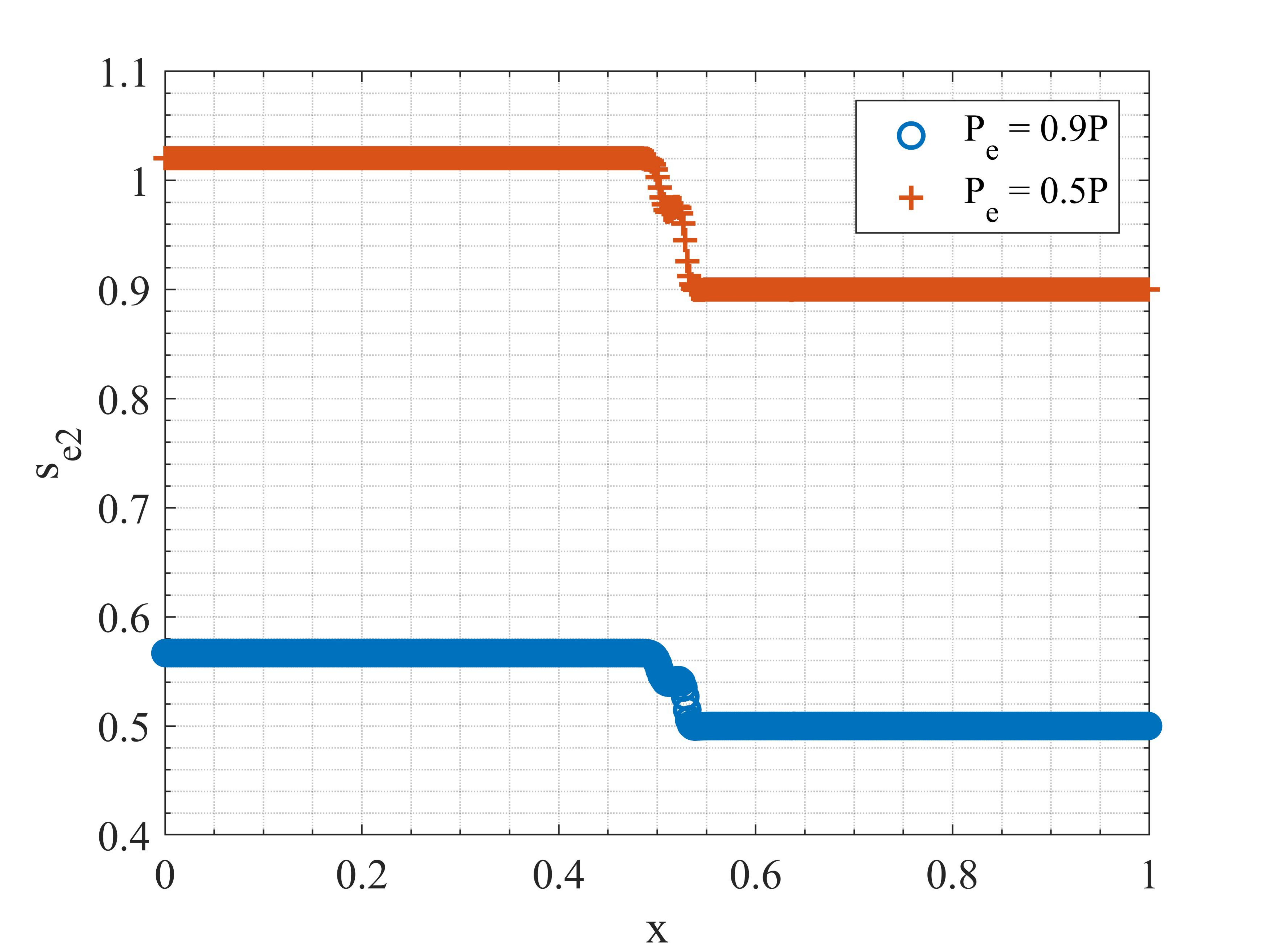}}
\caption{The numerical results for the multiphase shock problem. The circles and crosses represent the results with $P_e = 0.9P$ and $P_e = 0.5P$, respectively.}
\label{fig:BNP_compare_dumbster}
\end{figure}

The adiabatic coefficients $\gamma_{e1} = \gamma_{i1} = 3.0$ and $\gamma_{e2} = \gamma_{i2} = 1.4$.
For the first ion the Stiffened gas EOS is used with $p_{i1,\infty} = 100$.
We consider two cases $p_{ek} = 0.9p_k, \; p_{ik} = 0.1p_k$ and $p_{ek} = 0.5p_k, \; p_{ik} = 0.5p_k$. The obtained numerical results for densities are compared against the exact solutions in Figure \ref{fig:BNP_compare_dumbster}, which shows good consistency. The numerical results for these two cases coincide as expected. Moreover, the electron entropy $s_{ek}$ is continuous across the ion shocks, which indicates that the numerical methods well maintain the jump condition for electrons.

\subsection{The plasma shocktube problem}
\begin{figure} 
\centering
\subfloat[$\rho_1$]{\includegraphics[width=0.5\textwidth]{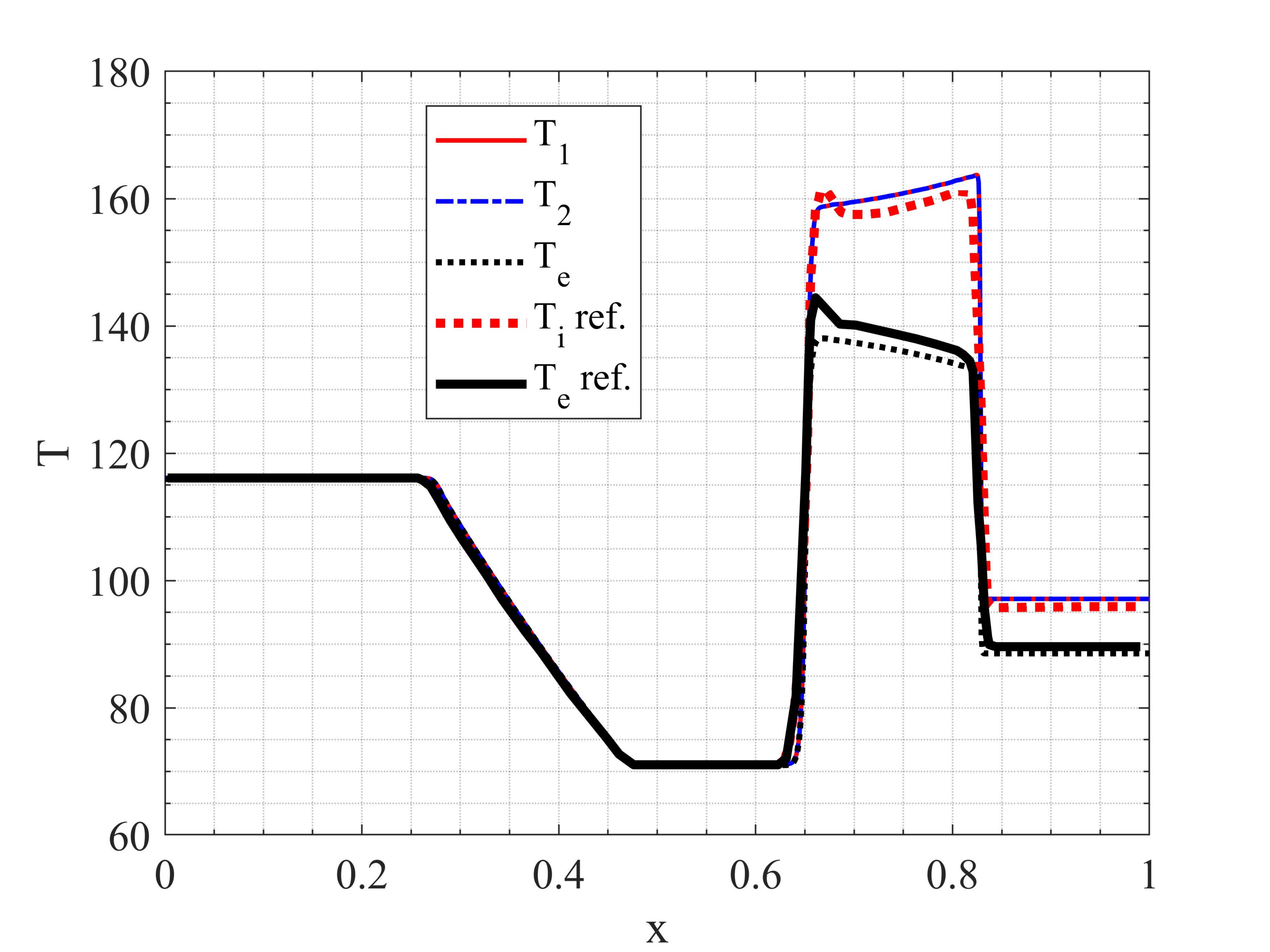}}
\caption{Comparison with the results of Sangum [1].}
\label{fig:Compare_sangum}
\end{figure}

In this section we consider the plasma shock tube problem with the relaxation mechanism. We first consider the one-component problem in Sangum [1].  Our numerical results are compared with those therein in Figure \ref{fig:Compare_sangum}. Although a slight difference from the solutions with the relaxation scheme is observed, the overall consistency is well.

Then we adjust this problem to a multi-component version to consider relaxation mechanisms in different mixing topology, including the atomic mixing and grain mixing at different length scales. We assume the components are carbon(C, $k=1$) and deuterium(D, $k=2$) to the left and right of the initial interface. The atomic relaxation rates are computed with eqs. (\ref{eq:nu_velRelax})(\ref{eq:nu_tempRelax}). For the grain mixing scenario, we prescribe the particle number $N$ and the size of the physical domain $L_{phys}$, then the particle size is $b = \sqrt{L_{phys}^2 / N}$ in 2D. The grain-scale relaxation coefficients are evaluated with eqs. (\ref{eq:mu_presRelax})(\ref{eq:mu_velRelax})(\ref{eq:mu_mechTempRelax}).

\begin{figure} 
\centering
\subfloat[$N_{grain}=50000$]{\includegraphics[width=0.5\textwidth]{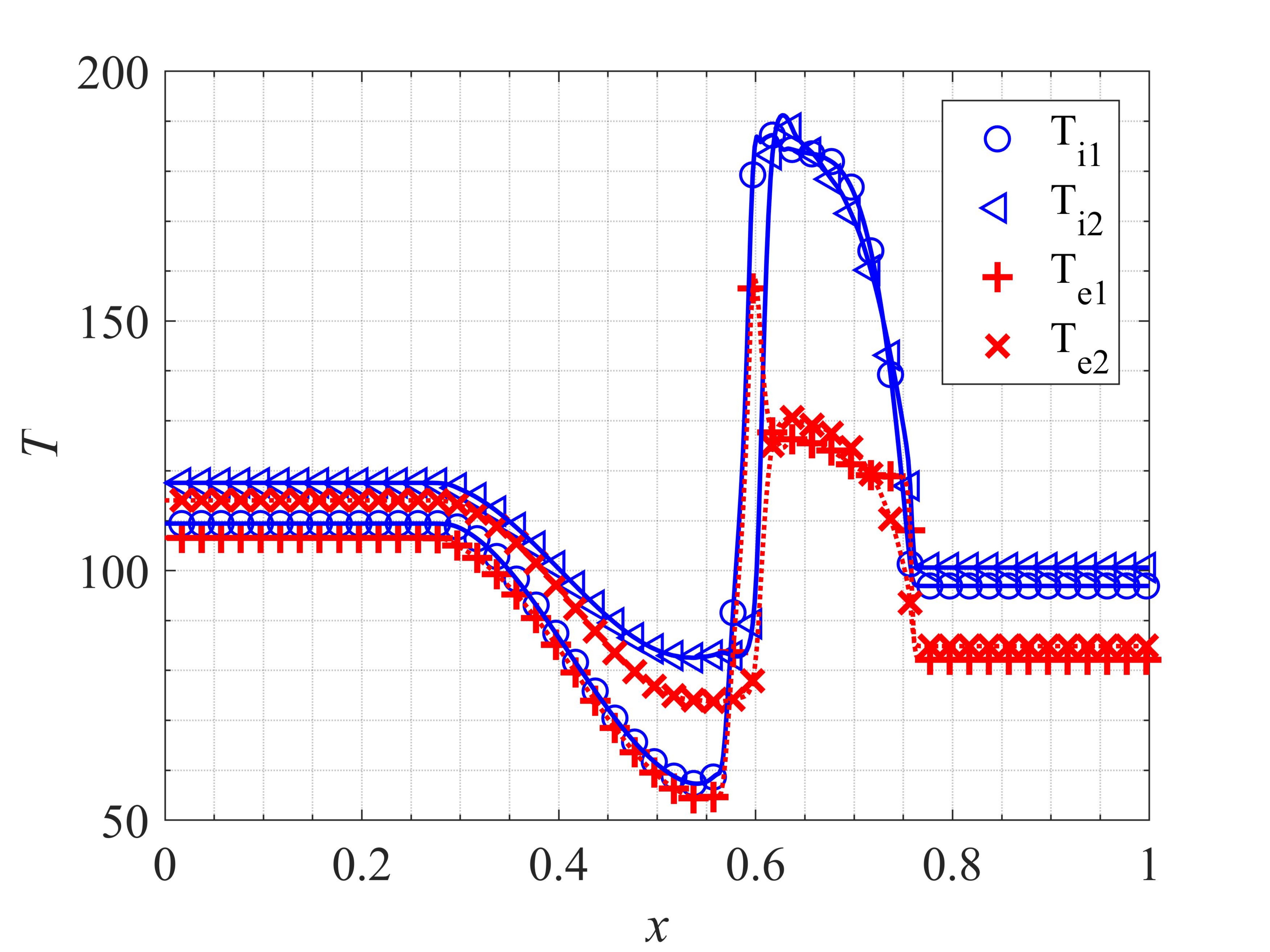}}
\subfloat[$N_{grain}=100000$]{\includegraphics[width=0.5\textwidth]{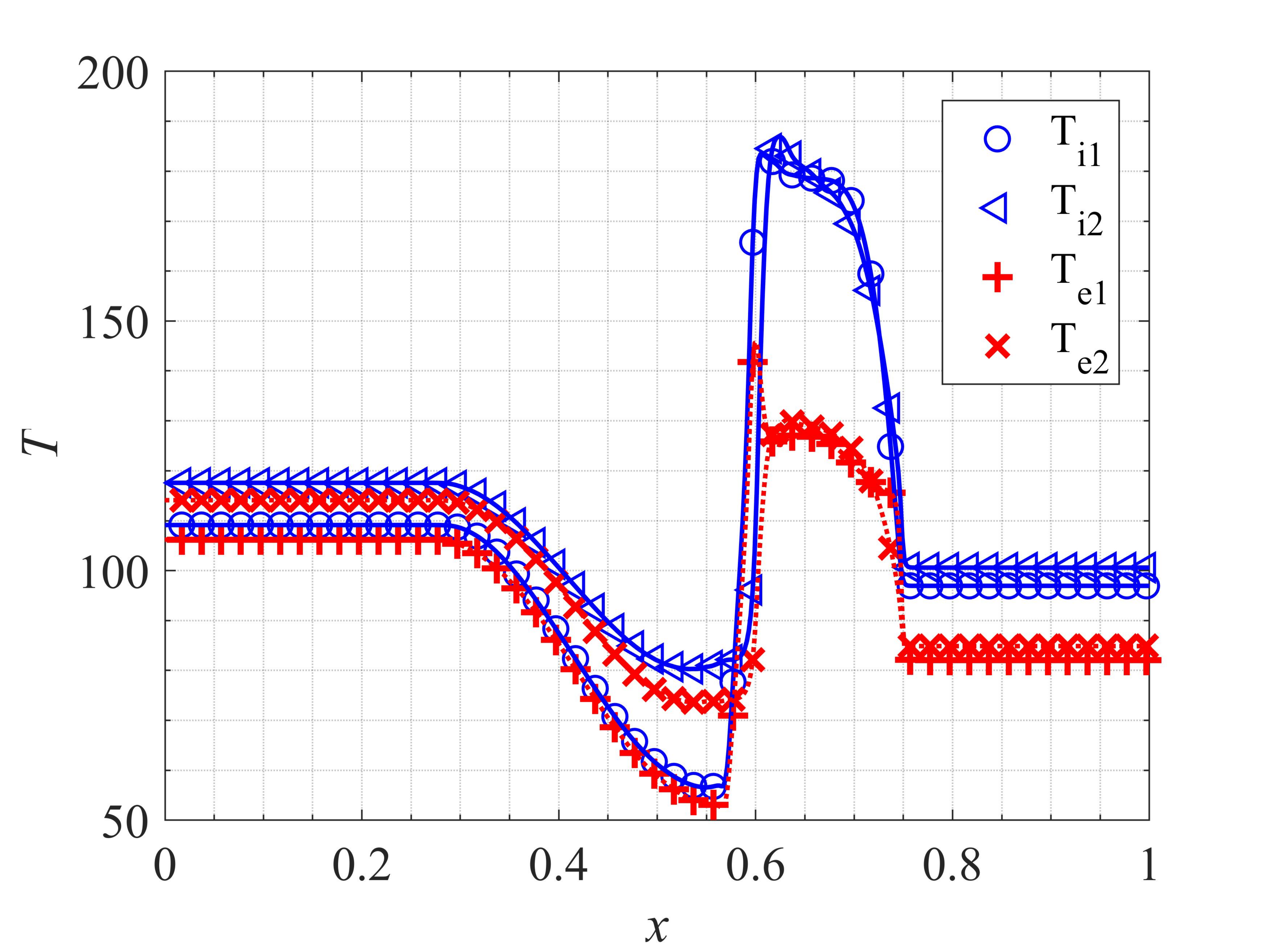}}\\
\subfloat[$N_{grain}=500000$]{\includegraphics[width=0.5\textwidth]{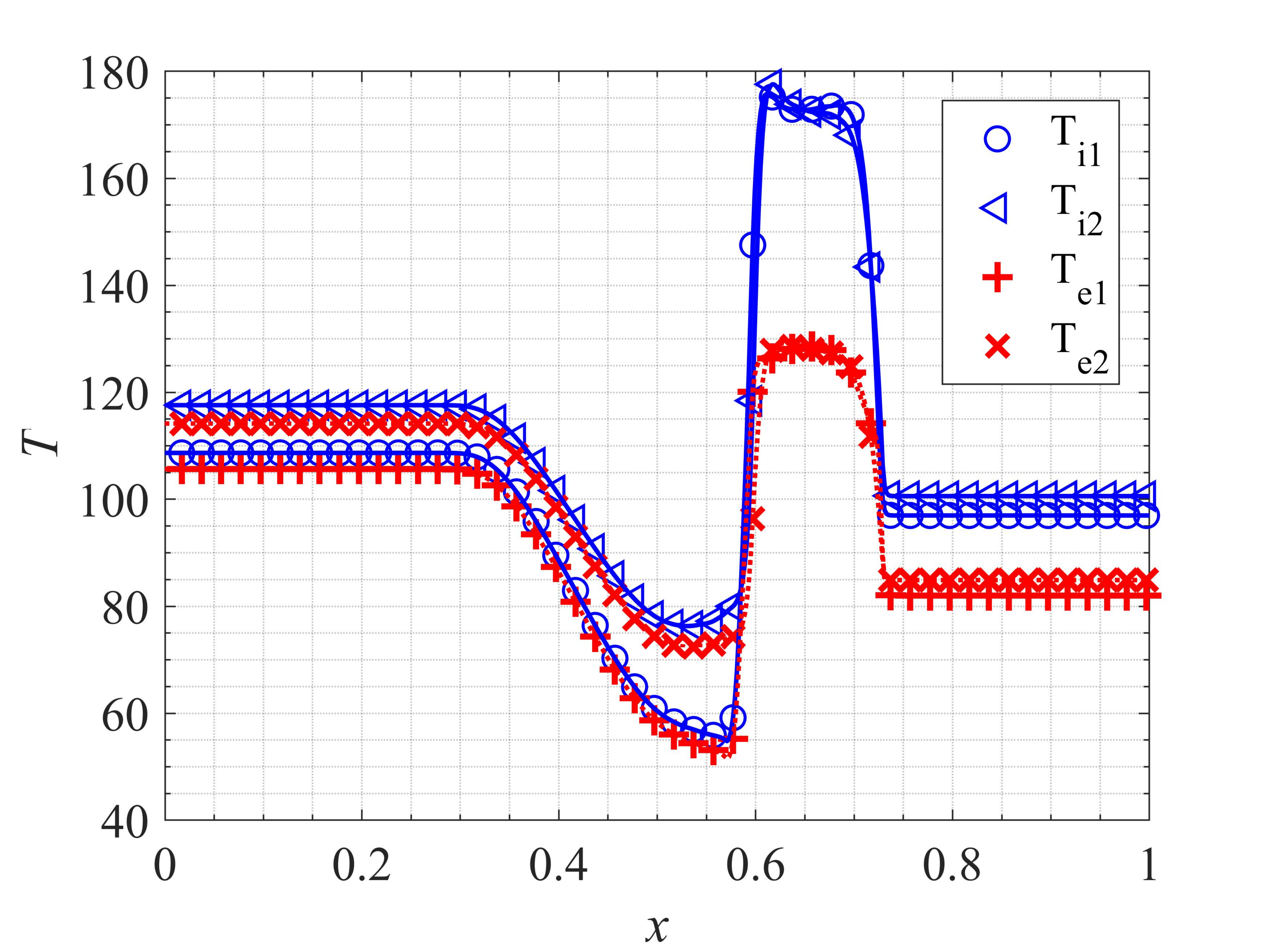}}
\subfloat[Atomic mix]{\includegraphics[width=0.5\textwidth]{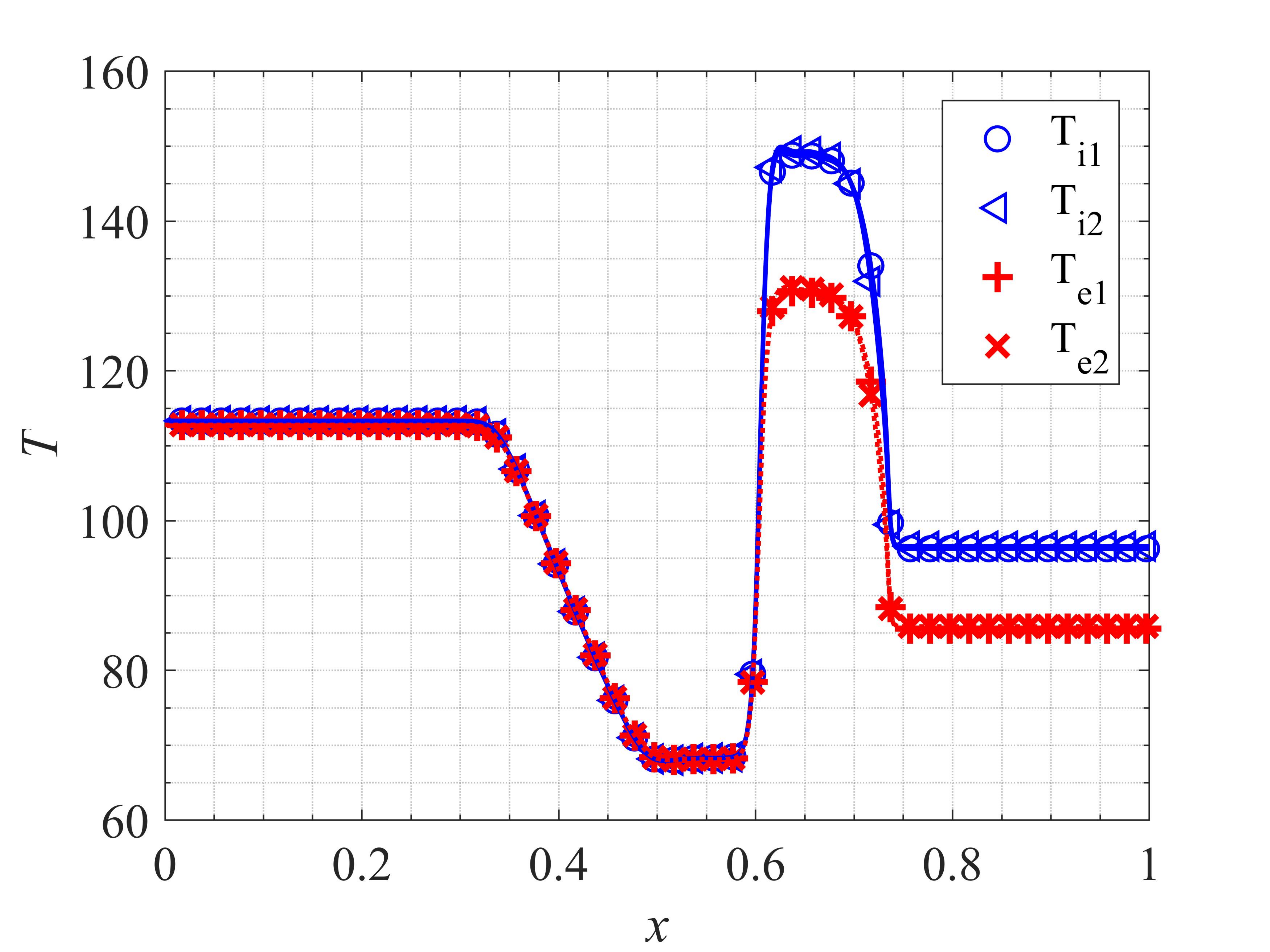}}
\caption{The numerical results for temperatures ($T_{i1}$, $T_{i2}$, $T_{e1}$, $T_{e2}$) in  grain mixing scenarios with different numbers of grains ($N_{grain}$) and atomic mixing.}
\label{fig:Shock_mixType_Temp}
\end{figure}

\begin{figure} 
\centering
\subfloat[$N_{grain}=50000$]{\includegraphics[width=0.5\textwidth]{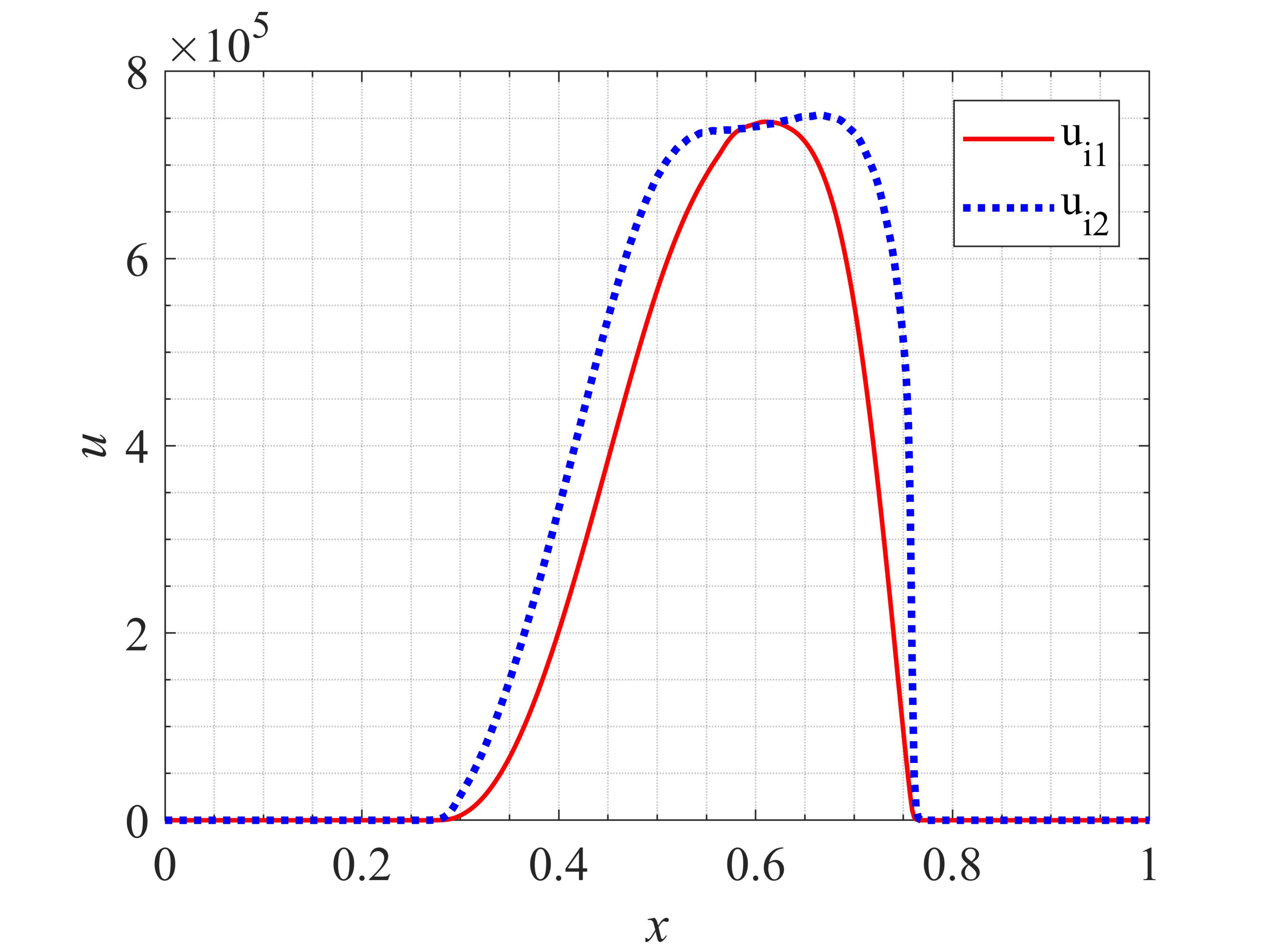}}
\subfloat[$N_{grain}=100000$]{\includegraphics[width=0.5\textwidth]{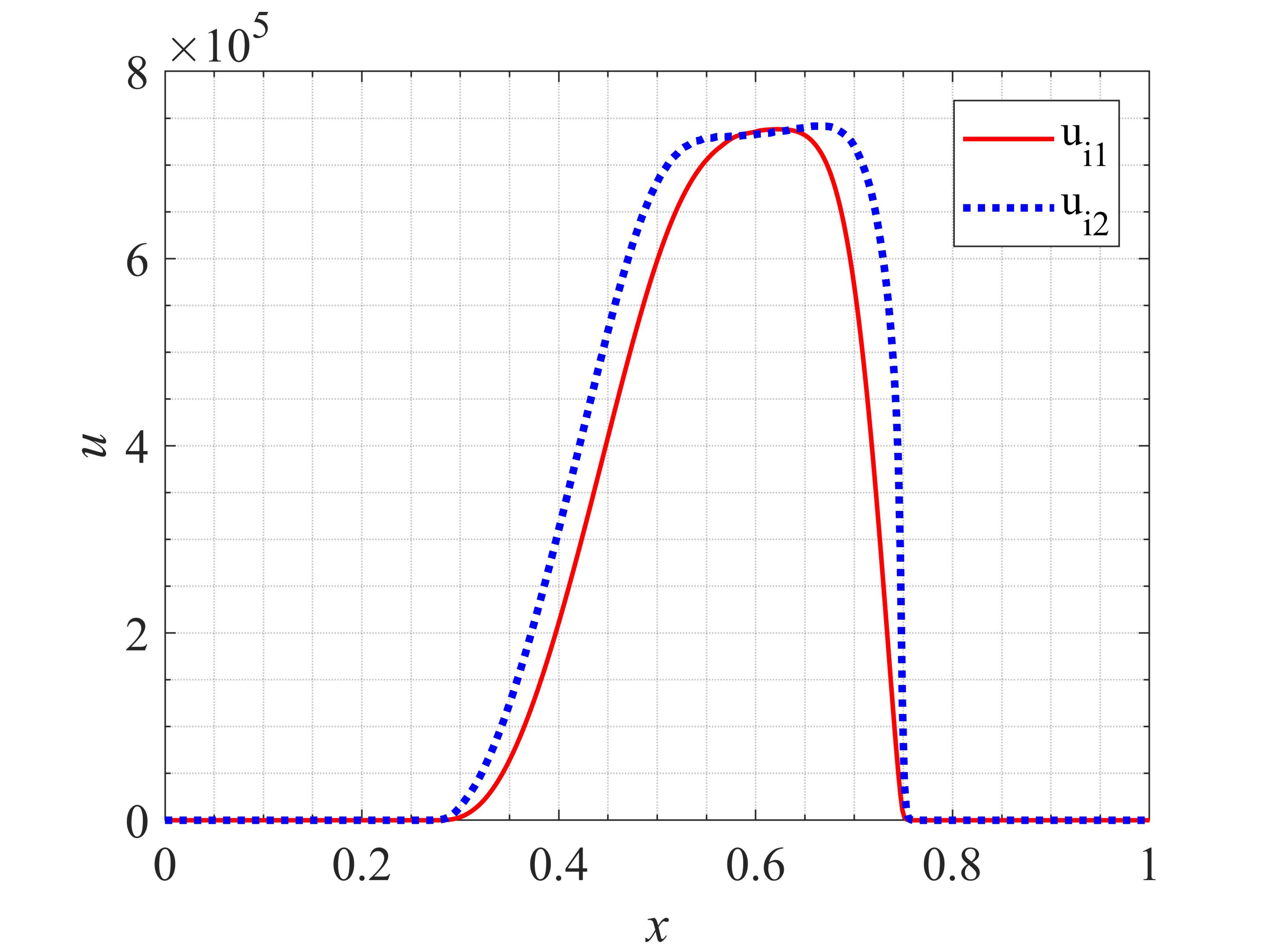}}\\
\subfloat[$N_{grain}=500000$]{\includegraphics[width=0.5\textwidth]{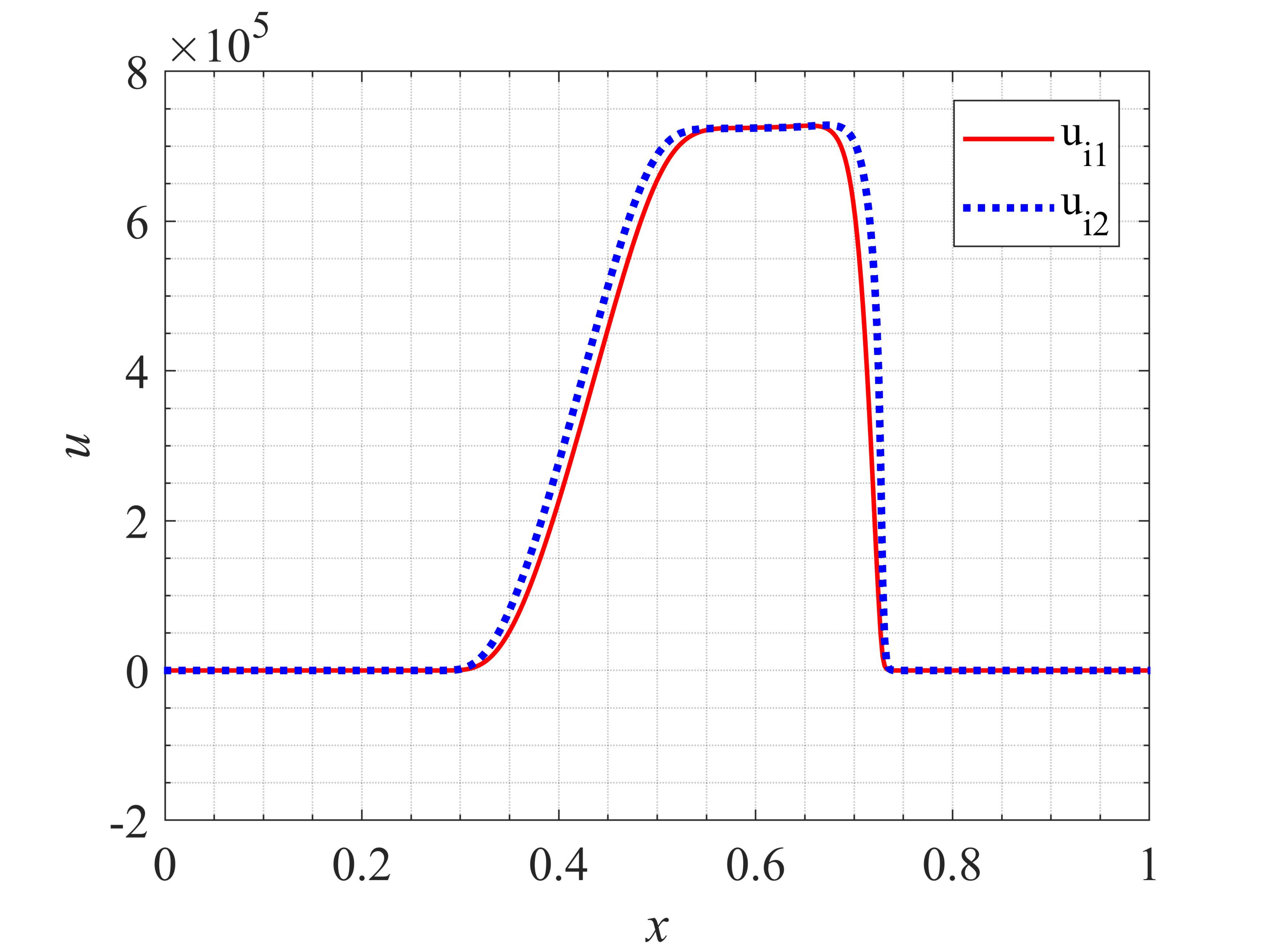}}
\subfloat[Atomic mix]{\includegraphics[width=0.5\textwidth]{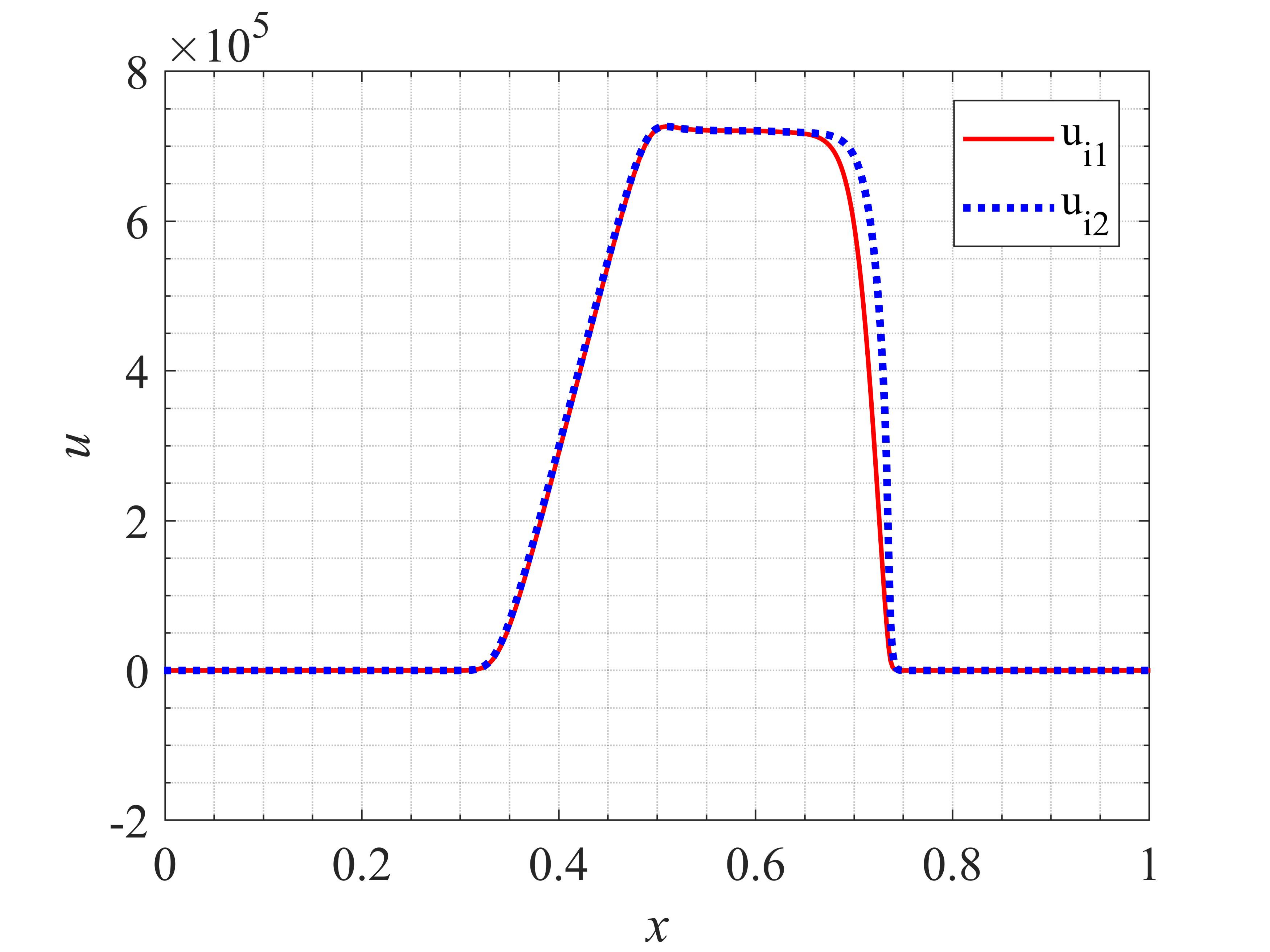}}
\caption{The numerical results for component velocities $u_{i1}, u_{i2}$  in grain mixing scenarios with different numbers of grains ($N_{grain}$) and atomic mixing.}
\label{fig:Shock_mixType_Vel}
\end{figure}

\begin{figure} 
\centering
\subfloat[$N_{grain}=50000$]{\includegraphics[width=0.5\textwidth]{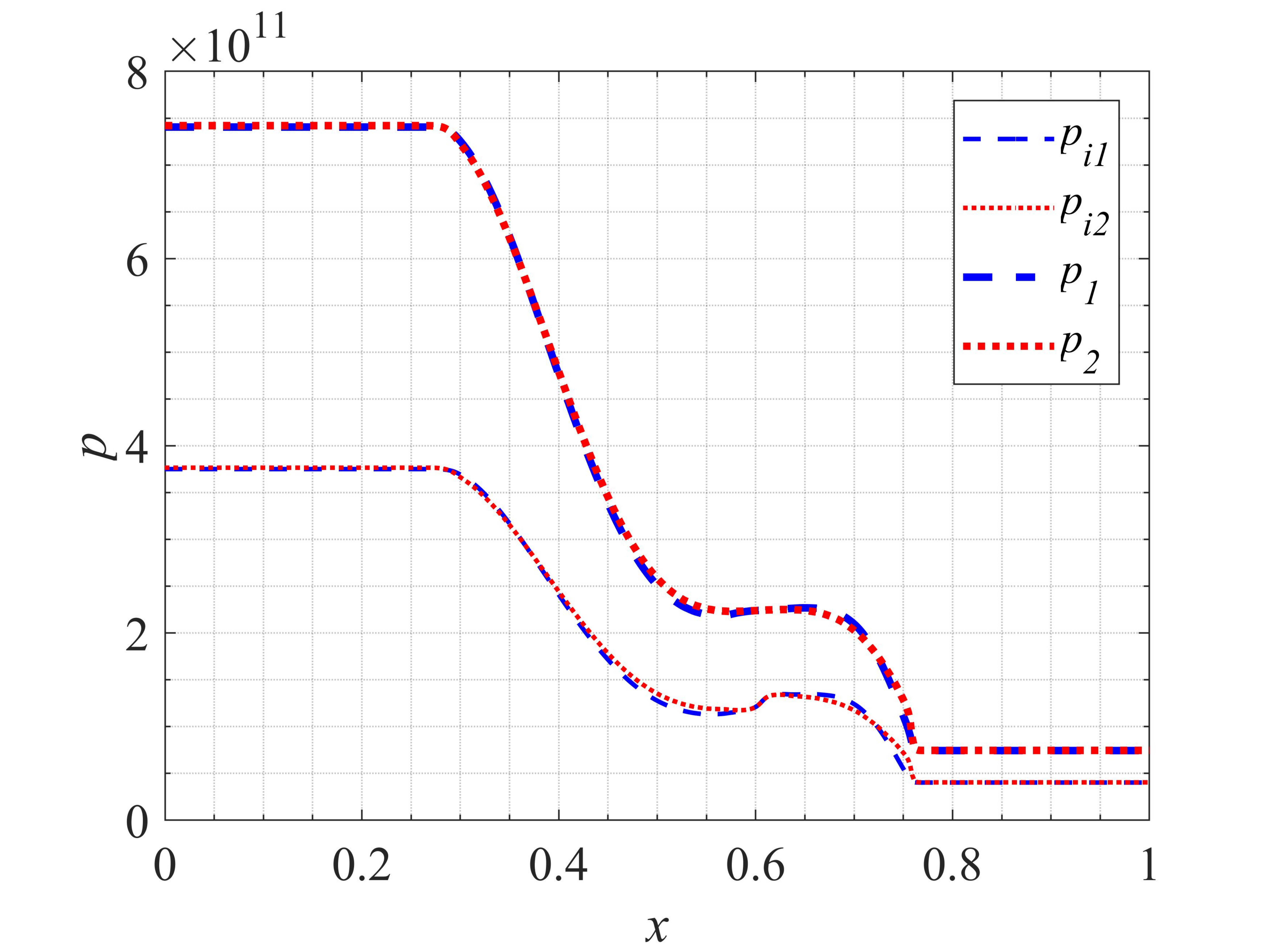}}
\subfloat[$N_{grain}=100000$]{\includegraphics[width=0.5\textwidth]{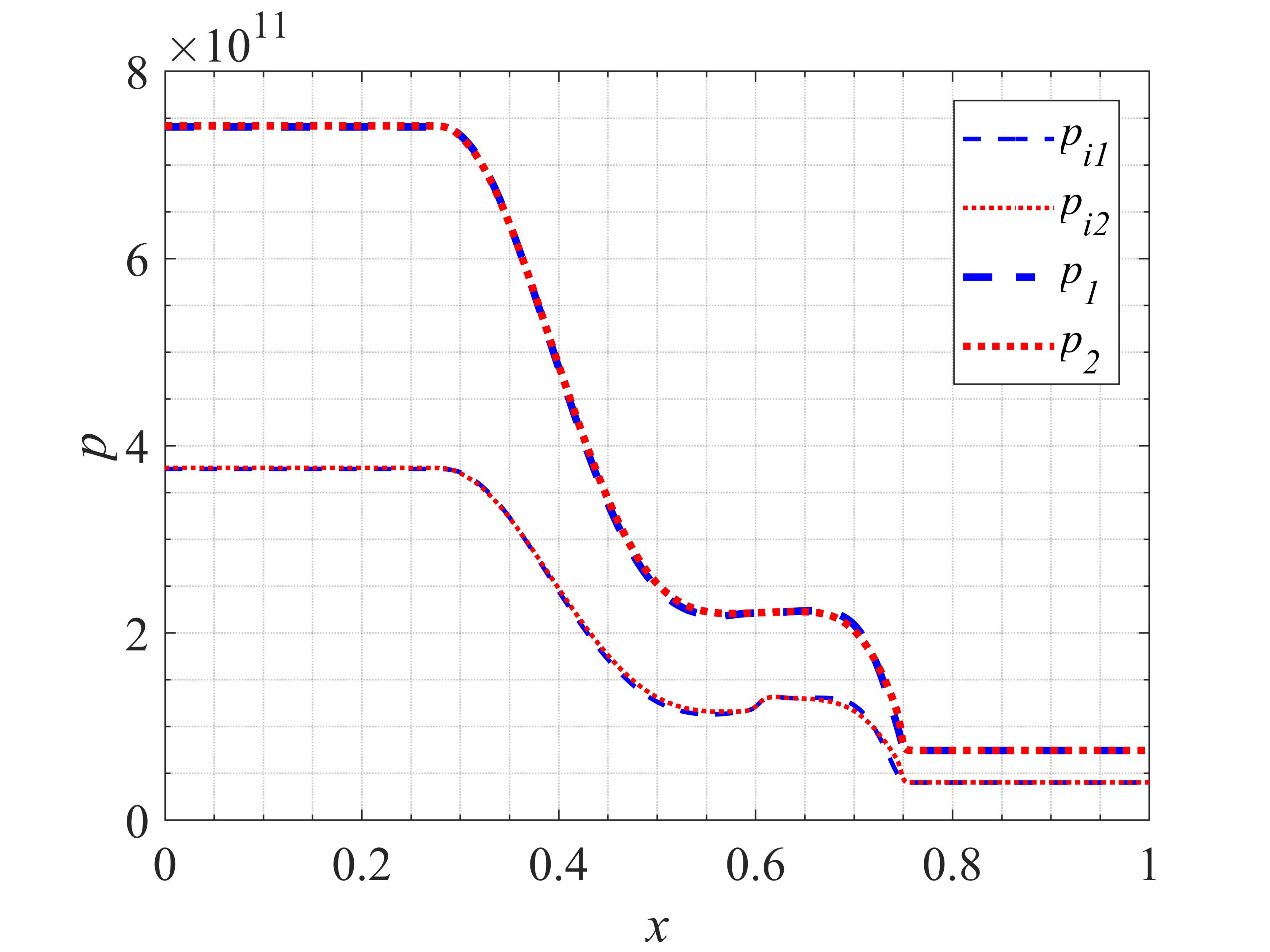}}\\
\subfloat[$N_{grain}=500000$]{\includegraphics[width=0.5\textwidth]{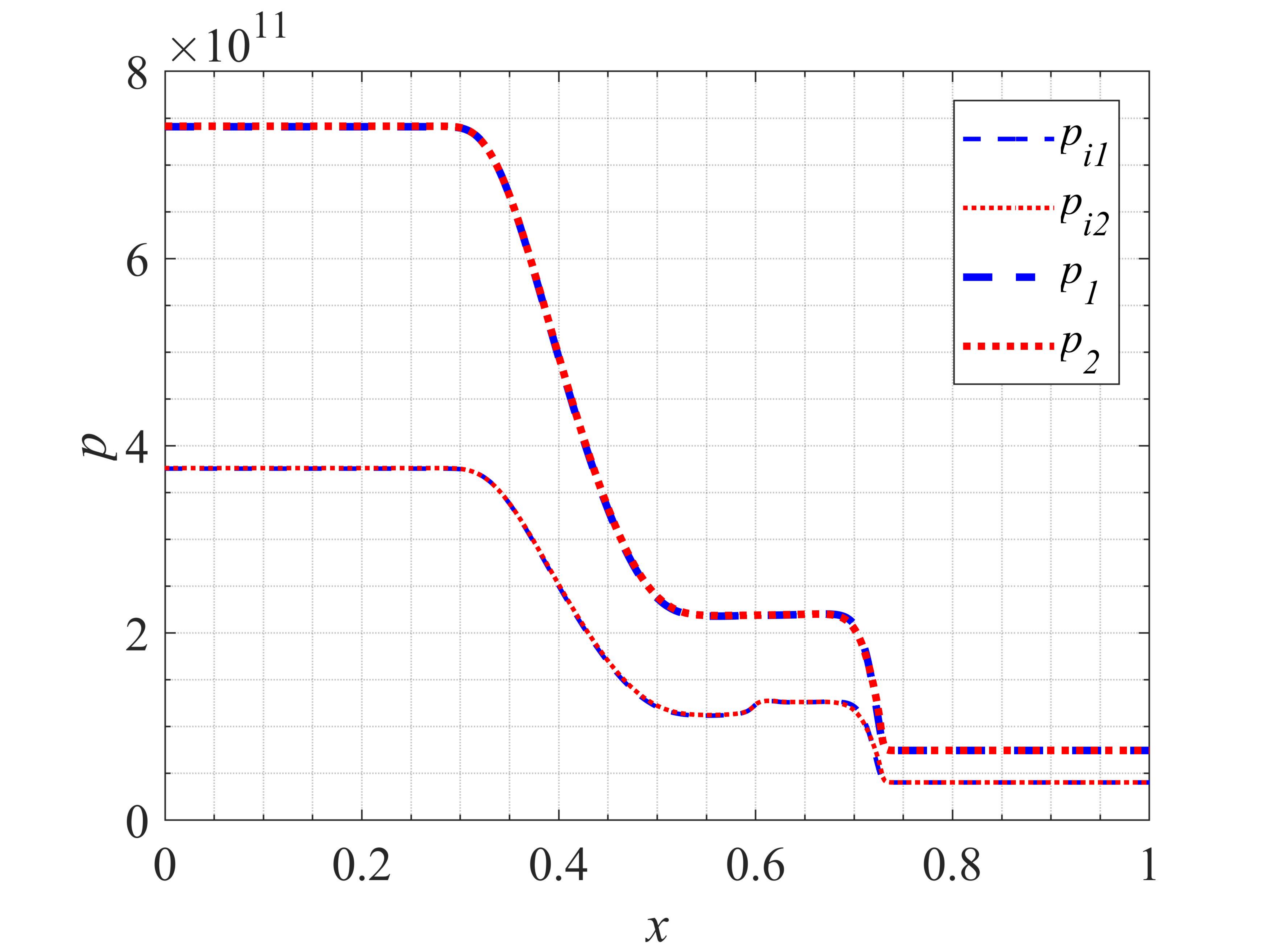}}
\subfloat[Atomic mix]{\includegraphics[width=0.5\textwidth]{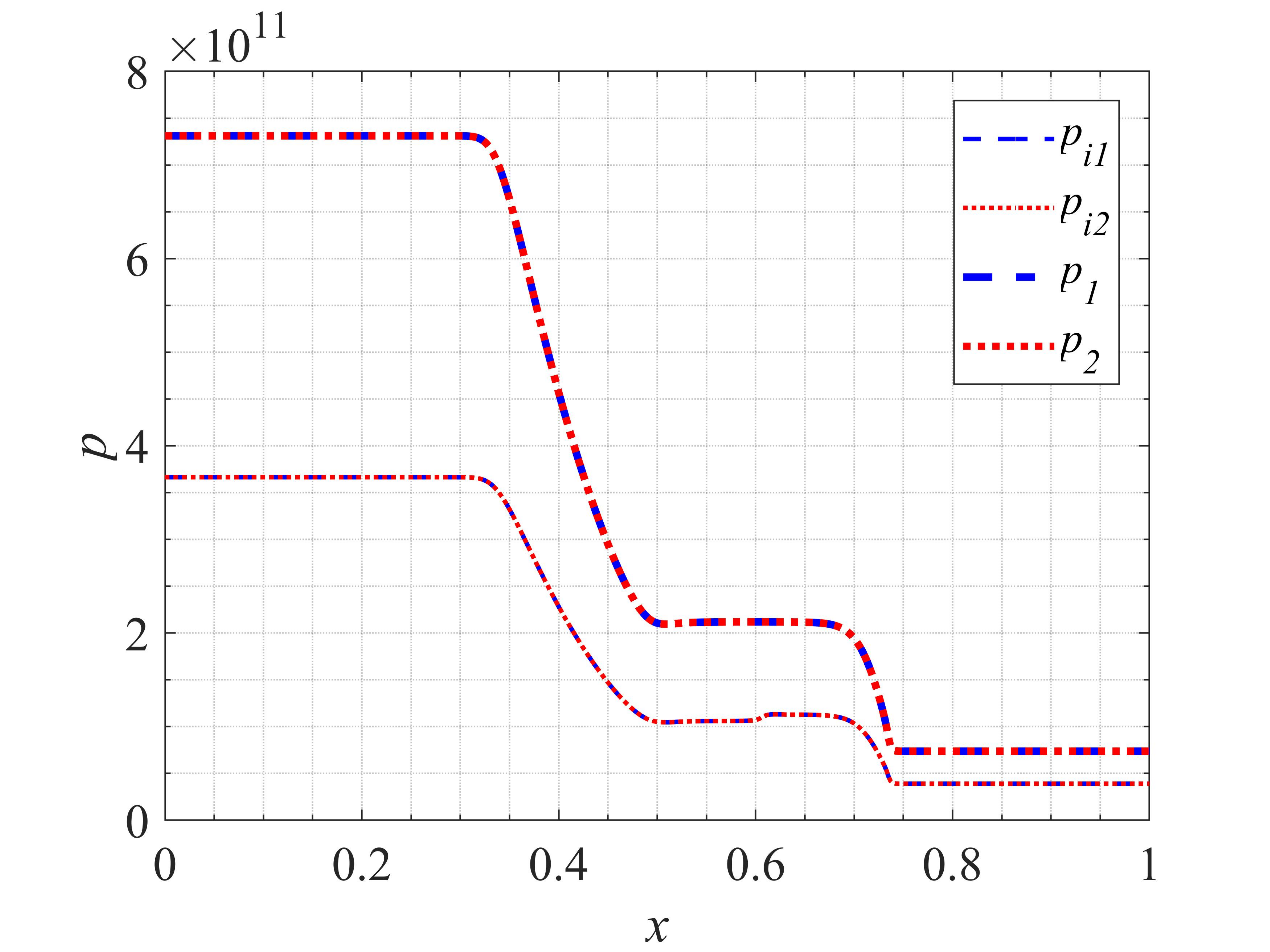}}
\caption{The numerical results for the ion pressure $p_{ik}$ and the component total pressure $p_{k}$ in different types of mixing.}
\label{fig:Shock_mixType_Pres}
\end{figure}

\begin{figure} 
\centering
\subfloat[$N_{grain}=50000$]{\includegraphics[width=0.5\textwidth]{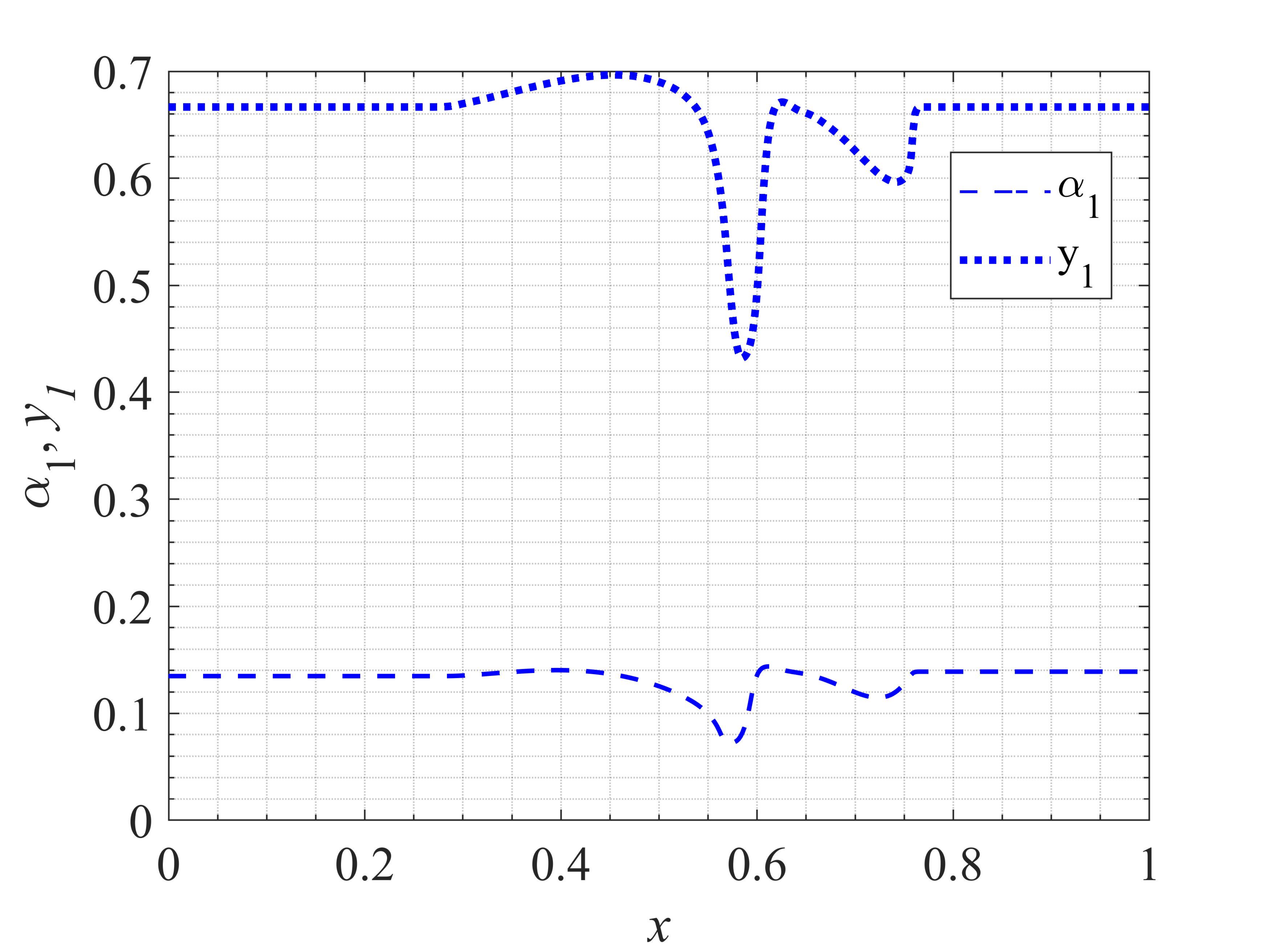}}
\subfloat[$N_{grain}=100000$]{\includegraphics[width=0.5\textwidth]{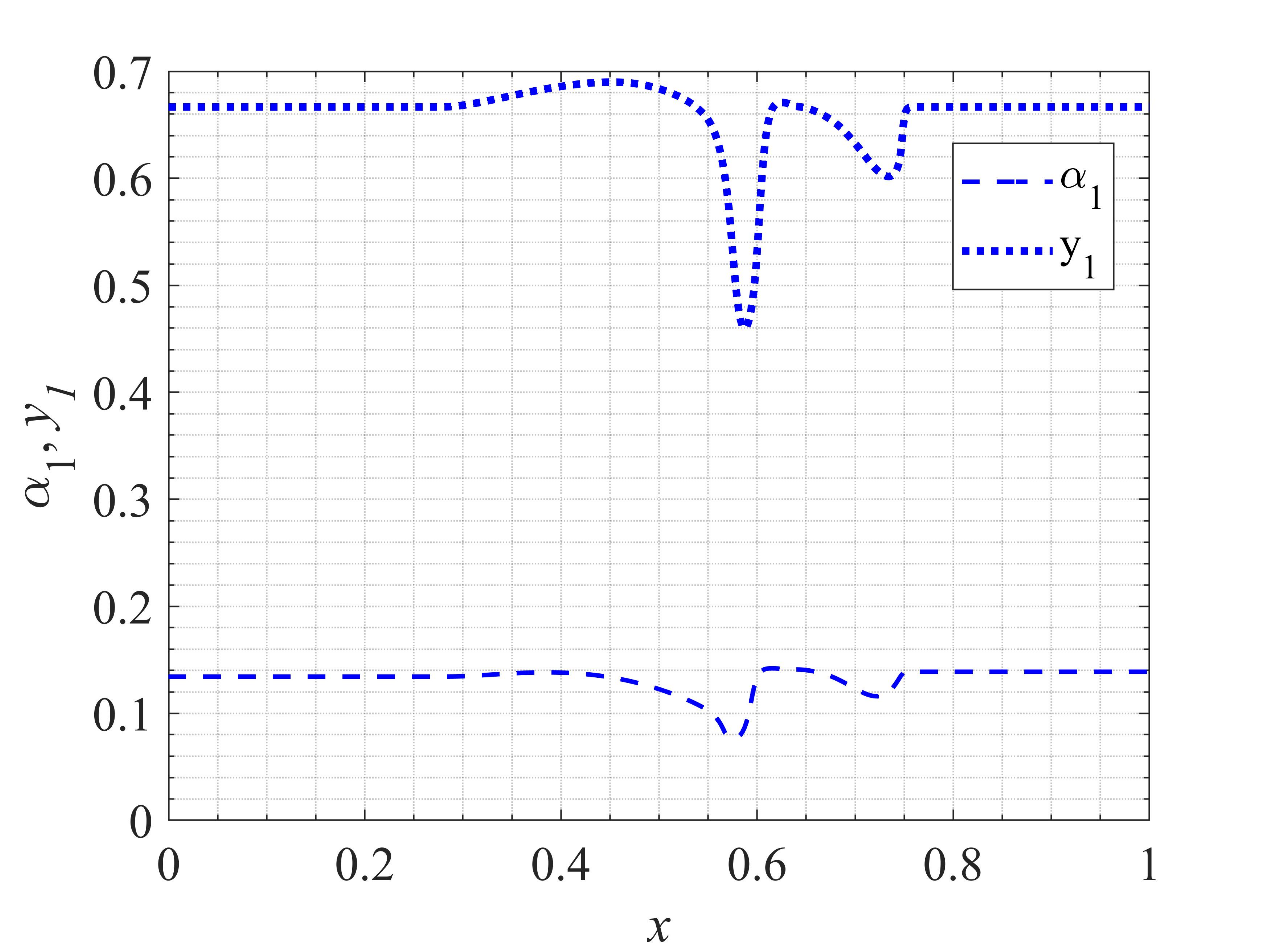}}\\
\subfloat[$N_{grain}=500000$]{\includegraphics[width=0.5\textwidth]{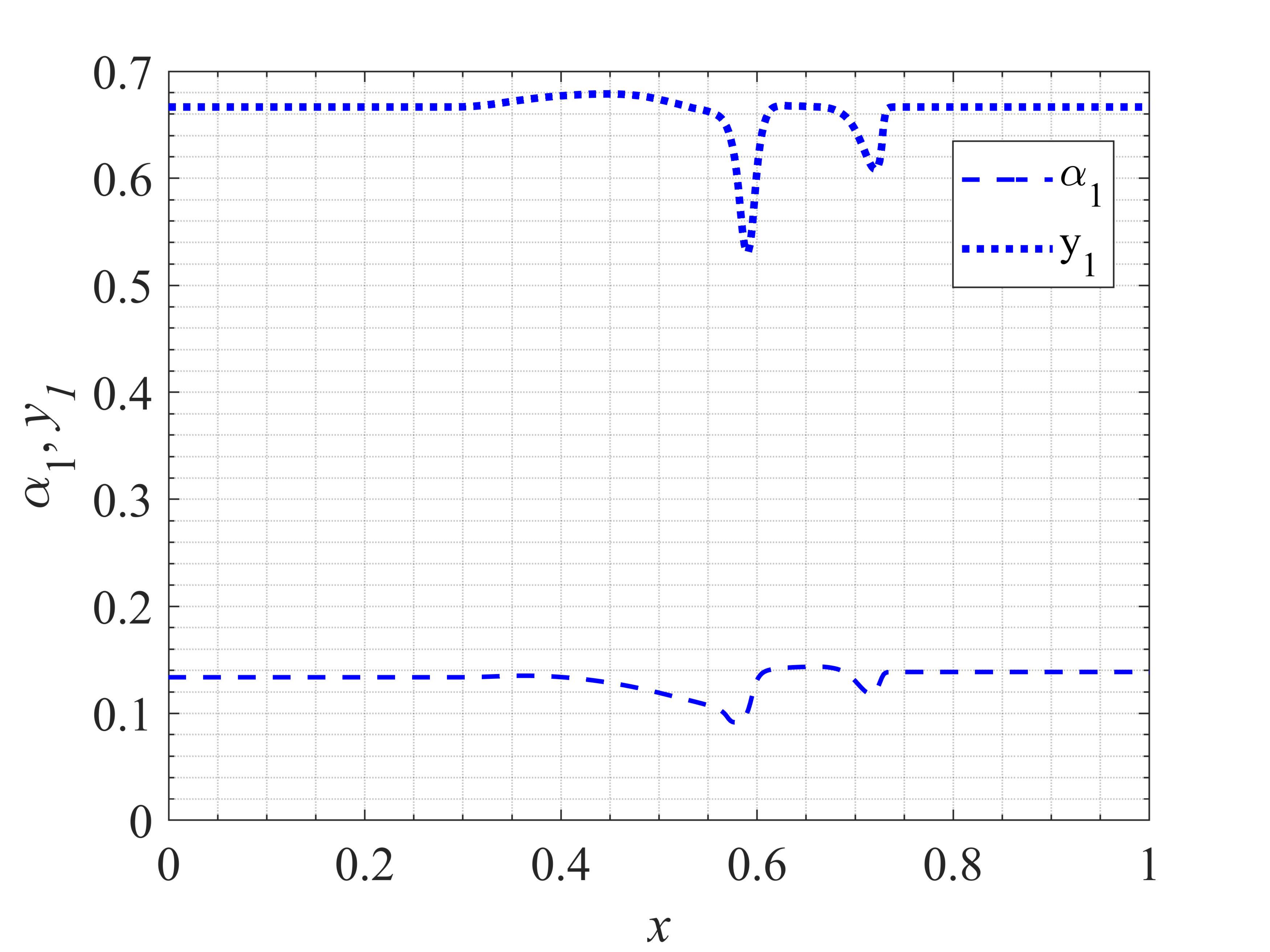}}
\subfloat[Atomic mix]{\includegraphics[width=0.5\textwidth]{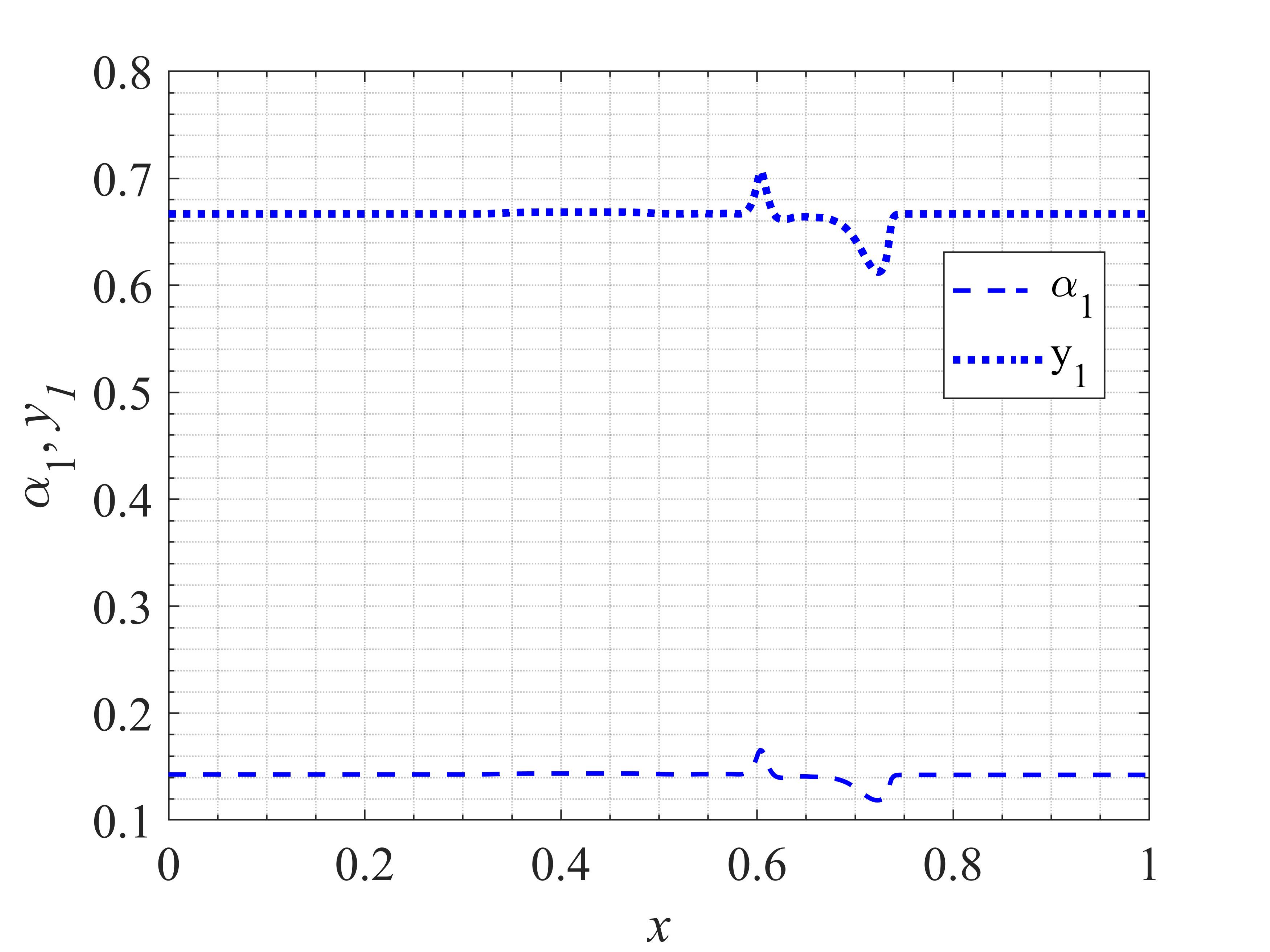}}
\caption{The numerical results for component mass fraction $y_{1}$, volume fraction $\alpha_1$  in grain mixing scenarios with different numbers of grains ($N_{grain}$) and atomic mixing.}
\label{fig:Shock_mixType_Alp}
\end{figure}

The numerical results for temperature, velocity, pressure and volume/mass fractions are displayed in Figures \ref{fig:Shock_mixType_Temp}, \ref{fig:Shock_mixType_Vel}, \ref{fig:Shock_mixType_Pres} and \ref{fig:Shock_mixType_Alp}, respectively. We see that with the increase of the particle number $N$, the temperatures, velocities and pressures tend to equilibrium. Meanwhile, with atomic relaxation rates the ions are at quasi-equilibrium state.  For the atomic relaxation, the velocities in the vicinity of the right shock are in more disequilibrium since the temperatures are higher than those in the left-going rarefaction waves.

The variation in mass fraction is directly driven by the velocity difference (or the drift effect), while the volume fraction by the pressure difference (or the compaction effect).
The right-going shock in the deuterium is faster than that in the carbon, thus the former is first compressed and resulting in a increase in $y_2$ and a decrease in $y_1$ (Figure \ref{fig:Shock_mixType_Alp}).   The situation is similar for the left-going rarefaction wave where the deuterium is first depressed, resulting in a smooth dive in $y_2$ and hump in $y_1$. The variations between these waves are caused by the pressure, velocity and temperature relaxations.

\begin{figure} 
\centering
\subfloat[$N_{grain}=10000$]{\includegraphics[width=0.5\textwidth]{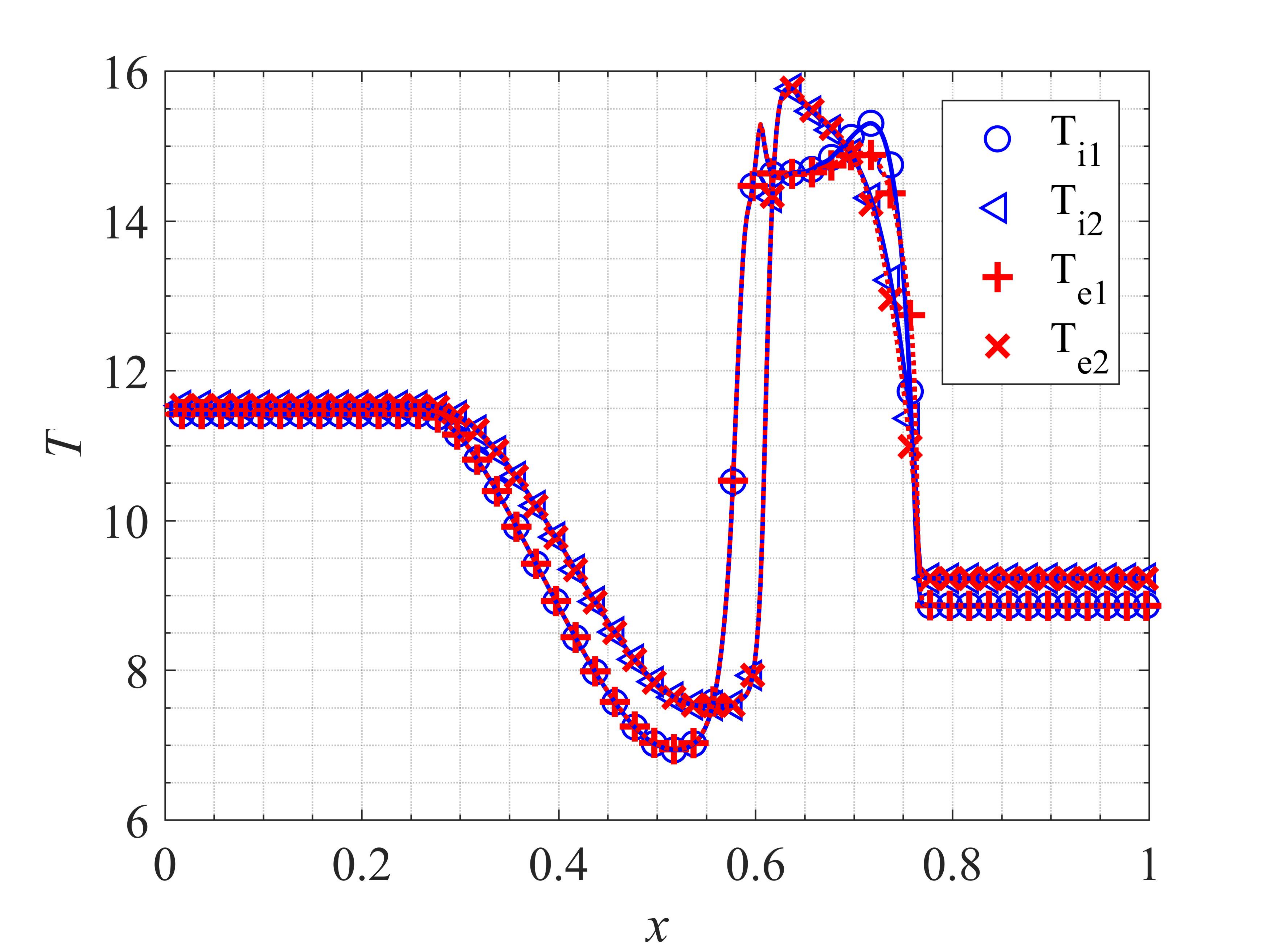}}
\subfloat[Atomic mix]{\includegraphics[width=0.5\textwidth]{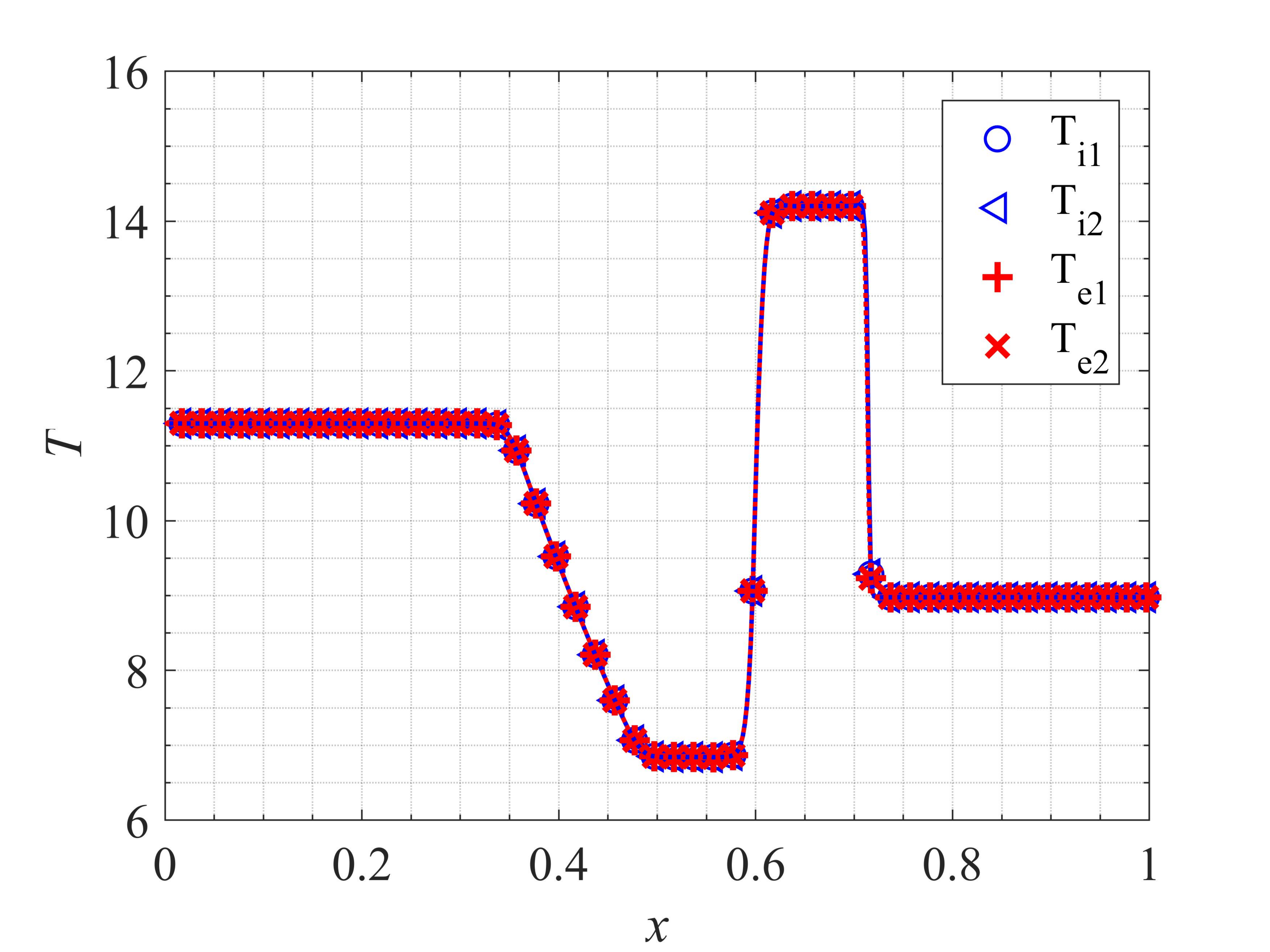}}
\caption{The numerical results for temperatures in different types of mixing at temperature smaller by one order.}
\label{fig:Shock_mixType_Temp_lowT}
\end{figure}

 
 If we decrease the initial temperature by one order, we obtain the results for temperature in Figure \ref{fig:Shock_mixType_Temp_lowT}, where the temperatures of different components are in obvious disequilibrium and the temperatures of ions and electrons in each component are almost in equilibrium. In contrast, in Figure \ref{fig:Shock_mixType_Temp} the ion temperature and electron temperature for each component are in obvious disequilibrium. This well agrees with the temperature dependence of the atomic and grain relaxation times. As can be seen Figures \ref{fig:relaxation_time_atomic} and \ref{fig:relaxation_time_grain}, with the temperature decrease, the temperature relaxation time in grain scale increases and that in atomic scale decreases. Thus, ion-electron temperature disequilibrium is observed in Figure \ref{fig:Shock_mixType_Temp} at a high temperature ($\sim$ 100eV), while ion-ion temperature disequilibrium in Figure \ref{fig:Shock_mixType_Temp_lowT} at low temperature ($\sim$ 10eV).

\subsection{Comparison between 1D and 2D numerical results}
\begin{figure} 
\centering
{\includegraphics[width=0.65\textwidth]{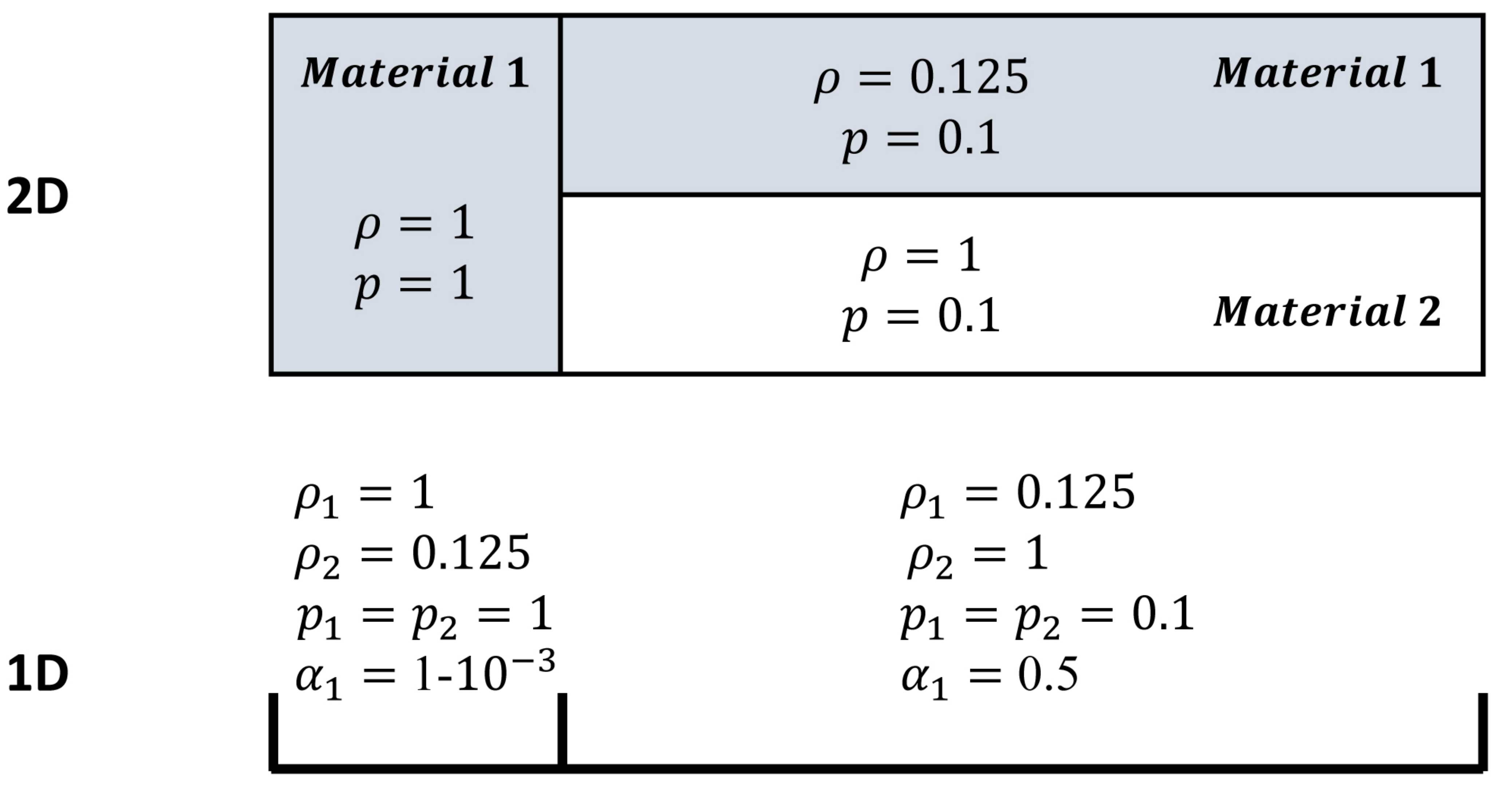}}
\caption{The triple point problem statement in 2D and 1D. 
$\mathbf{2D \; statement}$: The computational domain is of size 7$\times$3. The left sub-domain is occupied by Material 1. An horizontal interface separates the right sub-domain into two equal zones. The top and bottom half zones are filled up with Material 1 and 2, respectively. 
$\mathbf{1D \; statement}$: Assume the computational domain cannot be resolved by the grid and a averaging mapping into 1D is done. The left sub-domain is occupied by Material 1, the right sub-domain by both materials with volume fraction 0.5 for each material.}
\label{fig:2D_VS_1D_ProbStatement}
\end{figure}

The triple-point problem (Figure \ref{fig:2D_VS_1D_ProbStatement}) is usually used to test numerical methods for multi-material flows. Here, we use the 2D numerical results as a reference solution for the 1D results.
Assume that the 2D computational domain cannot be resolved by the grid, we average the 2D initial data into a 1D domain, as shown in Figure \ref{fig:2D_VS_1D_ProbStatement}. 

The 2D numerical results are obtained with a one-velocity Kapila's five-equation model on a uniform grid of 1400$\times$300 cells. The the 2D velocity is mapped into 1D in the following way:
\begin{equation}
    u_{i}^{2D} = \frac{\sum_{j} \rho_{i,j} u_{i,j}}{\sum_{j}\rho_{i,j}}, \nonumber
\end{equation}
where $i,j$ are the cell indexes in $x$ and $y$ direction, respectively.

The 1D numerical results is obtained on a uniform grid of 1400 cells with the averaged initial data in Figure \ref{fig:2D_VS_1D_ProbStatement}, the two-velocity BNZ model, and the proposed relaxation methods in Section \ref{sec:numer_method}.

\begin{figure} 
\centering
\subfloat[$N_{grain}=10$]{\includegraphics[width=0.5\textwidth]{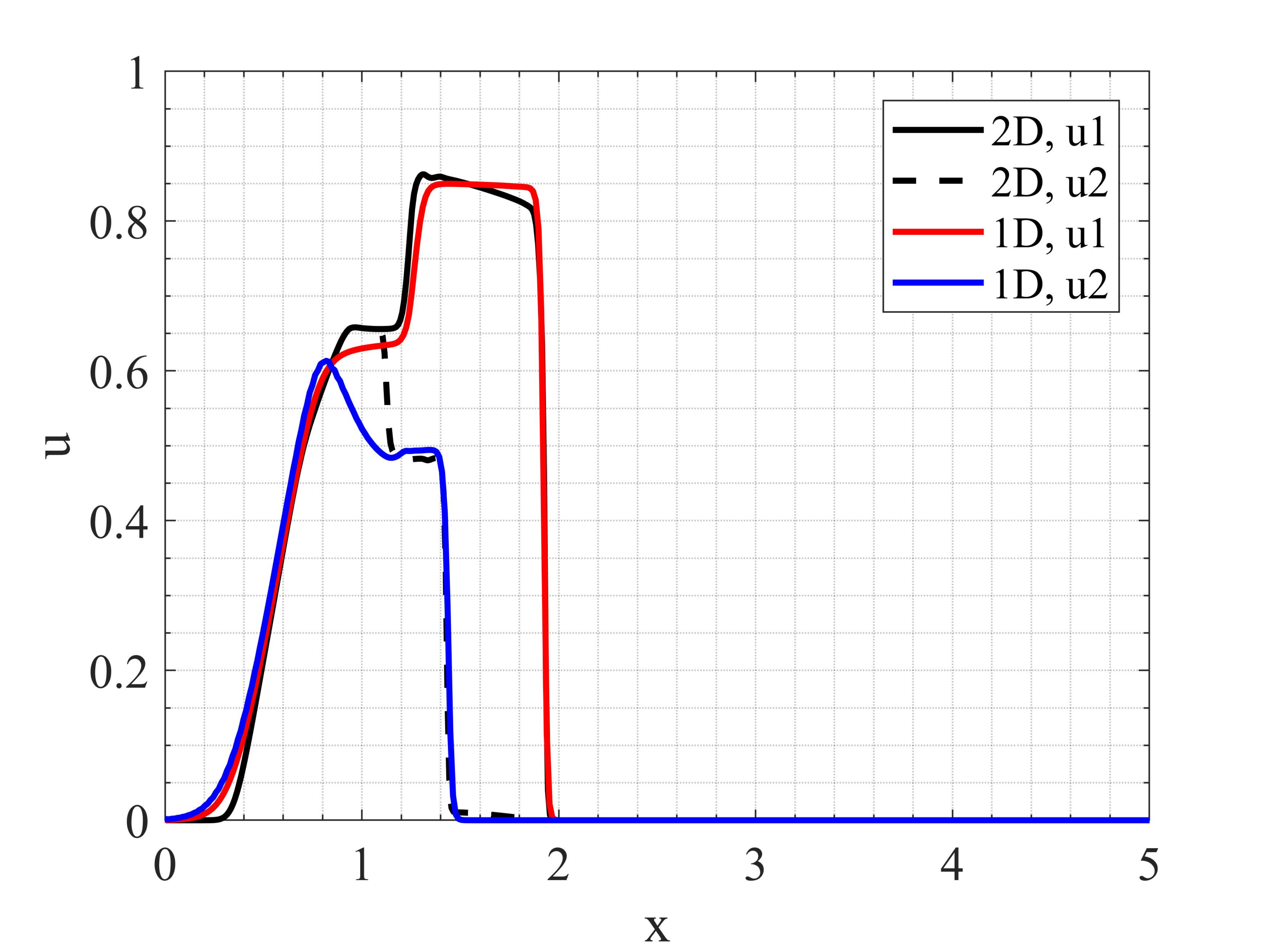}}
\subfloat[$N_{grain}=10^3$]{\includegraphics[width=0.5\textwidth]{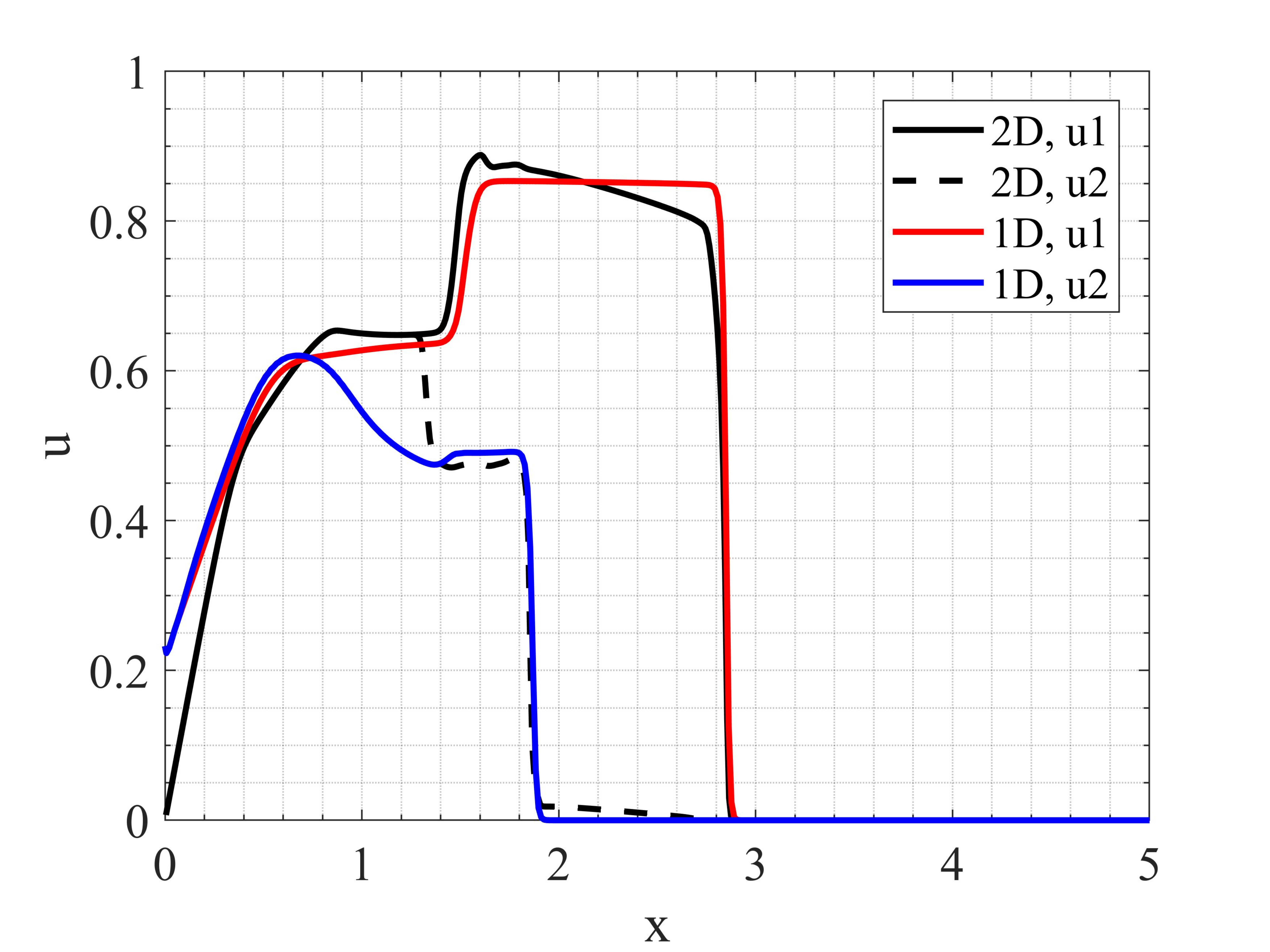}}
\caption{The comparison between 1D and 2D numerical results for component velocities for the triple point problem.}
\label{fig:1D_VS_2D_vel}
\end{figure}

The 1D and 2D numerical results for velocities are compared in Figure \ref{fig:1D_VS_2D_vel}. It can be seen that phase velocities from 1D/2D computations agree quite well. This indicates that the BNZ model is capable of dealing with temperature disequilibrium in grain scale induced by shocks.  However, with the development of the flows towards turbulence, a model with micro-inertia should be included as in \cite{gavrilyuk2002mathematical}.

\subsection{Passage of an ablation shock through the mixing zone}
\begin{figure} 
\centering
{\includegraphics[width=0.55\textwidth]{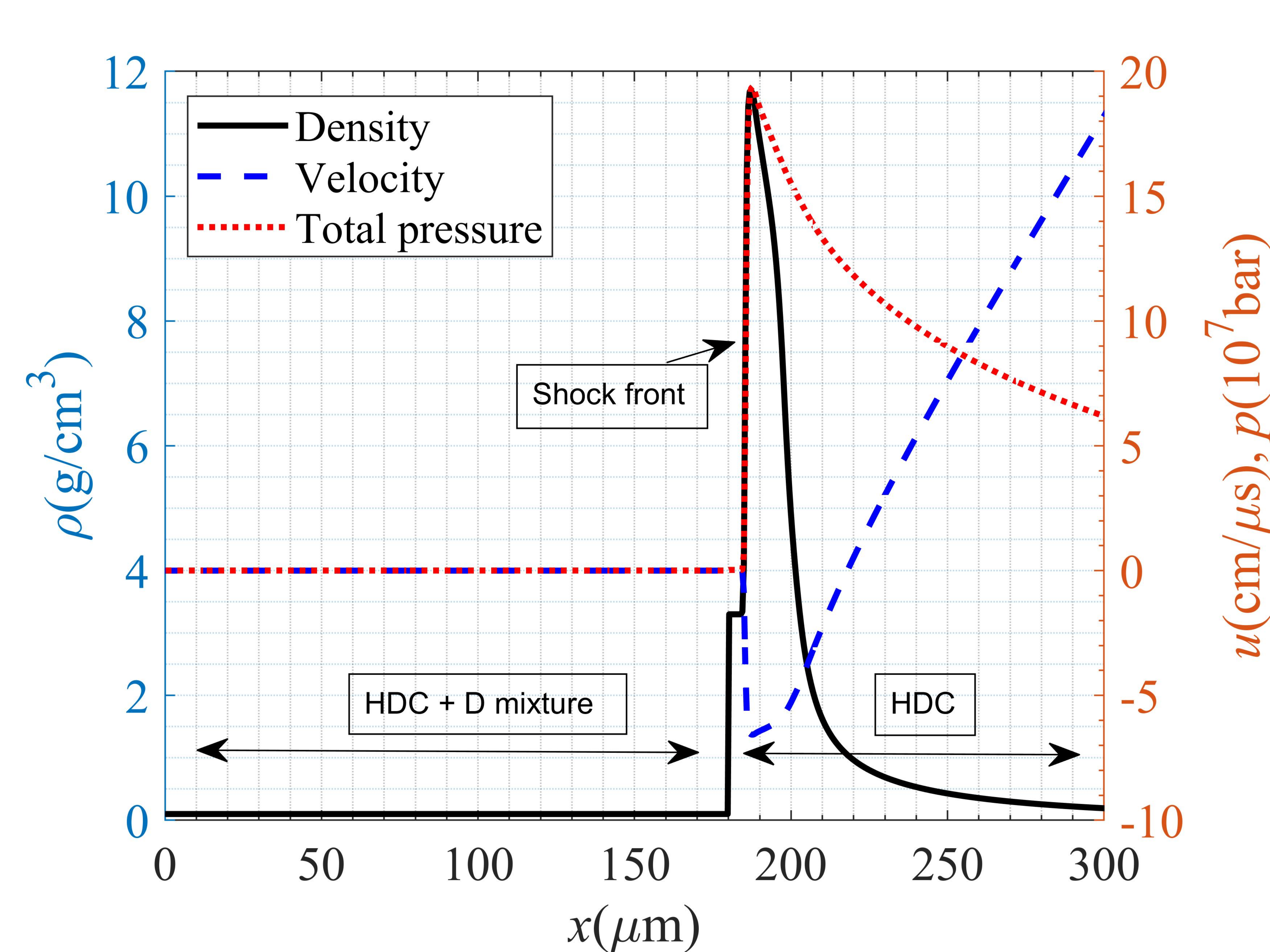}}
\caption{The initial data for the ablation shock - mixture interaction problem.}
\label{fig:Ablation1D_Initial}
\end{figure}

In this section we consider the passage of an ablation shock in high-density-carbon(HDC) through the carbon-deuterium mixture with the initial data shown in Figure \ref{fig:Ablation1D_Initial}. The ionization degree of both materials are assumed to be 1. This problem mimics the mixing at the acceleration stage of ICF. 

The disequilibrium of in the shocked zone depends on the characteristic length and time. From Figure \ref{fig:relaxation_time_grain}, one can determine the characteristic mechanical time $\tau^{u,p}$ and the temperature relaxation time $\tau^{T}$ corresponding to different length scales 0.1${\mu}m$, 1${\mu}m$ and 10${\mu}m$ as follows: 
\begin{enumerate}[label=\roman*.]
    \item $l = 0.1\mu$m,  $\tau^{u,p} = 1\times$10$^{-3}$ns, $\tau^{T} = 1\times 10$ns,
    \item $l = 1\mu$m,  $\tau^{u,p} = 1\times$10$^{-2}$ns, $\tau^{T} = 1\times 10^3$ns,
    \item $l = 10\mu$m,  $\tau^{u,p} = 1\times$10$^{-1}$ns, $\tau^{T} = 1\times 10^5$ns.
\end{enumerate}

\begin{figure} 
\centering
\subfloat[$L_{grain}=0.1{\mu}m$]{\includegraphics[width=0.5\textwidth]{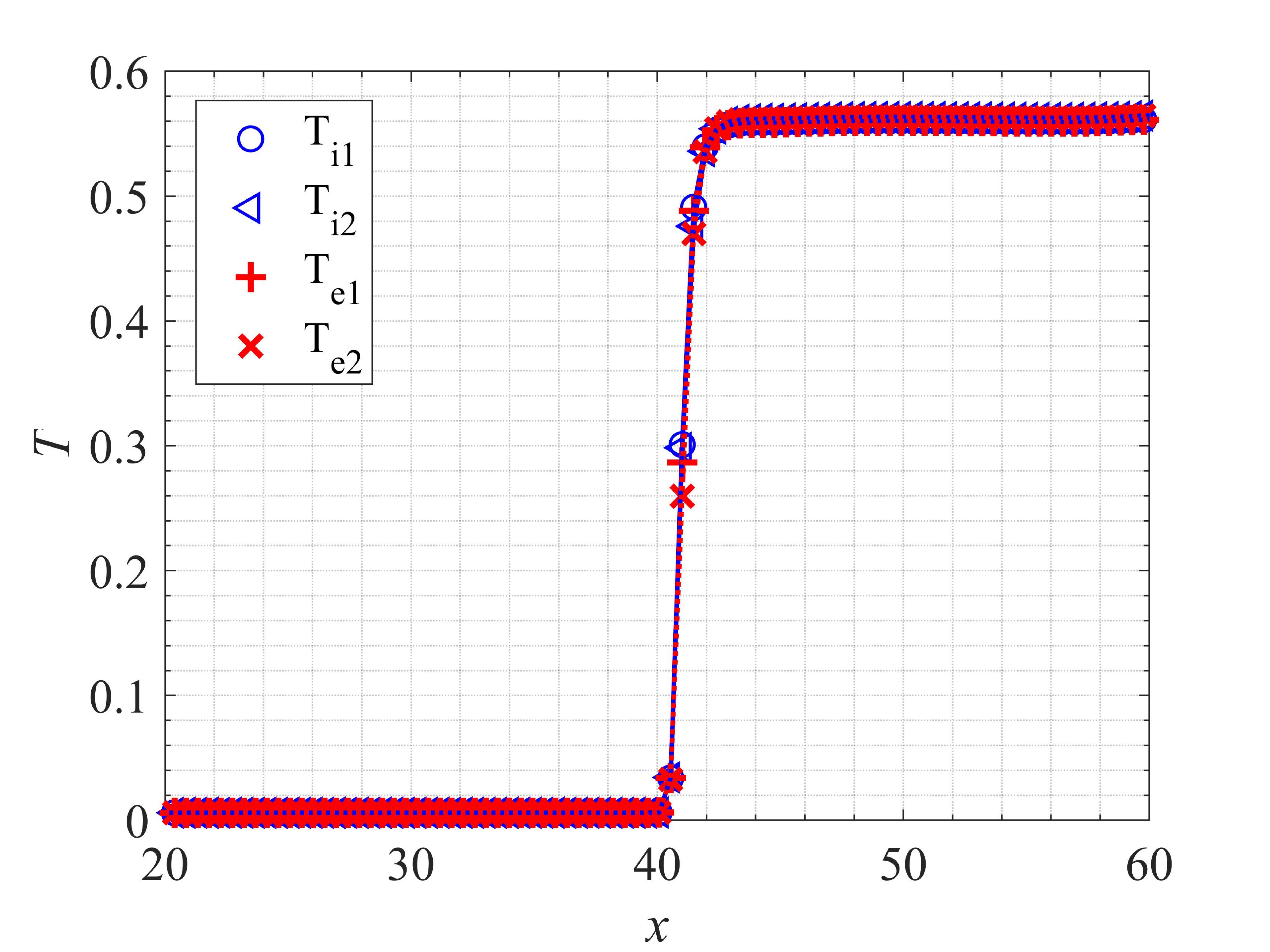}}
\subfloat[$L_{grain}=1{\mu}m$]{\includegraphics[width=0.5\textwidth]{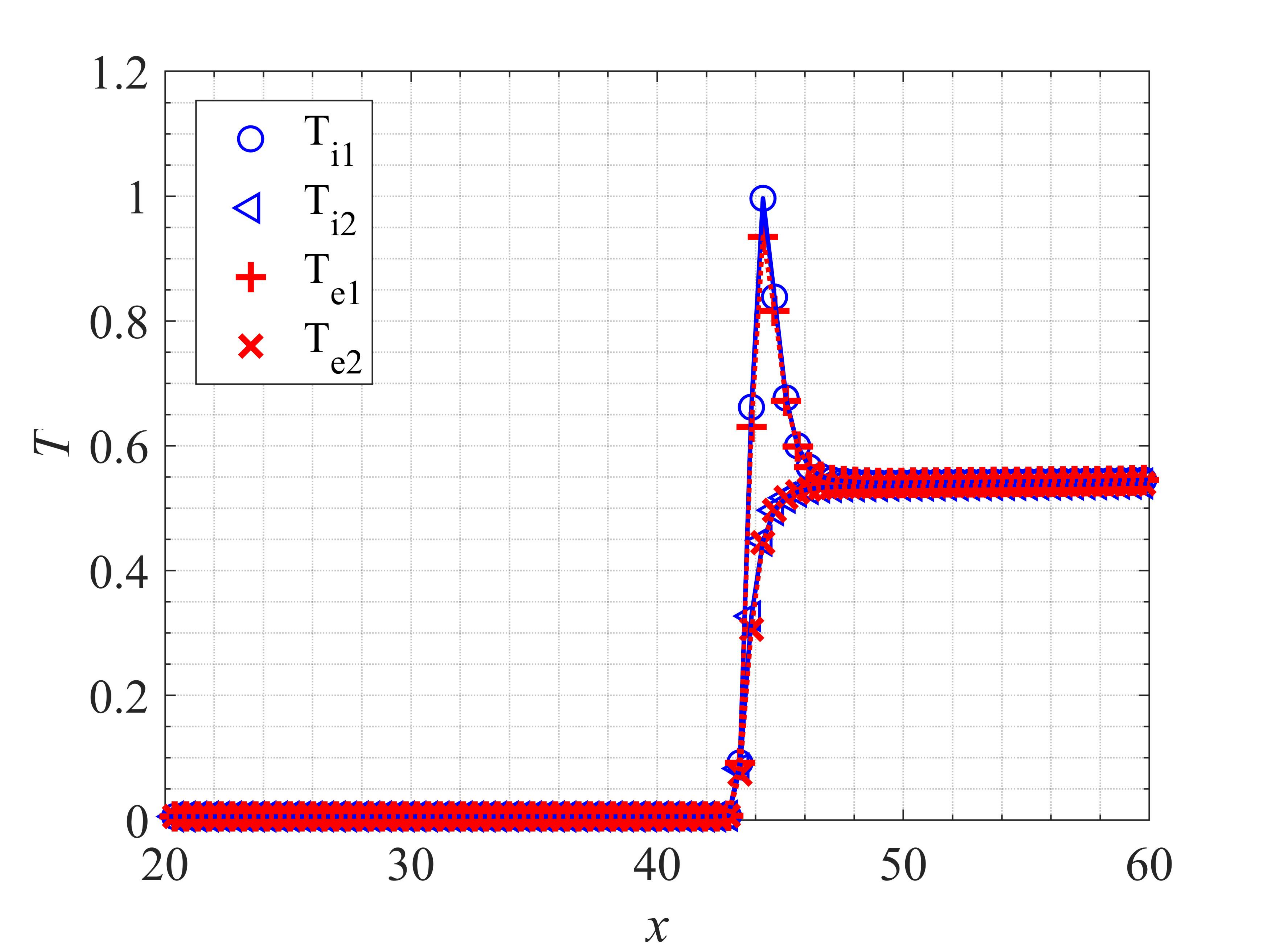}}\\
\subfloat[$L_{grain}=10{\mu}m$]{\includegraphics[width=0.5\textwidth]{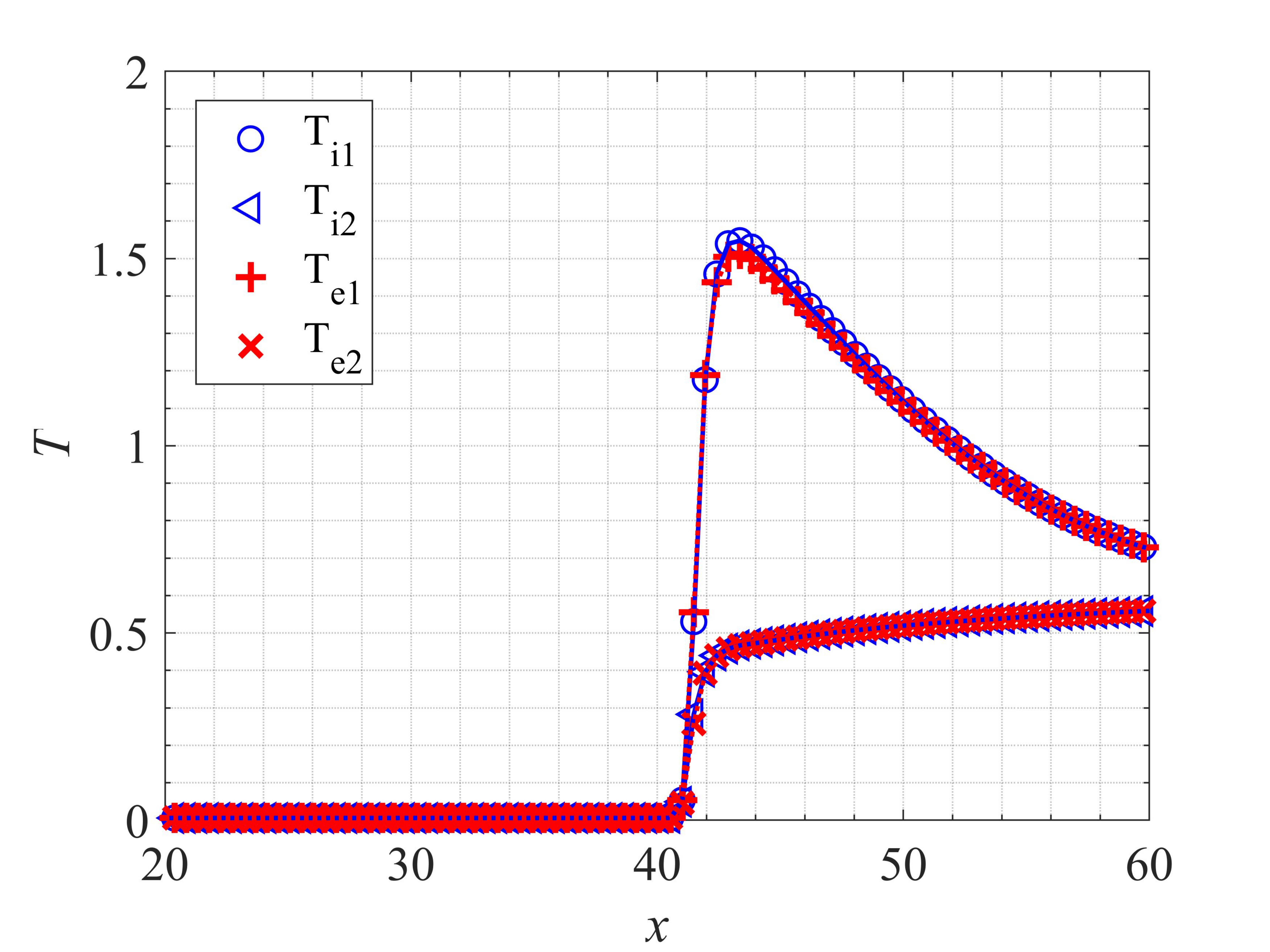}}
\caption{The numerical results for temperatures corresponding to different characteristic sizes.}
\label{fig:Ablation1D_Temp}
\end{figure}

\begin{figure} 
\centering
\subfloat[$L_{grain}=0.1{\mu}m$]{\includegraphics[width=0.5\textwidth]{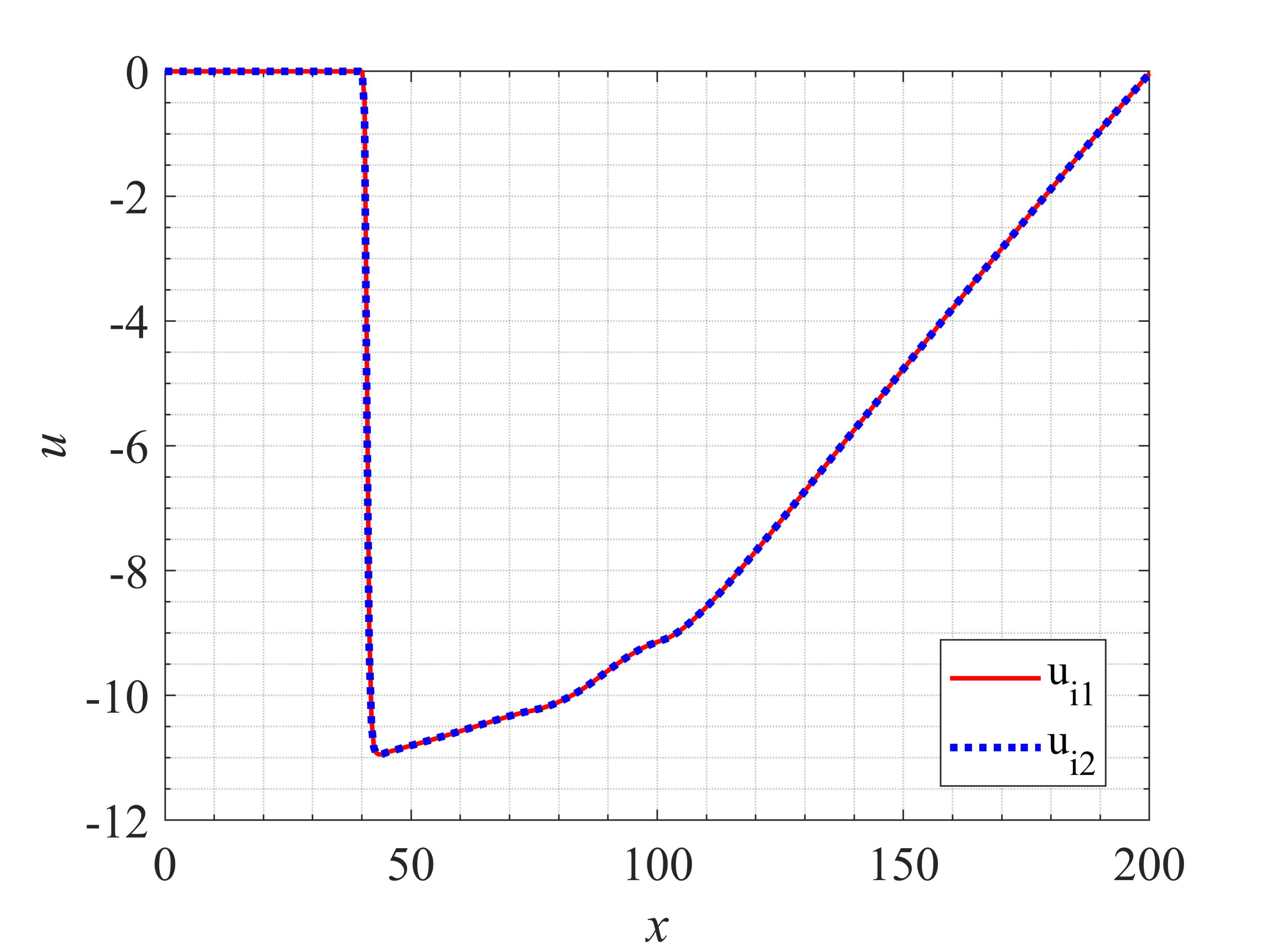}}
\subfloat[$L_{grain}=1{\mu}m$]{\includegraphics[width=0.5\textwidth]{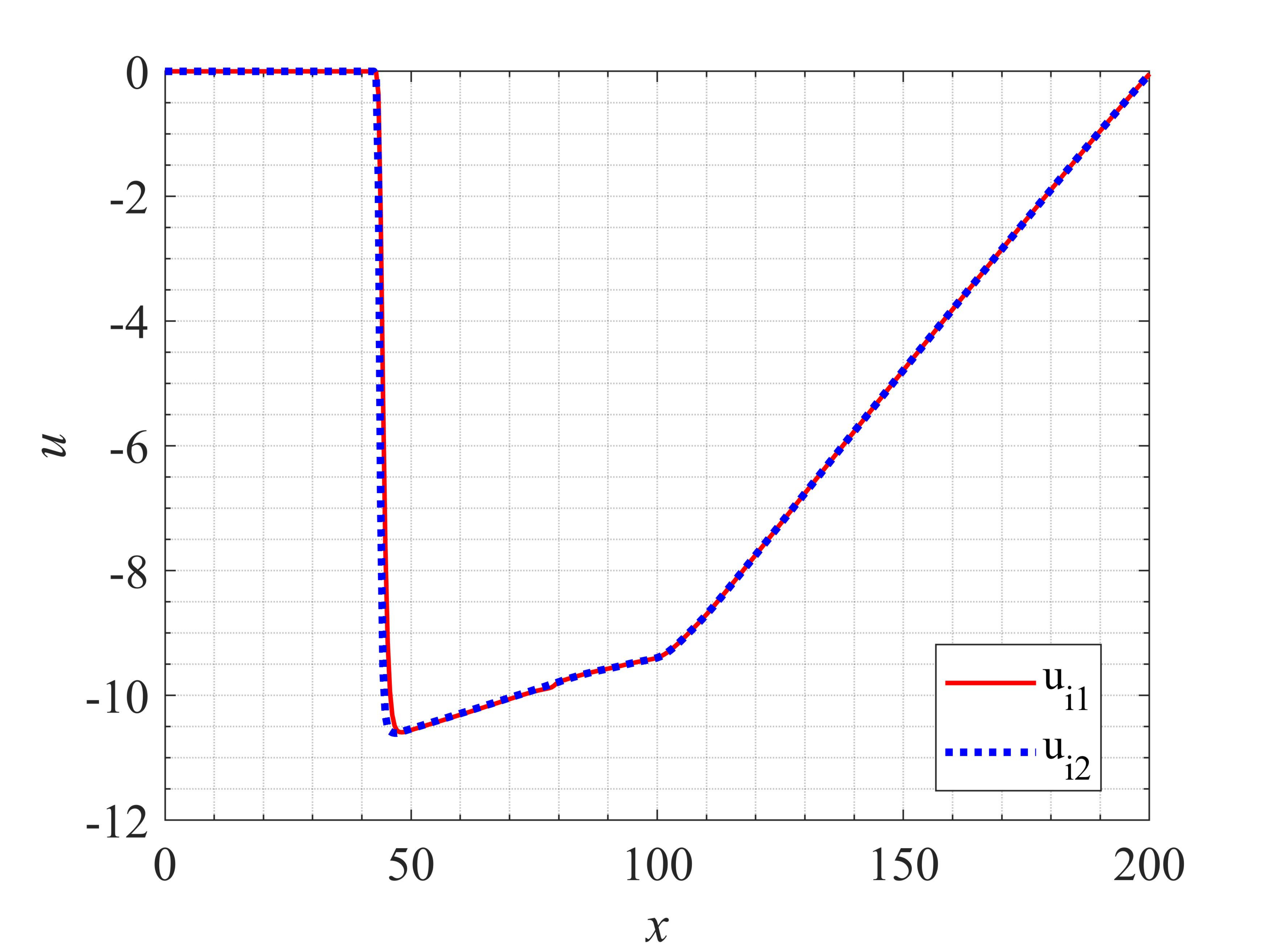}}\\
\subfloat[$L_{grain}=10{\mu}m$]{\includegraphics[width=0.5\textwidth]{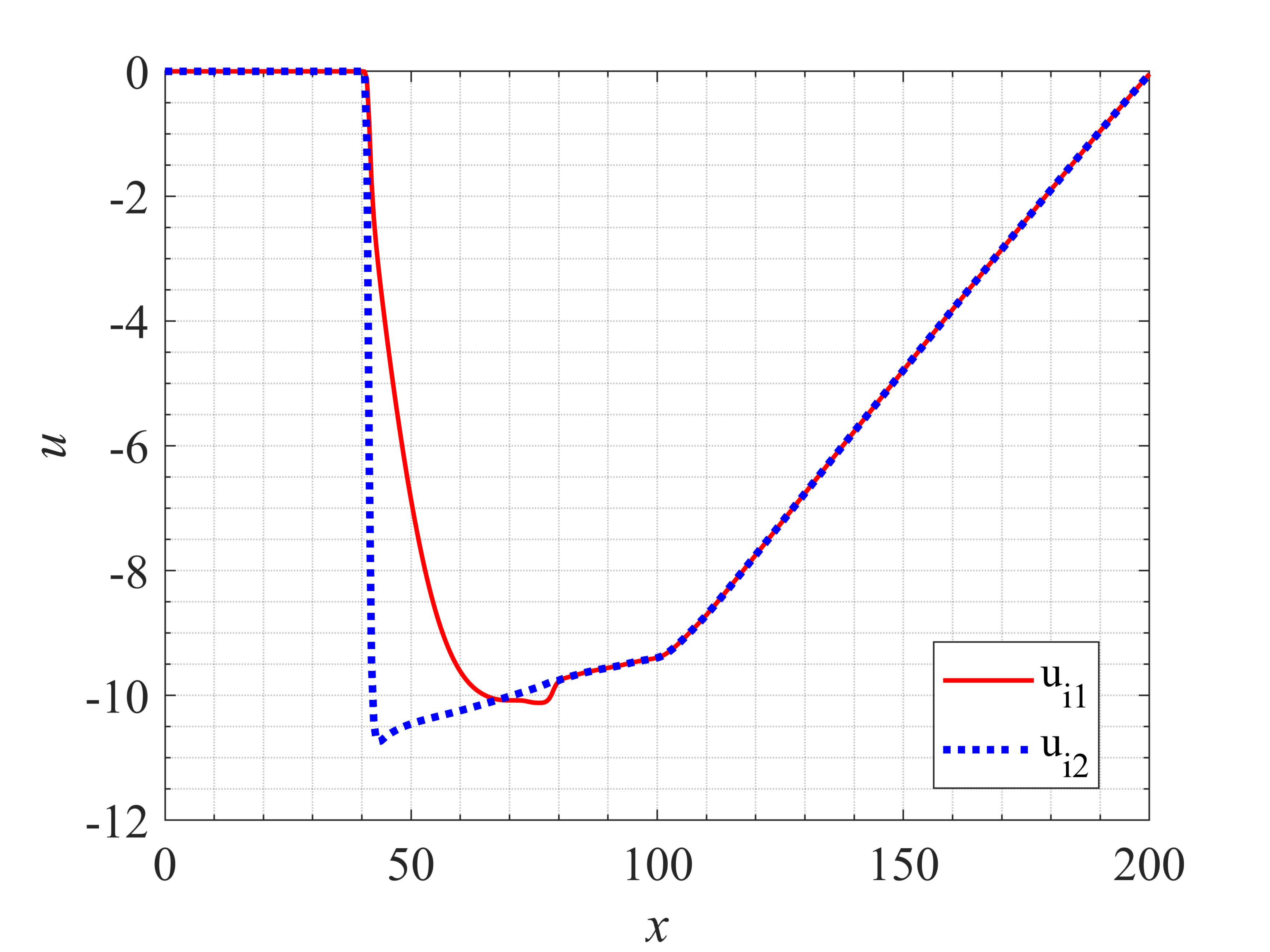}}
\caption{The numerical results for velocities corresponding to different characteristic sizes.}
\label{fig:Ablation1D_Vel}
\end{figure}

\begin{figure} 
\centering
\subfloat[$L_{grain}=0.1{\mu}m$]{\includegraphics[width=0.5\textwidth]{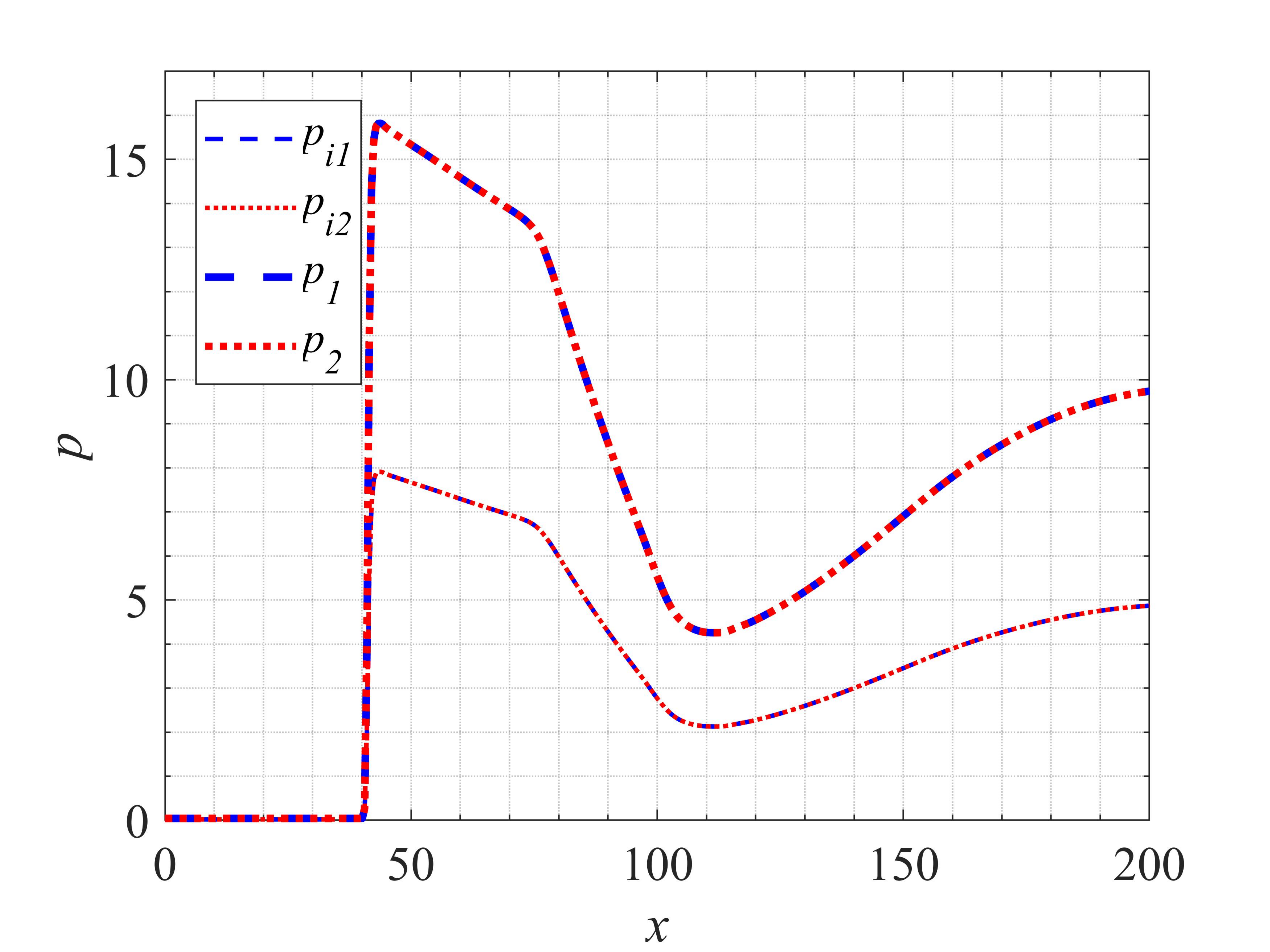}}
\subfloat[$L_{grain}=1{\mu}m$]{\includegraphics[width=0.5\textwidth]{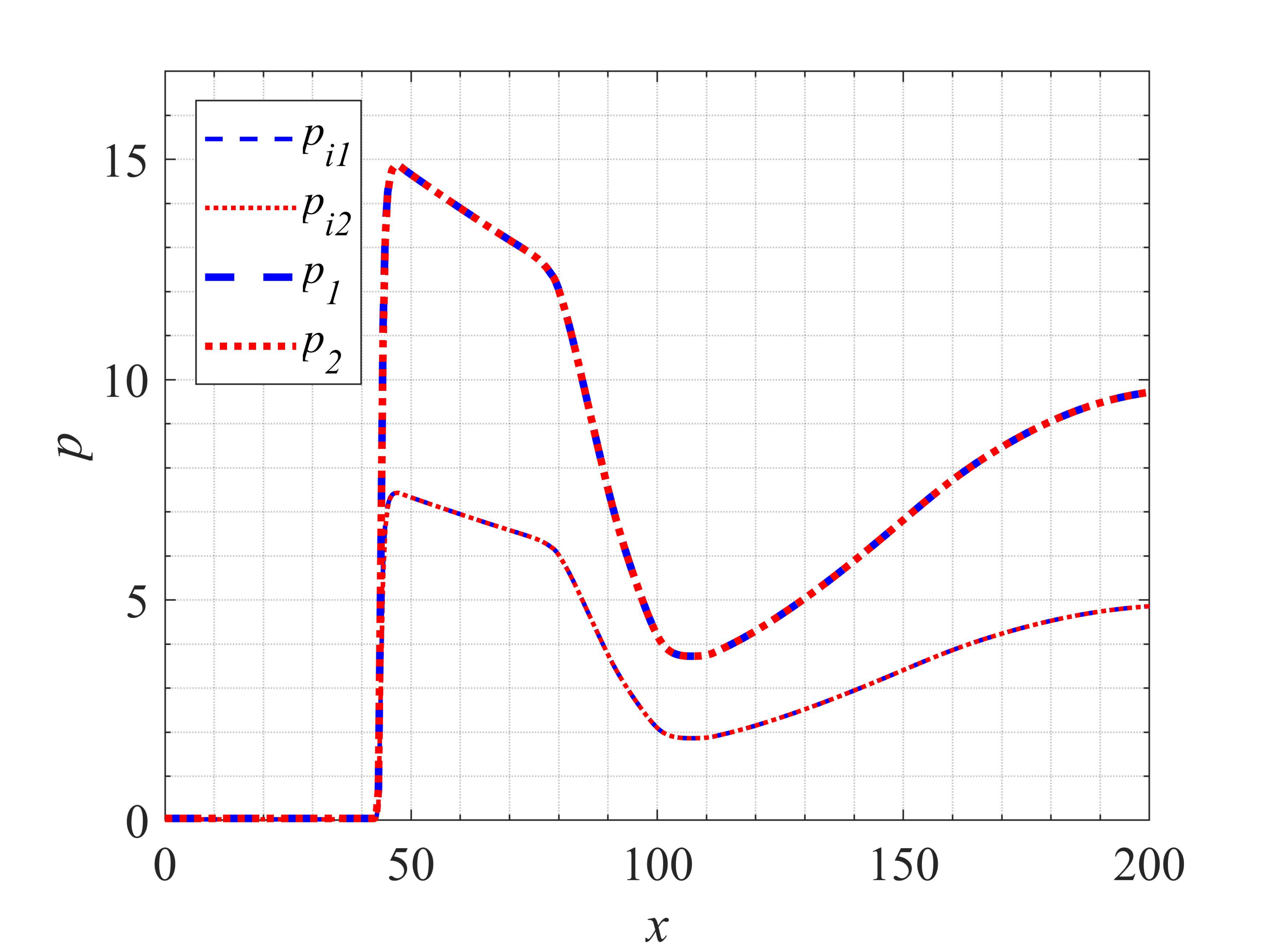}}\\
\subfloat[$L_{grain}=10{\mu}m$]{\includegraphics[width=0.5\textwidth]{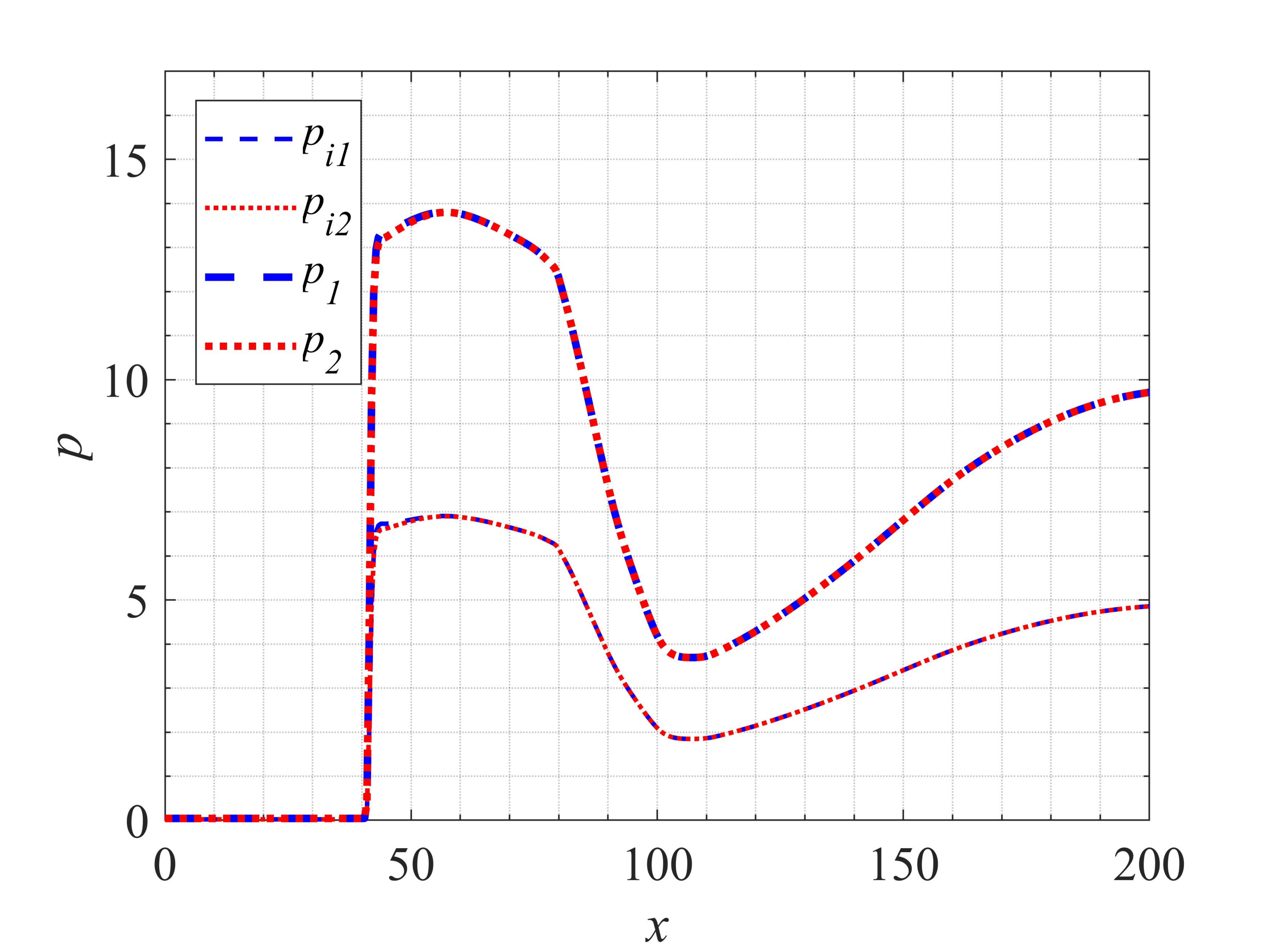}}
\caption{The numerical results for pressures corresponding to different characteristic sizes.}
\label{fig:Ablation1D_Pres}
\end{figure}

\begin{figure} 
\centering
\subfloat[$L_{grain}=0.1{\mu}m$]{\includegraphics[width=0.5\textwidth]{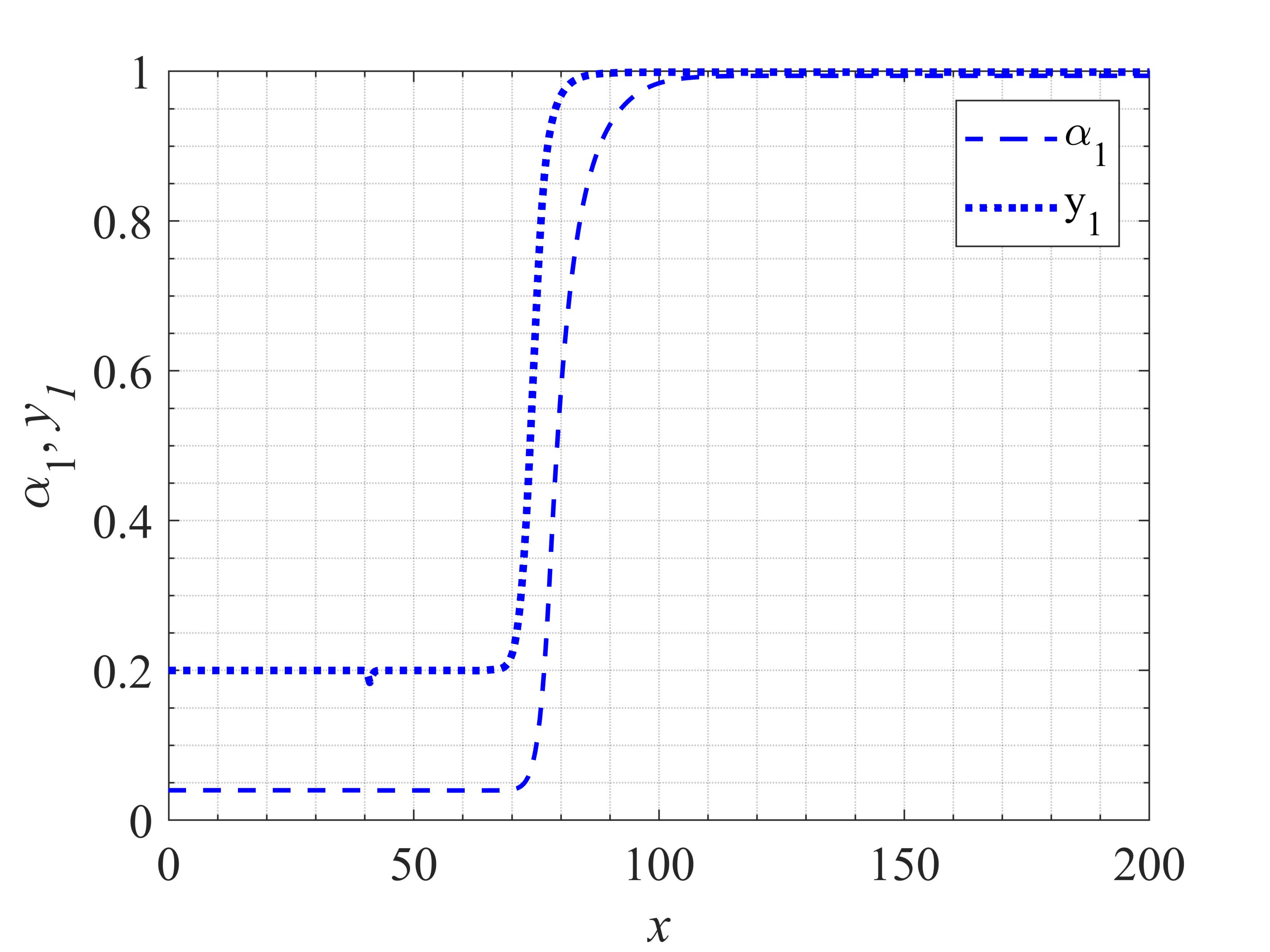}}
\subfloat[$L_{grain}=1{\mu}m$]{\includegraphics[width=0.5\textwidth]{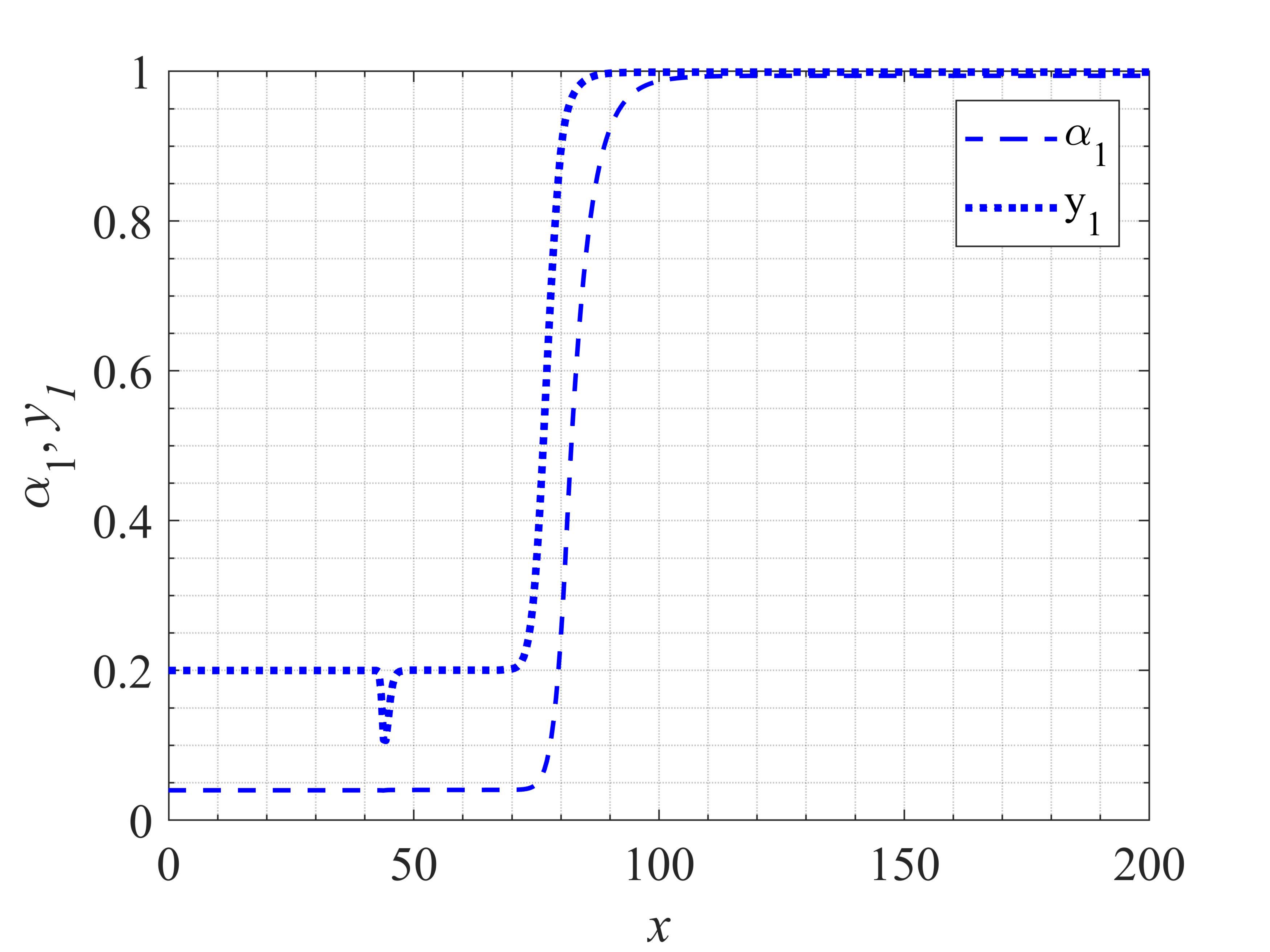}}\\
\subfloat[$L_{grain}=10{\mu}m$]{\includegraphics[width=0.5\textwidth]{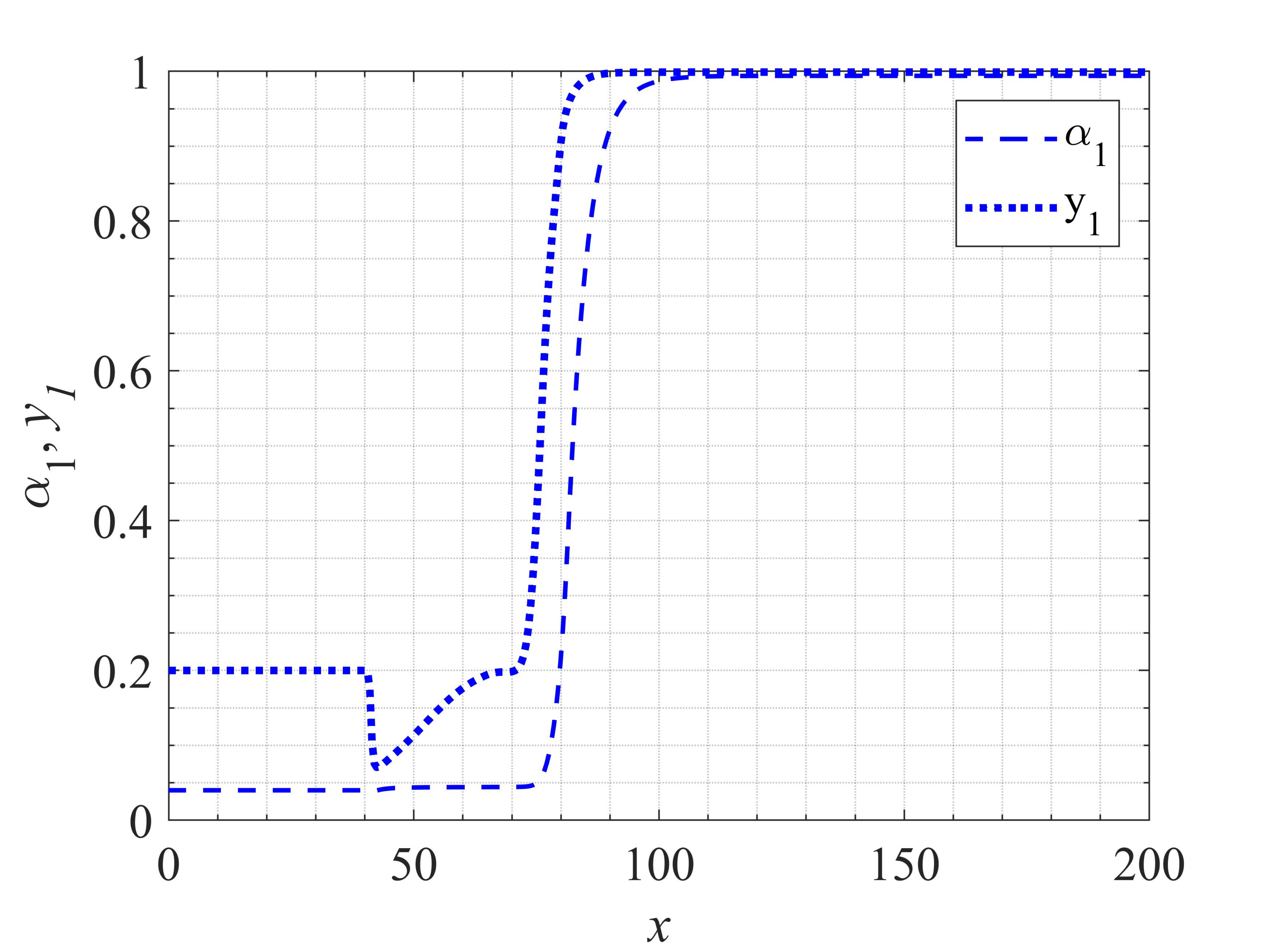}}
\caption{The numerical results for mass/volume fractions corresponding to different characteristic sizes.}
\label{fig:Ablation1D_Alp}
\end{figure}

Assuming these relaxation times, we compare the disequilibria induced by the ablation shock passage. The computations are performed without diffusion processes. The numerical results for temperature, velocity, pressure and volume/mass fractions are displayed in Figures \ref{fig:Ablation1D_Temp}, \ref{fig:Ablation1D_Vel}, \ref{fig:Ablation1D_Pres}, and \ref{fig:Ablation1D_Alp}, respectively. The temperatures relax much slower than pressures and velocities. Thus, the temperature separation can be clearly seen at length scales $1\mu$m and $10\mu$m. The velocities are at disequilibrium in the vicinity of the shock front. This velocity disequilibrium results in the enrichment of the material at the shock front, which can be seen from Figure \ref{fig:Ablation1D_Alp}. Similar phenomenon exist in multi-component flows in atomic mixing \citep[][]{zeldovich1967}. The ion pressures are almost at equilibrium for all three cases. 


\section{Conclusion}
The mixing of different plasmas, especially high-Z pusher into the hotspot, is an adverse factor for degrading the implosion performance of inertial confinement fusion (ICF). In order to simulate such high-energy-density mixing, we have proposed a series of hydrodynamic multiple-plasma flow models. The fully-disequilibrium model is derived from the simplified Boltzmann equation, i.e., the Bhatnagar–Gross–Krook (BGK) model.  For the sake of computational efficiency, the model is then reduced to more practical models. The reductions are performed on the basis of characteristic relaxation time evaluations and asymptotic analysis. Among these models the Baer-Nunziato-Zeldovich (BNZ) model is a combination of the classical Baer-Nunziato (BN) model and Zeldovich model. It consists of 9 equations, 4 temperatures (2 ion temperatures and 2 electron temperatures),  4 pressures (2 ion pressures and 2 electron pressures) and 2 ion velocities. 

The proposed BNZ model is compatible with first and second laws of thermodynamics and capable of dealing with mechanical and thermal disequilibria. The BNZ model is then reduced to the eight-equation model (4 pressures, 4 temperatures, and 1 velocity), the six equation model (2 pressures, 4 temperatures, and 1 velocity) an the five-equation model (2 pressures, 2 temperatures, and 1 velocity). Moreover, in order to simulate the inter-penetration mixing due to velocity disequilibrium, we have derived one-velocity models with diffusion fluxes.  These reduced models are also thermodynamically compatible.

We have performed mathematical analysis of the characteristic structure of the hydrodynamic subsystem of the proposed models. The analysis has confirmed the hyperbolicity of these models. Furthermore, the hyperbolic-parabolic-relaxation PDE system is solved with the fractional step method. The hydrodynamic subsystem is solved with the Godunov finite volume method with the path-conservative HLLEM approximate Riemann solver.   The model and numerical methods are verified against several benchmark problems and  used for simulating the mechanical and thermal disequilibria induced by the ablation shock.

\appendix
\section{The entropy equation of different reduced models}\label{appA}
\subsection{The nine-equation BNZ model}\label{subsec:entropy_BNZ}
For the nine-equation BNZ model, the mixture entropy equation is reduced to the following form
\begingroup
\allowdisplaybreaks
\begin{align}\label{eq:s_9eqn_model}
    && \sum_{l,k} \left[ \dudx{\alpha_{k} \rho_{ik} s_{lk}}{t} + \nabla\cdot\left( \alpha_{k} \rho_{ik} s_{lk} \vc{u}_k \right) + \nabla\cdot\left( \frac{\mathbf{q}_{lk}}{T_{lk}} \right) \right]  \notag\\
    &&= \underbrace{\frac{\mu^{T}_{i1,i2}\left( T_{i1} - T_{i2} \right)^2}{T_{i1} T_{i2}}}_{(1)} + \underbrace{\frac{\mu^{T}_{e1,e2}\left( T_{e1} - T_{e2} \right)^2}{T_{e1} T_{e2}}}_{(2)} + \underbrace{\sum_k \frac{\alpha_{1}\alpha_{k}\widehat{\mu}^{T}_{i1,ek}\left( T_{i1} - T_{ek} \right)^2}{T_{i1} T_{ek}}}_{(3)}  \notag\\
    && + \underbrace{\sum_k \frac{\alpha_{2}\alpha_{k}\widehat{\mu}^{T}_{i2,ek}\left( T_{i2} - T_{ek} \right)^2}{T_{i2} T_{ek}}}_{(4)} + \underbrace{\sum_{l,k} \frac{1}{T_{lk}}\left[ \left(\tilde{\vc{u}}_{I}-\vc{u}_{k}\right) \cdot \mathbf{M}_{lk} + \left(p_{lk}-\tilde{p}_{lI}\right) {\mathcal{P}}_{lk} \right]}_{(5)} \notag\\
    &&+ \underbrace{\sum_{l,k}  \frac{\left(p_{lI}-p_{lk}\right)\left(\vc{u}_{I}-\vc{u}_{k}\right) \cdot \nabla \alpha_{k}}{T_{lk}}}_{(6)}  + \underbrace{\sum_{l,k} \frac{\left( \vc{u}_{I} -\vc{u}_{k} \right) \cdot \left(  \overline{\overline{\Pi}}_{lI} \cdot \nabla \alpha_k \right)}{T_{lk}}}_{(7)}  \notag\\
&&+ \underbrace{\sum_{l,k}   \mathbf{q}_{lk} \cdot  \nabla \left(\frac{1}{T_{lk}} \right)}_{(8)} +  \underbrace{\sum_{l,k}\frac{\mathcal{S}_{lk}}{T_{lk}}}_{(9)}
  +  \underbrace{\sum_{l,k}\frac{\alpha_{k} {Q}_{lk}^{f}}{T_{lk}}}_{(10)}.
\end{align}
\endgroup

Note that here and in the following we have omitted the superscript prime over $s_{ek}^{\prime}$ without causing confusion.

The (1)-(4) terms related to temperature relaxations are obviously non-negative. The (5)-(6) term is non-negative with the BN-type definitions
(eq. (\ref{eq:BN_pI_uI})).  The (7) term is non-negative with the constitutive equation (\ref{eq:PI_I}). The classical Fourier heat flux and Newton visous stress ensure the non-negativity of (8)-(9). The term (10) is non-negative according to eq. (\ref{eq:def_Qrf}).

As for the eight-equation interpenetration model (\ref{eq:reduced_avbn}), we define $\vc{u}_k = \vc{u} + \vc{w}_k$, where $\vc{w}_k$ is determined with the closure law such as Fick's law. Then its entropy equation  is of the same form as (\ref{eq:s_9eqn_model}), only with the terms of order $|\vc{w}_k|^2$ (the first term in (5), term (6) and term (7)) are omitted. The non-negativity of the right hand side terms retains.

\subsection{The eight-equation model}
For the eight-equation model, the mixture entropy equation is reduced to the following form
\begingroup
\allowdisplaybreaks
\begin{eqnarray}
&& \sum_{l,k} \left[ \dudx{\alpha_{k} \rho_{ik} s_{lk}}{t} + \nabla\cdot\left( \alpha_{k} \rho_{ik} s_{lk} \vc{u} \right) + \nabla\cdot\left( \frac{\mathbf{q}_{lk}}{T_{lk}} \right) \right] = \frac{\mu^{T}_{i1,i2}\left( T_{i1} - T_{i2} \right)^2}{T_{i1} T_{i2}}   \nonumber\\
&& + \frac{\mu^{T}_{e1,e2}\left( T_{e1} - T_{e2} \right)^2}{T_{e1} T_{e2}} + \sum_k \frac{\alpha_{1}\alpha_{k}\widehat{\mu}^{T}_{i1,ek}\left( T_{i1} - T_{ek} \right)^2}{T_{i1} T_{ek}} + \sum_k \frac{\alpha_{2}\alpha_{k}\widehat{\mu}^{T}_{i2,ek}\left( T_{i2} - T_{ek} \right)^2}{T_{i2} T_{ek}}  \nonumber\\
&&+ \sum_{l,k} \left[ \frac{\left(p_{lk}-\tilde{p}_{lI}\right) {\mathcal{P}}_{lk}}{T_{lk}} + \mathbf{q}_{lk} \cdot  \nabla \left(\frac{1}{T_{lk}} \right) +  \frac{\mathcal{S}_{lk}+ \alpha_{k} {Q}_{lk}^{f}}{T_{lk}}   \right].
\end{eqnarray}
\endgroup

The non-negativity of the right-hand side terms in the above equation can be proved as in subsection \ref{subsec:entropy_BNZ}.

\subsection{The six-equation KZ model}
For the six-equation KZ model, the mixture entropy equation is reduced to the following form
\begin{eqnarray}
&& \sum_{l,k} \left[ \dudx{\alpha_{k} \rho_{ik} s_{lk}}{t} + \nabla\cdot\left( \alpha_{k} \rho_{ik} s_{lk} \vc{u} \right) + \nabla\cdot\left( \frac{\mathbf{q}_{lk}}{T_{lk}} \right) \right]  \nonumber\\ 
&&= \frac{\mu^{T}_{i1,i2}\left( T_{i1} - T_{i2} \right)^2}{T_{i1} T_{i2}} + \frac{\mu^{T}_{e1,e2}\left( T_{e1} - T_{e2} \right)^2}{T_{e1} T_{e2}} + \sum_k \frac{\alpha_{1}\alpha_{k}\widehat{\mu}^{T}_{i1,ek}\left( T_{i1} - T_{ek} \right)^2}{T_{i1} T_{ek}} \nonumber\\
&&+ \sum_k \frac{\alpha_{2}\alpha_{k}\widehat{\mu}^{T}_{i2,ek}\left( T_{i2} - T_{ek} \right)^2}{T_{i2} T_{ek}} + \sum_{l,k} \left[  \mathbf{q}_{lk} \cdot  \nabla \left(\frac{1}{T_{lk}} \right) +  \frac{\mathcal{S}_{lk} + \alpha_{k} {Q}_{lk}^{f}}{T_{lk}}   \right].
\end{eqnarray}

The non-negativity of the right-hand side terms in the above equation can be proved as in subsection \ref{subsec:entropy_BNZ}.

\section{The Jacobian of hydrodynamic sub-system of different reduced models, and their eigenvalues}\label{appB}
The hydrodynamic part of the derived models can be recast into the following form
\begin{equation}
    \dudx{\vc{Z}}{t} + \vc{A}\dudx{\vc{Z}}{x} = \vc{0},
\end{equation}
where $\vc{Z}$ is the vector of unknowns, and $\vc{A}$ is the Jacobian. 

\subsection{The nine-equation BNZ model}
With respect primitive variables
\[ \vc{Z} = \left[ \alpha_1 \;\; \rho_1 \;\; u_1 \;\; p_{i1} \;\; \rho_2 \;\; u_2 \;\; p_{i2} \;\; p_{e1} \;\; p_{e2} \right], \]
the Jacobian for the nine-equation BNZ model is
\begin{equation}\label{eq:aaa}
 \vc{A} = \left[
  \begin{array}{ccccccccc}
    {u}_I & 0 & 0 & 0 & 0 & 0 & 0 & 0 & 0 \\
    \frac{\rho_1\left(u_I - u_1\right)}{\alpha_1} & u_1 & \rho_1 & 0 & 0 & 0 & 0 & 0 & 0 \\
    \frac{p_{i1}+p_{e1}-p_{iI}-p_{eI}}{\alpha_1 \rho_1} & 0 & u_1 & 1/\rho_1 & 0 & 0 & 0 & 1/\rho_1 & 0 \\
    \frac{\rho_1 a_{I1}^2 \left( u_I - u_1 \right)}{\alpha_1} & 0 & \rho_1 a_{i1}^2 & 0 & 0 & 0 & 0 & 0 & 0 \\
    \frac{\rho_2\left(u_I - u_2\right)}{\alpha_2} & 0 & 0 & 0 & u_2 & \rho_2 & 0 & 0 & 0 \\
    \frac{p_{i2}+p_{e2}-p_{iI}-p_{eI}}{\alpha_2 \rho_2} & 0 & 0 & 0 & 0 & u_2 & 1/\rho_2 & 0 & 1/\rho_2 \\
    \frac{\rho_2 c_{I2}^2 \left( u_I - u_2 \right)}{\alpha_2} & 0 & 0 & 0 & 0 & \rho_2 a_{i2}^2 & u_2 & 0 & 0 \\
    \frac{\rho_1 a_{eI1}^2 \left( u_I - u_1 \right)}{\alpha_1} & 0 & \rho_1 a_{e1}^2 & 0 & 0 & 0 & 0 & 0 & 0 \\
    \frac{\rho_2 c_{eI2}^2 \left( u_I - u_2 \right)}{\alpha_2} & 0 & 0 & 0 & 0 & \rho_2 a_{e2}^2 & 0 & 0 & u_2 \\
  \end{array}
\right].
\end{equation}

The eigenvalues of the above Jacobian
\begin{equation}\label{eq:bbb}
  \left[ u_1, \; u_1, \; u_2, \; u_2, \; u_1\pm\sqrt{a_{i1}^2 + a_{e1}^2}, \; u_2\pm\sqrt{a_{i2}^2 + a_{e2}^2}, \; u_I   \right].
\end{equation}

\subsection{The eight-equation model}
With respect primitive variables
\[ \vc{Z} = \left[ \alpha_1 \;\; \rho_1 \;\; \rho_2 \;\; u \;\; p_{i1} \;\; p_{i2} \;\; p_{e1} \;\; p_{e2} \right], \]
the Jacobian for the eight-equation model is
\begin{equation}\label{eq:aaaa2}
\left(
  \begin{array}{cccccccc}
    u & 0 & 0 & 0 & 0 & 0 & 0 & 0 \\
    0 & u & 0 & \rho_1 & 0 & 0 & 0 & 0 \\
    0 & 0 & u & \rho_2 & 0 & 0 & 0 & 0 \\
    \frac{p_{i1}+p_{i2}-p_{e1}-p_{e2}}{\rho} & 0 & 0 & u & \frac{\alpha_1}{\rho} & \frac{\alpha_2}{\rho} & \frac{\alpha_1}{\rho} & \frac{\alpha_2}{\rho} \\
    0 & 0 & 0 & \rho_1 a_{i1}^2 & u & 0 & 0 & 0 \\
    0 & 0 & 0 & \rho_2 a_{i2}^2 & 0 & u & 0 & 0 \\
    0 & 0 & 0 & \rho_1 a_{e1}^2 & 0 & 0 & u & 0 \\
    0 & 0 & 0 & \rho_2 a_{e2}^2 & 0 & 0 & 0 & u
  \end{array}
  \right).
\end{equation}

The eigenvalues of the above Jacobian
\begin{equation}\label{eq:bbb3}
  \left[ u, \; u, \; u, \; u, \;u, \;u, \; u\pm\sqrt{y_1 a_{i1}^2 + y_2 a_{i2}^2 + y_1 a_{e1}^2 + y_2 a_{e2}^2}  \right].
\end{equation}

\subsection{The six-equation KZ model}
With respect primitive variables
\[ \vc{Z} = \left[ \alpha_1 \;\; \rho_1 \;\; \rho_2 \;\; u \;\; p_{i} \;\; p_{e} \right], \]
the Jacobian for the six-equation model is
\begin{equation}\label{eq:ddd}
  \left(
    \begin{array}{cccccc}
      u & 0 & 0 & \alpha_1 \frac{A_{i} - A_{i1}}{A_{i1}} & 0 & 0 \\
      0 & u & 0 & \rho_1 & 0 & 0 \\
      0 & 0 & u & \rho_2 & 0 & 0 \\
      0 & 0 & 0 & u & 1/\rho & 1/\rho \\
      0 & 0 & 0 & \rho c_{i}^2 & u & 0 \\
      0 & 0 & 0 & \rho c_e^2 & 0 & u \\
    \end{array}
  \right).
\end{equation}

The eigenvalues of the above Jacobian
\begin{equation}\label{eq:bbb4}
  \left[ u, \; u, \; u, \; u, \; u\pm\sqrt{c_{i}^2 + c_{e}^2}  \right].
\end{equation}

\section*{Acknowledgement}
This work was performed under the auspices of the National Natural Science Foundation of China [grant number 12205022].

\bibliography{references}
\bibliographystyle{elsarticle-num}


\end{document}